\documentclass[twocolumn,
nofootinbib,superscriptaddress,aps,10pt,longbibliography]{revtex4-1}

\usepackage[dvipsnames]{xcolor}
\definecolor{colorforS}{rgb}{0.1, 0.5, 0.3.}
\definecolor{colorforU}{rgb}{0.05, 0, 1}
\definecolor{colorforW}{rgb}{0.9, 0, 0}

\def\colU{colorforU!70}

\def\minsize{18}
\def\innersep{2}

\usepackage{amsmath,braket,amssymb}
\usepackage[final]{graphicx}
\usepackage{subfigure}
\usepackage[english]{babel}
\usepackage[utf8]{inputenc}
\usepackage{mathtools}
\usepackage{empheq}

\usepackage{bbm}
\usepackage{times}

\usepackage{enumitem}
\usepackage{slashed}
\usepackage{graphicx}
\usepackage{tikz}

    \usetikzlibrary{decorations.markings}

\usepackage{bm}
\usepackage[colorlinks=true,citecolor=blue,linkcolor=blue,urlcolor=blue]{hyperref}
\usepackage{dsfont}
\usepackage{changepage}
\usepackage{array}
\usepackage{soul}
\usepackage[capitalise]{cleveref}
\crefname{section}{Sec.}{Secs.}
\Crefname{section}{Sec.}{Secs.}
\usepackage{xifthen}
\usepackage{xargs}

\newcommand{\Eq}[1]{Eq.~(\ref{#1})}
\newcommand{\Eqs}[1]{Eqs.~(\ref{#1})}
\newcommand{\eqK}[1]{(\ref{#1})}
\newcommand{\ds}[1]{\displaystyle }
\newcommand{\half}{\frac12}
\newcommand{\bea}{\begin{eqnarray}}
\newcommand{\eea}{\end{eqnarray}}
\newcommand{\beq}{\begin{equation}}
\newcommand{\eeq}{\end{equation}}

\newcommand{\nn}{\nonumber}

\newcommand{\nott}[1]{}
\newcommand{\ca}[1]{{\cal #1}}
\newcommand{\be}{\begin{equation}}
\newcommand{\ee}{\end{equation}}

\usepackage{bm}
 \renewcommand{\vec}[1]{\boldsymbol{\mathbf{#1}}}

\graphicspath{{./}{./figures/}}

\newcommand{\bit}{\begin{itemize}}
\newcommand{\eit}{\end{itemize}}

\newcommand{\f}{\frac}
\renewcommand{\>}{\right\rangle}
\newcommand{\<}{\left\langle}
\newcommand{\ba}{\begin{align}}
\newcommand{\ea}{\end{align}}
\newcommand{\bi}{\begin{itemize}}
\newcommand{\ei}{\end{itemize}}
\newcommand{\lf}{\left(}
\newcommand{\ri}{\right)}

\newcommand{\Fig}[1]{\includegraphics[width=\columnwidth]{./#1}}
\newcommand{\fig}[2]{\includegraphics[width=#1\columnwidth]{./#2}}

\def\a{\alpha}
\def\b{\beta}

\DeclareMathAlphabet{\mymathbb}{U}{BOONDOX-ds}{m}{n}

\renewcommand{\log}{\ln}

\usepackage{comment}

\usepackage{braket}
\usepackage[normalem]{ulem}

\newcommand{\nocontentsline}[3]{}
\newcommand{\tocless}[2]{\vspace{2em}\bgroup\let\addcontentsline=\nocontentsline#1{#2}\egroup}

\newcommand{\triagdiag}{{\parbox{0.8cm}{{\begin{tikzpicture}[scale=1]
\coordinate (x1) at (0,0) ;
\coordinate (x2) at  (0.7,0) ;
\coordinate (x3) at  (0.35,0.5) ;
\fill (x1) circle (1.5pt);
\fill (x2) circle (1.5pt);
\fill (x3) circle (1.5pt);
\draw [black] (x1) -- (x2) -- (x3) -- (x1);
\end{tikzpicture}}}}}
\newcommand{\smalltriagdiag}{\parbox{0.4cm}{\scalebox{0.5}{\triagdiag}}}

\newcommand{\propdiag}{{{\parbox{1.cm}{{\begin{tikzpicture}[scale=1]
\coordinate (x1) at (0,0) ;
\coordinate (x2) at  (0.6,0) ;
\coordinate (x1p) at  (-.2,0) ;
\coordinate (x2p) at  (0.8,0) ;
\fill (x1) circle (1.5pt);
\fill (x2) circle (1.5pt);
\draw (.3,0) circle (3mm);
\draw [black] (x1) -- (x1p);
\draw [black] (x2) -- (x2p);
\end{tikzpicture}}}}}}
\newcommand{\smallpropdiag}{\parbox{0.5cm}{\scalebox{0.5}{\propdiag}}}

\newcommand{\diagDone}{{{\parbox{0.8cm}{{\begin{tikzpicture}[scale=1]
\coordinate (x1) at (0,0) ;
\coordinate (x2) at  (0.7,0) ;
\coordinate (x3) at  (0,0.5) ;
\coordinate (x4) at (0.7,0.5) ;
\coordinate (x5) at (0.35,1) ;
\fill (x1) circle (1.5pt);
\fill (x2) circle (1.5pt);
\fill (x3) circle (1.5pt);
\fill (x4) circle (1.5pt);
\fill (x5) circle (1.5pt);
\draw [black] (x1) -- (x2) -- (x4) -- (x3) -- (x1);
\draw [black] (x4)--(x5)--(x3);
\end{tikzpicture}}}}}}
\newcommand{\smalldiagDone}{\parbox{0.4cm}{\scalebox{0.5}{\diagDone}}}

\newcommand{\diagDthree}{{{\parbox{1.4cm}{{\begin{tikzpicture}[scale=1]
\coordinate (x0) at (-.2,0.35);
\coordinate (x1) at (0,0.35) ;
\coordinate (x2) at  (0.5,0) ;
\coordinate (x3) at (0.5,0.7) ;
\coordinate (x4) at (1,0.35) ;
\coordinate (x5) at (1.2,0.35) ;
\fill (x1) circle (1.5pt);
\fill (x2) circle (1.5pt);
\fill (x3) circle (1.5pt);
\fill (x4) circle (1.5pt);
\draw [black] (x0)-- (x1)-- (x2) -- (x3) --(x1) ;
\draw [black] (x2)--(x4)--(x3);
\draw [black] (x4)--(x5);
\end{tikzpicture}}}}}}
\newcommand{\smalldiagDthree}{\parbox{0.7cm}{\scalebox{0.5}{\diagDthree}}}

\newcommand{\diagDfour}{{\parbox{1.6cm}{{\begin{tikzpicture}[scale=1]
\coordinate (x0) at (-.2,0);
\coordinate (x1) at (0,0) ;
\coordinate (x2) at  (0.3,0) ;
\coordinate (x3) at (0.9,0) ;
\coordinate (x4) at (1.2,0) ;
\coordinate (x5) at (1.4,0) ;
\fill (x1) circle (1.5pt);
\fill (x2) circle (1.5pt);
\fill (x3) circle (1.5pt);
\fill (x4) circle (1.5pt);
\draw [black] (x0)-- (x1)-- (x2)  ;
\draw [black] (x3)--(x4)--(x5);
\draw (.6,0) circle (3mm);
\draw (1.2,0) arc (0:180:6mm);
\end{tikzpicture}}}}}
\newcommand{\smalldiagDfour}{\parbox{0.8cm}{\scalebox{0.5}{\diagDfour}}}

\newcommand{\diagDfiveAmp}{{\parbox{1.6cm}{{\begin{tikzpicture}[scale=1]
\coordinate (x0) at (-.2,0);
\coordinate (x1) at (0,0) ;
\coordinate (x2) at  (0.3,0) ;
\coordinate (x3) at (0.9,0) ;
\coordinate (x4) at (1.2,0) ;
\coordinate (x5) at (1.4,0) ;
\coordinate (x6) at (0.6,0.6) ;
\coordinate (x7) at (0.6,0.8) ;
\fill (x1) circle (1.5pt);
\fill (x2) circle (1.5pt);
\fill (x3) circle (1.5pt);
\fill (x4) circle (1.5pt);
\fill (x6) circle (1.5pt);
\draw [black] (x0)-- (x1)-- (x2)  ;
\draw [black] (x3)--(x4)--(x5);
\draw (.6,0) circle (3mm);
\draw (1.2,0) arc (0:180:6mm);
\end{tikzpicture}}}}}

\newcommand{\diagDsix}{{\parbox{1.1cm}{{\begin{tikzpicture}[scale=1]
\coordinate (x1) at (0,0) ;
\coordinate (x2) at  (0.5,0) ;
\coordinate (x3) at  (1,0) ;
\coordinate (x4) at (0.5,0.5) ;
\coordinate (x5) at (0.5,-0.5) ;
\coordinate (o1) at (0.5,0.2) ;
\coordinate (o2) at (0.7,-0.5) ;
\coordinate (o3) at (0.3,0.5) ;
\fill (x1) circle (1.5pt);
\fill (x2) circle (1.5pt);
\fill (x3) circle (1.5pt);
\fill (x4) circle (1.5pt);
\fill (x5) circle (1.5pt);
\draw  (x1) -- (x2) -- (x3) -- (x4) -- (x1)--(x5)--(x3);
\draw  (o1) -- (x2);
\draw  (o3) -- (x4);
\draw  (o2) -- (x5);
\end{tikzpicture}}}}}
\newcommand{\smalldiagDsix}{\parbox{0.55cm}{\scalebox{0.5}{\diagDsix}}}

\newcommand{\triagdiagCTtwoLegs}{{\parbox{1.1cm}{{\begin{tikzpicture}[scale=1]
\coordinate (x1) at (0,0) ;
\coordinate (x1p) at (-.2,0) ;
\coordinate (x2) at  (0.7,0) ;
\coordinate (x2p) at  (0.9,0) ;
\coordinate (x3) at  (0.35,0.5) ;
\fill (x1) circle (1.5pt);
\fill (x2) circle (1.5pt);
\fill ([xshift=-2pt,yshift=-2pt]x3) rectangle ++ (4pt,4pt);
\draw [black] (x1p) -- (x1) -- (x2) -- (x3) -- (x1);
\draw [black] (x2) -- (x2p);
\end{tikzpicture}}}}}

\newcommand{\diagDoneTwoLegs}{{{\parbox{1.2cm}{{\begin{tikzpicture}[scale=1]
\coordinate (x1) at (0,0) ;
\coordinate (x1p) at (-.2,0) ;
\coordinate (x2) at  (0.7,0) ;
\coordinate (x2p) at  (0.9,0) ;
\coordinate (x3) at  (0,0.5) ;
\coordinate (x4) at (0.7,0.5) ;
\coordinate (x5) at (0.35,1) ;
\coordinate (x5p) at (0.35,1.2) ;
\fill (x1) circle (1.5pt);
\fill (x2) circle (1.5pt);
\fill (x3) circle (1.5pt);
\fill (x4) circle (1.5pt);
\fill (x5) circle (1.5pt);
\draw [black] (x1) -- (x2) -- (x4) -- (x3) -- (x1);
\draw [black] (x4)--(x5)--(x3);
\draw [black] (x1)--(x1p);
\draw [black] (x2)--(x2p);
\end{tikzpicture}}}}}}

\newcommand{\diagDfive}{{\parbox{1.6cm}{{\begin{tikzpicture}[scale=1]
\coordinate (x0) at (-.2,0);
\coordinate (x1) at (0,0) ;
\coordinate (x2) at  (0.3,0) ;
\coordinate (x3) at (0.9,0) ;
\coordinate (x4) at (1.2,0) ;
\coordinate (x5) at (1.4,0) ;
\coordinate (x6) at (0.6,0.6) ;
\coordinate (x7) at (0.6,0.8) ;
\fill (x1) circle (1.5pt);
\fill (x2) circle (1.5pt);
\fill (x3) circle (1.5pt);
\fill (x4) circle (1.5pt);
\fill (x6) circle (1.5pt);
\draw [black] (x0)-- (x1)-- (x2)  ;
\draw [black] (x3)--(x4)--(x5);
\draw [black] (x6)-- (x7)  ;
\draw (.6,0) circle (3mm);
\draw (1.2,0) arc (0:180:6mm);
\end{tikzpicture}}}}}
\newcommand{\smalldiagDfive}{\parbox{0.8cm}{\scalebox{0.5}{\diagDfive}}}

\newcommand{\triagdiagCTWF}{{\parbox{0.8cm}{{\begin{tikzpicture}[scale=1]
\coordinate (x1) at (0,0) ;
\coordinate (x2) at  (0.7,0) ;
\coordinate (x3) at  (0.35,0.5) ;
\coordinate (x4) at  (0.35,0) ;
\coordinate (x5) at  (0.85,0) ;
\coordinate (x6) at  (-0.15,0) ;
\fill (x1) circle (1.5pt);
\fill (x2) circle (1.5pt);
\fill (x3) circle (1.5pt);
\draw ([xshift=-2pt,yshift=-2pt]x4) rectangle ++ (4pt,4pt);
\draw [black] (x6) -- (x1) -- (x2) -- (x3) -- (x1);
\draw [black]  (x2) -- (x5);
\end{tikzpicture}}}}}

\newcommand{\propdiagCTm}{{{\parbox{1.cm}{{\begin{tikzpicture}[scale=1]
\coordinate (x1) at (0,0) ;
\coordinate (x2) at  (0.6,0) ;
\coordinate (x1p) at  (-.2,0) ;
\coordinate (x2p) at  (0.8,0) ;
\coordinate (x3) at (0.3,-0.3) ;
\fill (x1) circle (1.5pt);
\fill (x2) circle (1.5pt);
\draw ([xshift=-2pt,yshift=-2pt]x3) rectangle ++ (4pt,4pt);
\draw (.3,0) circle (3mm);
\draw [black] (x1) -- (x1p);
\draw [black] (x2) -- (x2p);
\end{tikzpicture}}}}}}

\usepackage[dvipsnames]{xcolor}
\definecolor{colorforS}{rgb}{0.858, 0.188, 0.8}
\definecolor{colorforS}{rgb}{0.1, 0.5, 0.3.}
\definecolor{colorforU}{rgb}{0.05, 0, 1}
\definecolor{colorforW}{rgb}{0.9, 0, 0}

\def\colU{colorforU!70}

\def\minsize{18}
\def\innersep{2}

\usepackage{hyperref}
\hypersetup{
    colorlinks,
    linkcolor={blue!70!black},
    citecolor={blue!75!},
    urlcolor={blue!80!black}
}

    \usepackage{tikz}

    \usetikzlibrary{decorations.markings}
    \usetikzlibrary{calc}
    \usepackage{graphicx}
    \usepackage{xcolor}
    \usepackage{bbold}

\begin{document}

\title{Bayesian phase transition for the critical Ising model:\\ Enlarged replica symmetry in  the epsilon expansion and in 2D}
\author{Kay J\"org Wiese}
\author{Alapan Das}
\author{Adam Nahum}
\affiliation{Laboratoire de Physique de l’\'Ecole Normale Sup\'erieure, CNRS, ENS \& Universit\'e PSL, Sorbonne Universit\'e, Universit\'e Paris Cit\'e, 75005 Paris, France}

\date{\today}

\begin{abstract}
A process that images or measures bond energies in the critical Ising model can be in distinct measurement ``phases'', depending on the precision of measurement. We study the transition into the strong-measurement phase using replica field theory (an epsilon expansion around six dimensions), a symmetry analysis, and numerical simulations in two dimensions. The results show that certain inference problems possess unexpected structure associated with an  ``enlarged'' symmetry of the replica description. The symmetry imposes constraints on correlation functions, and is an analog of the Nishimori phenomenon in the Ising spin glass, but in a distinct replica limit. It determines the exact critical exponent for an Edwards-Anderson-like correlator in the measured Ising model, both in 2D and near the upper critical dimension. The symmetry is present microscopically for certain measurement protocols, and it emerges in the infrared for  generic protocols at the critical measurement strength. We also examine the epsilon expansion for models with power-law interactions and/or long-range measurement. Finally, we describe an analogous replica symmetry enlargement for measurement in Wegner's lattice gauge theory.
\end{abstract}

\maketitle

\section{Introduction}\label{sec:intro}

How easy is it to extract the ``long-distance'' information about a configuration of the critical Ising model from local measurements? 
Prior to any measurement, our statistical knowledge of the spin configuration is expressed through expectation values $\<S(x) S(y) \cdots\>$ whose asymptotic properties are well understood \cite{KardarBook,Mussardo2010}. If we receive additional information about the spin configuration, in the form of the outcomes $M=\{M(x) \}_{x}$ for a set of measurements of some local observable $\mathcal{O}(x)$ at various positions $x$ throughout the system,
then our statistical knowledge of the configuration is improved~\cite{NahumJacobsen2025,PutzGarrattNishimoriTrebstZhu2026}.

Depending on the measured observable, 
and the precision of the measurement process, 
the improvement in our knowledge may be limited to short-scale features of the configuration;   alternatively, our estimates of long-distance correlation functions
$\<S(x) S(y) \cdots\>_M$ may radically change after conditioning on the  outcomes $M$ obtained in a given instance of the measurement protocol.
If we vary the precision of the measurements, we may drive a phase transition in the ``conditioned ensemble'' and its  correlation functions.
A  variety of such phase transitions are possible in the critical Ising model and other simple lattice models \cite{NahumJacobsen2025,PutzGarrattNishimoriTrebstZhu2026,Iba1999}.
Nontrivial inference phase transitions are possible even 
for infinite temperature configurations \cite{Iba1999} 
(a wide range of inference phase transitions have also been studied in mean-field disordered systems~\cite{ZdeborovaKrzakala2016, Nishimori2001, RicciTersenghiSemerjianZdeborova2019,GamarnikMooreZdeborova2022}, and there are some rigorous results in finite dimensions \cite{ AbbeMassoulieMontanariSlySrivastava2018, GarbanSepulveda2020}; there are also inference transitions for dynamical monitoring processes \cite{NahumJacobsen2025,GopalakrishnanMcCullochVasseur2026}).
But whereas inference at infinite temperature can be mapped to well-studied transitions in disordered magnets, 
inference in critical states leads to new phase transitions and renormalization group fixed points that have not yet been much explored.

Phase transitions in the conditioned ensemble also have a surprising number of other physical interpretations, separate from the idea of measurement. Let us list a few.
First, the same formalism can be used to describe certain types of interaction quench, in which a subset of the lattice degrees of freedom are 
instaneously frozen \cite{NahumJacobsen2025}. 
This freezing or ``partial quenching'' can change the thermodynamic state of the system, 
just as the thermodynamic state of a liquid can be altered by instantaneously pinning some of the particles \cite{CammarotaBiroli2012} or by instantaneously introducing new bonds \cite{DeamEdwards1976, GoldbartGoldenfeld2004}.
The  remaining non-frozen degrees of freedom are governed by the conditioned ensemble, with the frozen degrees of freedom playing the role of the measurements.

Separately, the formalism can be used to 
describe  processes in which an initial configuration $\{S(x)\}_x$ is transformed 
into a new configuration $\{S'(x)\}_x$ by an error channel (e.g.\ by flipping some of the spins). The new configuration $\{S'(x)\}_x$ 
can play the same mathematical role as the measurement outcomes $\{M(x)\}_x$, so that  a Bayesian phase transition in the inference problem can translate to  a phase transition in the nature of the distribution of $\{S'(x)\}$  (diagnosable e.g.\ using the conditional mutual information, which is closely related to conditioned correlators). 
By embedding the classical probability distribution in a quantum density matrix, \textit{classical} Bayesian inference transitions also give examples of ``strong-to-weak'' symmetry breaking in a quantum system \cite{lee2023quantum,chen2024symmetry,lessa2025strong,wang2026strong,hauser2026strong}: an intuitive way to understand many standard examples of strong-to-weak symmetry breaking, which involve diagonal density matrices (i.e.\ classical probability distributions) is by thinking of them as classical problems of the above type.

In another realm, it is  possible to relate the success or failure of  a real-space RG treatment of a statistical system to the properties of the conditioned ensemble
\cite{NahumJacobsen2025}.

Measurement phase transitions also have applications to error correction. Measurement transitions in the Ising model are  related by duality (orbifolding) to measurement phase transitions in $\mathbb{Z}_2$ lattice gauge theory; in turn, these translate to important error correction phase transitions in the toric code~\cite{DennisKitaevLandahlPreskill2002, WangHarringtonPreskill2003}.
Finally we note that while we will only discuss measurement of ``static''  configurations,  it is also interesting to consider monitoring --- or error correction ---  as an ongoing process during time evolution \cite{NahumJacobsen2025,GopalakrishnanMcCullochVasseur2026, JinMartin2022,KimLamacraft2025,GerbinoGiachettiLeDoussalDeLuca2025, AgrawalZabaloChenWilsonPotterPixleyGopalakrishnanVasseur2022,KimLamacraft2025,PatilLudwig2025,KimvonKeyserlingkLamacraft2026}.

In this paper we address measurement of the  energy density on the bonds of a critical or near-critical Ising model \cite{NahumJacobsen2025,PutzGarrattNishimoriTrebstZhu2026}.
We may imagine that the  measurement device detects the energy for each bond, in a given configuration, with a measurement error of variance ${\propto 1/\Gamma}$.
We use lattice symmetry arguments, 
an epsilon expansion near an upper critical dimension, and numerical simulations
to address the critical point and the structure of the phase diagram.
The results show that  inference problems can exhibit a hidden symmetry that strongly constrains correlation functions and ``a posteriori'' estimates. 
(The connections mentioned above mean that this hidden symmetry also has implications for various other kinds of problem which at first glance are unrelated to  inference.)

The phase diagram for measurement of an Ising model at \textit{infinite} temperature can be mapped to the Nishimori 
line \cite{Nishimori1980} in the phase diagram of a random magnet
\cite{Iba1999}. 
For the \textit{critical} Ising model, 
the phase diagram 
(as a function of measurement precision $\Gamma$ and spatial dimensionality $d$)  turns out to be more complex.
Previous work has described the phase diagram  in 2D \cite{NahumJacobsen2025,PutzGarrattNishimoriTrebstZhu2026} and various aspects of the phase diagram for $d>2$ \cite{NahumJacobsen2025}. 
However, the evolution of the RG flow as a function of dimension is nontrivial and even some qualitative aspects were not previously resolved. One aim of this work is to shed light on this. We will also give quantitative results based on the emergent symmetry mentioned above and on the renormalization group (RG).

Our main findings are as follows.
First, we demonstrate the existence of a  fixed point in the $\epsilon$ expansion that describes a transition between ``weak'' and ``strong'' measurement of the critical Ising model, 
and determine its critical exponents. 
For short-range models, the dimension is ${d=6-\epsilon}$;
we also make an $\epsilon$-like expansion for low-dimensional models with power-law interactions and/or long-range measurements.  
These analyses require a two-loop calculation, because the one-loop contributions to the beta function vanish on the attractive submanifold of higher symmetry discussed below. 

Second, we demonstrate that, in any dimensionality, an enlarged replica symmetry 
plays an important role for the phase transition into the strong-measurement phase. 
Here, ``enlarged'' means that the  symmetry  is larger than would be expected a priori for a generic measurement protocol or a generic inference problem.
This enlarged symmetry is present  microscopically for some special choices of measurement protocol,
but we will also show, using both the epsilon expansion and 2D simulations, that it can \textit{emerge} in the infra-red even when it is not present microscopically.
The enlarged symmetry is a novel feature  of these Bayesian inference problems, and 
has striking consequences.

For example, let $\{S_\text{true}(x)\}_x$ be the ``true'' configuration that we measured, and let $\{S(x)\}_x$ be a configuration sampled from the posterior measure, using the noisy measurement data.
That is, loosely speaking, $\{S(x)\}_x$  is a possible ``guess'' at what the true configuration might have been.
Enlarged symmetry means that the ``overlap'' configuration, 
$\{\Phi(x)\}_x$, 
where ${\Phi(x) = S_\text{true}(x) S(x)}$,
and $\Phi(x)=\pm 1$,
which gives the spatial pattern of errors in our guess,
has precisely the same statistics as a configuration drawn from the conventional  Ising ensemble (!).
That is, 
\be\label{eq:fullconfigidentity}
P\lf \{\Phi(x)\}_x \ri = e^{-\mathcal{H}[\Phi]}/Z,
\ee
where $\mathcal{H}$ is the Ising Hamiltonian (and $Z$ is the Ising partition function).
More generally, enlarged symmetry imposes identities on correlation functions.
(The enlarged symmetry can be achieved even for non-critical Ising,
but its consequences are most interesting in the critical case.)

Finally, in 2D, we demonstrate  numerically the existence of a fixed point with this enlarged symmetry, and with nontrivial ``multiscaling''  exponents for moments ${\mathbb{E} \<S(x)S(y)\>_M^{l}}$ of conditioned correlation functions.
We  give strong numerical evidence that this fixed point describes a stable universality class for the Bayesian phase transition in more general microscopic models (as in $6-\epsilon$ dimensions),
implying that the enlarged symmetry
emerges at  the critical point
(without the need for further fine-tuning) not only in high dimensions but also in two dimensions.

The enlarged replica symmetry is analogous to that which arises for the random-bond Ising model on the Nishimori line  \cite{LeDoussalHarris1988,LeDoussalHarris1989,ChenLubensky1977,GruzbergReadLudwig2001}: loosely speaking, it involves an enlargement of $S_n$ replica symmetry to an $S_{n+1}$  symmetry
which mixes the replicated spin field with an Edwards-Anderson-like ``overlap'' field (we will be more precise below;  we also  clarify some common confusions about the Nishimori symmetry in Sec.~\ref{sec:symmclar}).
The replica limit of interest in the problem we study is ${n\to 1}$,
while in the  random-bond Ising model the replica limit is  ${n\to0}$. In both cases the enlarged symmetry imposes identities between correlation functions. 
In the present case ($n\to 1$), the symmetry also forces the exponents for some nontrivial correlation functions,
including the Edwards-Anderson-like correlator, 
to take simple, exactly known values, both in 2D and in ${6-\epsilon}$.

Related emergent symmetries are relevant to a range of other measurement or conditioned-ensemble problems. Towards the end of the paper we also describe the example of 3D $\mathbb{Z}_2$ gauge theory, which allows for a variety of Bayesian phase transitions, some of which exhibit enlarged replica symmetry.

\subsection{Overview of this work}

Let us now give a schematic picture of the ``Landau-Ginsburg'' field theory that is useful for the epsilon expansion.

While ${d=4}$ is the upper critical dimension in the usual Ising model, 
the upper critical dimension for the measurement problem is ${d=6}$. 
In other words, the conditioned correlators 
$\<S(x) S(y)\cdots\>_M$ remain nontrivial  in dimensions between ${d=4}$ and  ${d=6}$, even though standard unconditioned correlators 
$\<S(x) S(y)\cdots\>$ are Gaussian above four dimensions.

Near the upper critical dimension we use a replica field theory whose Lagrangian ${\mathcal{L}=\mathcal{L}[\phi, \Phi]}$ 
includes both a field  $\phi_\alpha$ representing 
the  lattice spin $S_\alpha$
(here ${\alpha = 1,\ldots, n}$ is a replica index),
and a field $\Phi_{\alpha\beta}$
representing lattice replica 
overlaps $S_\alpha S_\beta$ \cite{NahumJacobsen2025} 
(i.e.\ the Edwards-Anderson order parameter). 
The form of the Lagrangian is familiar from spin glasses \cite{LeDoussalHarris1988,LeDoussalHarris1989,ChenLubensky1977}, although the required replica limit will be different:
\ba
\label{eq:introintroducecubicL}\notag
\mathcal{L}
& =  
 \frac1{2} \left[ (\nabla \phi_{\alpha})^2  + r_1 \phi_\alpha^2 \right]
+
 \frac1{4}\left[ (\nabla \Phi_{\alpha\beta})^2 + r_2 \Phi_{\alpha\beta}^2 \right]
\\
   & +    \frac{\lambda_1}{2} \phi_{\alpha}  \phi_{ \beta } \Phi_{ \alpha \beta } 
+ 
 \frac{\lambda_2}{6} \Phi_{\alpha \beta}\Phi_{ \beta\gamma }\Phi_{\gamma \alpha }
\end{align}
(all indices are summed, and $\Phi_{\a \b}=\Phi_{\b\a }$; precise definitions below).
If we consider the Ising model at its critical temperature, then the first mass coefficient, $r_1$, is zero; 
the second mass coefficient, $r_2$, varies as a function of the measurement strength $\Gamma$.
The measurement phase transition that we study 
is the condensation of  $\Phi_{\alpha\beta}$ at a critical measurement strength $\Gamma_c$
(as reviewed below).
Long-range ordering of $\Phi$ implies that 
$|\<S(x) S(y)\>_M|$  does not vanish at large distance,
indicating that the  measurements 
carry significant information about the relative orientation 
even of arbitrarily distant spins. 

The ${n\to 0}$ limit of the field theory defined by $\mathcal{L}$  has been used to study the Nishimori line in disordered magnets \cite{LeDoussalHarris1988,LeDoussalHarris1989,ChenLubensky1977},
so that the one-loop RG equations are known for general $n$. However, we require the field theory in the ${n\to 1}$ limit that properly accounts for the statistics of measurement outcomes.
It turns out that in this limit, the one-loop RG equations fail to determine the topology of RG  flows, necessitating a two-loop calculation.
We find that for ${d\lesssim 6}$ the Bayesian phase transition is governed by a nontrivial critical point with the emergent symmetry mentioned above.
(The expansion will be developed to further loop orders in Ref.~\cite{DavletbaevaKompanietsWiese2026}.)

Formally,
while the generic Bayesian inference problem has a replica symmetry group which we denote ${\mathcal{G}_n \equiv\mathbb{Z}_2^n\rtimes S_n}$,
the emergent global symmetry for  the Bayesian transition 
is ${\mathcal{G}_n^+\equiv \mathbb{Z}_2^n\rtimes S_{n+1}}$. 
This symmetry unites the fields $\phi$ and $\Phi$ into a single irreducible representation, implying that (surprisingly) the Edwards Anderson order  parameter shares the standard Ising anomalous dimension (which vanishes above 4D).

If we directly try to continue the critical exponents found in $6-\epsilon$ dimensions down to (say) two dimensions, 
we encounter a nonanalyticity in 4D,
as the result of a quartic term that ``turns on'' there.
However, we expect that critical exponents nonetheless evolve \textit{continuously} as we pass through 4D (Sec.~\ref{sec:RG}).
Therefore we expect that various features of the high-dimensional theory are relevant to low dimensions. 
These include the emergence of the higher symmetry, 
and also ``multiscaling'' of correlators. 
The existence of a stable symmetric fixed point in $6-\epsilon$ dimensions that survives down to $d<4$ also clarifies the phase diagram in general $d$, ruling out one of the two possibilities speculated about in Ref.~\cite{NahumJacobsen2025}.

As in the ${n \to 0}$ (spin glass) problem, 
it is also possible to realize the enlarged symmetry  $\mathcal{G}_n^+$ microscopically.
We show that this occurs --- 
at the critical temperature and critical measurement precision ---
for a simple protocol where  bond energies are measured 
with normally-distributed measurement errors
(and also for some other special protocols).
However, the presence of \textit{microscopic} $\mathcal{G}_n^+$ symmetry is a form of fine-tuning.
The epsilon expansion shows that, in high dimensions,  $\mathcal{G}_n^+$ emerges in the infra-red (IR) even in the absence of this fine tuning. 

Our simulations 
compare different lattice protocols (with and without the enlarged symmetry at the microscopic level) 
giving strong evidence that the enlarged symmetry  emerges in the IR without fine-tuning also in 2D.
The simulations are based on a two-step protocol, first simulating the critical Ising model in order to extract a measurement realization, then performing a second simulation 
to access conditioned correlators. 
(We use ``real replicas'' \cite{MarinariMartinMayorParisiRicciTersenghiRuizLorenzo2024} to access the spectrum of multiscaling exponents.)

We also consider models in low dimensions but with nonlocal interactions 
${\sim \int_{x,y}[S(x)-S(y)]^2 |x-y|^{-d-\alpha_J}}$
and/or with measurements of long-range spin pairs $S(x)S(y)$ whose informativeness decays like
$|x-y|^{-d-\alpha_M}$. 
These long-range models yield epsilon expansion parameters 
\begin{align}
    \epsilon_1 &  = 2 \alpha_J + \alpha_M -d,
    & 
    \epsilon_2 &  = 3 \alpha_M -d.
\end{align}
Since there  is no physical requirement for $\epsilon_1$, $\epsilon_2$ to be integer-valued, these models could be simulated with Monte Carlo for any values of $\epsilon_{1,2}$.
(As for the short-range models, many of our RG results are also applicable to spin glass problems, by considering the ${n=0}$ limit instead of ${n=1}$ limit.)
Depending on the values of $\alpha_J$ and $\alpha_M$, 
both, one, or neither of the cubic couplings in Eq.~\ref{eq:introintroducecubicL}
may be relevant.
We find that in the case where ${\alpha_J=\alpha_M}$, the flows in the  long-range model are qualitatively like those of the short-range model in ${6-\epsilon}$ dimensions, showing a nontrivial fixed point with emergent symmetry. 

Finally, we discuss measurement of Wegner's ``Ising gauge theory'' ($\mathbb{Z}_2$ gauge theory) in three dimensions. This model has a thermodynamic Higgs transition and a thermodynamic confinement transition, both of which have 3D Ising exponents. We find that  measurement of these two ensembles gives rise to two inequivalent kinds of inference problem, and that either can show enlarged replica symmetry.

\tableofcontents

\section{Measuring short-range Ising models}

\subsection{Setup} \label{sec:setup}

We begin with a brief review of the bond-energy measurement problem (see Refs.~\cite{NahumJacobsen2025,PutzGarrattNishimoriTrebstZhu2026} for more information).

\begin{figure}
\fig{0.95}{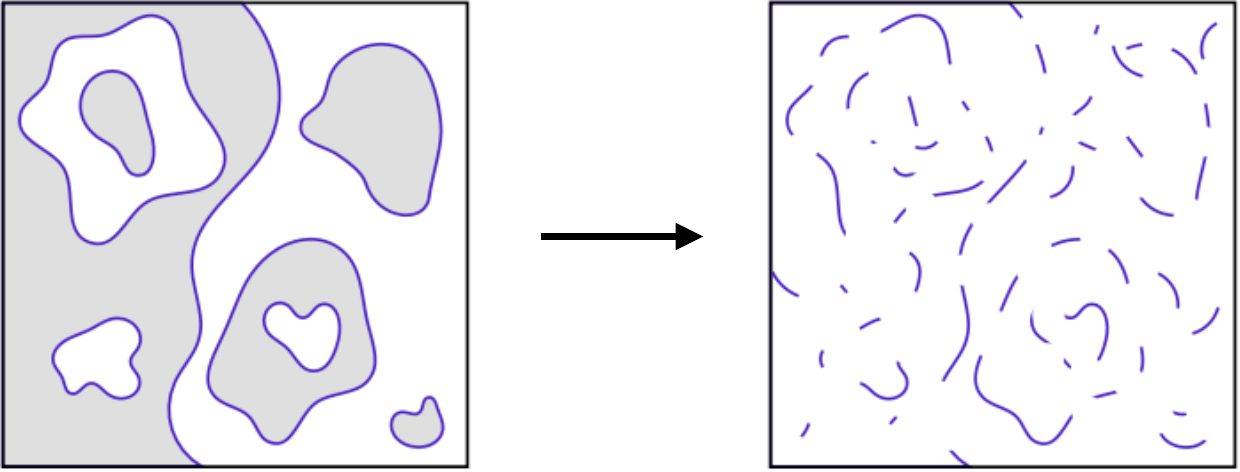}
\caption{Schematic illustration of part of a critical Ising configuration (Left) and the measurement data (Right), which can be thought of as a noisy image of the domain wall configuration.
See Fig.~\ref{fig:Snapshot} for a more accurate depiction of one of the specific lattice measurement protocols discussed in the main text. (Figure from \cite{NahumJacobsen2025}.)}
\label{isingcartoon}
\end{figure}

Consider an Ising model on the hypercubic lattice in   dimension $d$ (we absorb the inverse temperature into $J$, so that the Boltzmann weight is  $e^{-\mathcal{H}}$):
\ba
\mathcal{H} & = - J \sum_{\<ij\>} S_i S_j. 
\end{align}
We will almost always focus on the critical temperature, i.e. 
\be
J=J_c.
\ee
We make measurements of bond energies homogeneously throughout space.  These measurements are assumed to be imperfect.  We may imagine that the measurements give us an incomplete or noisy image of the configuration of  the Ising domain walls, as shown in Fig.~\ref{isingcartoon}:  
 we ask how much we learn about the underlying configuration from such a noisy image.

This question can also be asked at infinite temperature \cite{Iba1999}  (or in the paramagnetic phase \cite{NahumJacobsen2025,PutzGarrattNishimoriTrebstZhu2026}), where there are applications to error correction in the toric code \cite{DennisKitaevLandahlPreskill2002,WangHarringtonPreskill2003}. This infinite-temperature problem is well understood by a mapping to the standard  Ising spin glass on the Nishimori line \cite{Iba1999}.  
We focus on the critical Ising model, which does not reduce to the conventional disordered systems problem and where there are new kinds of critical behavior (and, in general $d$,  a richer phase diagram).

One simple measurement protocol, 
which we   discuss in Sec.~\ref{sec:latticeprotocols},
is that each measurement $M_{ij}\in \mathbb{R}$ gives a noisy readout of ${S_i S_j}$ on a bond $\<ij\>$, with a Gaussian error of variance $\Delta^2$.
An alternative protocol involves  binary measurements of all  bonds, but with some probability $p_\text{error}$  of error. 
As a final example, we could make  precise binary measurements, but only of a random fraction $p_\text{measure}$ of the bonds. 
The naive expectation is that all  these protocols  show a similar phase diagram,
as a function of increasing {\em measurement strength} 
(i.e.\ as a function of  increasing   $1/\Delta^2$, or $1/p_\text{error}$, or $p_\text{measure}$); we will discuss this more carefully below. For the case of Gaussian errors, we define the measurement strength as $\Gamma= 1/2\Delta^2$.

We imagine that we have been given the measurement information $\{M_{ij}\}$ and we wish to infer information about the ``true'' spin configuration $\{S^\text{true}_i\}$ that these measurements came from. 
We are assuming that we know both the initial Boltzmann distribution $e^{- \mathcal{H}[S]}/Z$ and  the measurement protocol.
The latter is  encoded in a conditional distribution $P_\text{measure}(\{M_{ij}\}|\{S_i\})$, which specifies the probability that a measurement of  configuration $\{S_i\}$ yields measurement outcomes $\{M_{ij}\}$. 
For example, when each measurement is centered on its true value, with a Gaussian error, $P_\text{measure}$ is a product of Gaussian factors ${\propto e^{- \Gamma (M_{ij} - S_i S_j)^2}}$ on the bonds of the lattice.

\begin{figure}[t]\centerline{\fig{1.0}{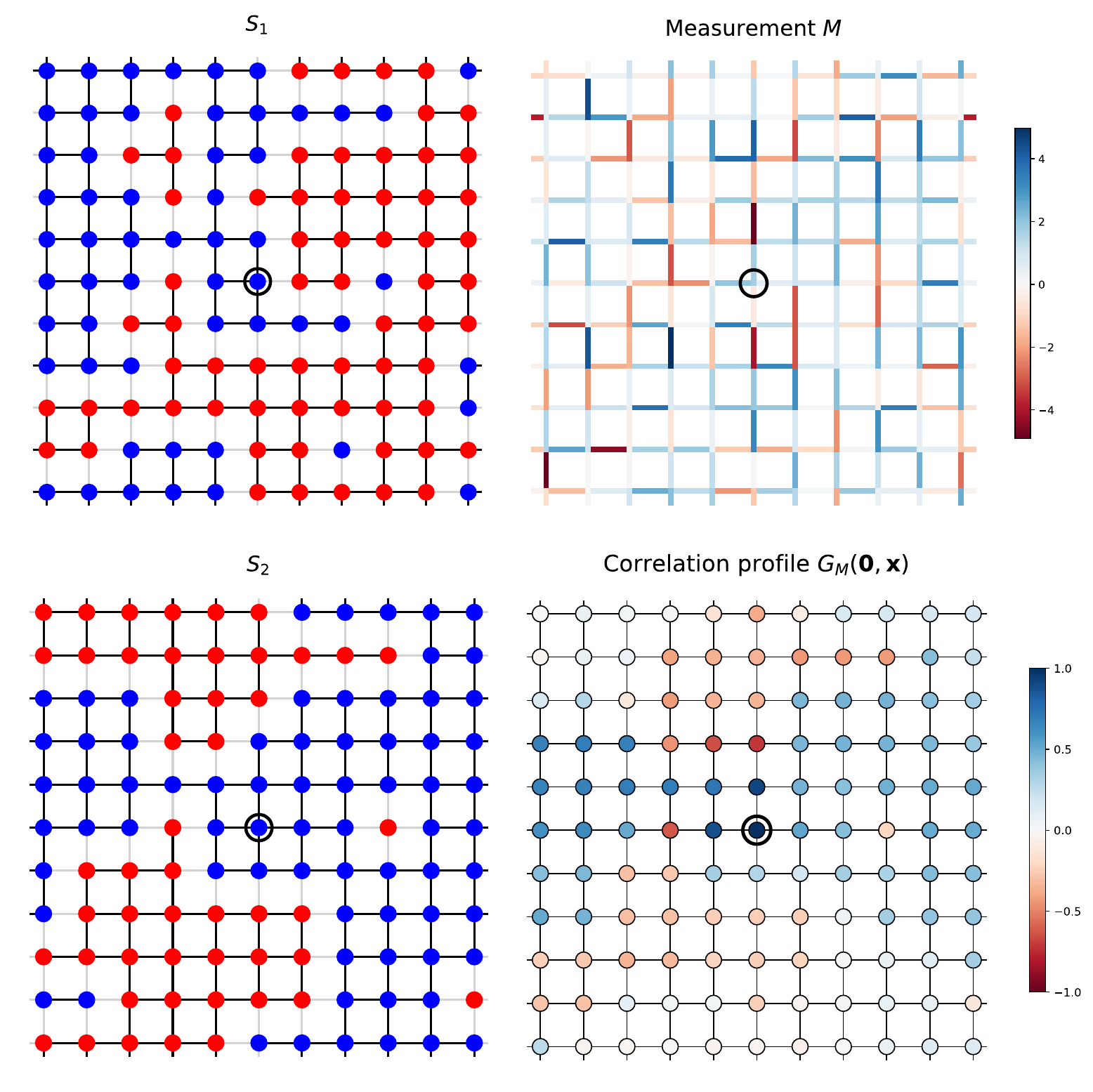}}
    \caption{A snapshot of the process in an $11\times 11$ lattice with periodic boundary conditions.
    Top left: We start with a configuration $S_1=S^\text{true}$ of the pure critical Ising model (blue sites are spin up,  red ones spin down). 
    Top right: A collection $M$ of bond measurements is made, 
    with normally-distributed errors 
    (the Gaussian protocol). 
    Bottom left: A spin configuration $S_2$ may now be sampled  from the a posteriori measure $P(\bullet|M)$. Bottom right: 
    The measurement outcomes $M$ define a conditioned  two-point correlation ${G_M(\mathbf{0}, \mathbf{x})=\<S({\mathbf 0})S({\mathbf x})\>_M}$. The origin $\mathbf 0$ is indicated by the black circle. 
    (These snapshots were obtained  at $J=J_c$, $\Gamma=\Gamma_c$, but no claim is made for their typicality.)}
    \label{fig:Snapshot}
\end{figure}

Given this information, Bayes' rule specifies a {\em conditioned} or {\em posterior} probability distribution $P(\{S_i\} |\{M_{ij}\})$, which represents our knowledge of the spin state, given the measurements (see Ref.~\cite{ZdeborovaKrzakala2016} for a review of Bayesian inference in a statistical physics context). Formally, this conditioned distribution resembles the Boltzmann distribution for a disordered system, with the measurement outcomes playing the role of the bond disorder. For example, for Gaussian measurements:
\be\label{eq:HeffSM}
P\lf\{S_i\} |\{M_{ij}\}\ri \propto 
e^{ - \mathcal{H}[S] - \Gamma \sum_{\<ij\>} ( M_{ij} - S_i S_j)^2}.
\ee
(We have omitted an $S$--independent normalization factor.) The difference from a standard disordered system is that the measurement outcomes $M_{ij}$
have highly nontrivial correlations that were inherited from the measured critical configuration. 
Just as in disordered systems, we may ask about the nature of correlation functions
$\<\cdots\>_M$
in the ensemble for a typical instance of $M=\{M_{ij}\}$.

Fig.~\ref{fig:Snapshot} is intended to give a heuristic idea of this process.
The top-left panel shows the ``true'' configuration ${S_1 = S^\text{true}}$ that was measured.
The  top-right panel shows a realization of Gaussian measurements $M$ taken from this configuration. 
The bottom-left panel shows a configuration $S_2$ sampled from the conditioned ensemble $P(\bullet|M)$.
Finally, the bottom-right panel shows the correlations, conditioned on measurements, between the spin at the origin and the other spins.

\subsection{Measurement phases for critical Ising}
\label{sec:phases}

The phase diagram of the measurement problem, as  a function of  measurement strength $\Gamma$,
evolves in a nontrivial way as the spatial dimensionality $d$ is varied \cite{NahumJacobsen2025},
but it has a similar, simple structure  in $d=2$ and in $d\lesssim 6$.  
We are currently restricting to the Ising model \textit{at its critical temperature}, 
so the phase diagram under discussion is one-dimensional and labelled by  $\Gamma$.\footnote{When we vary both the physical temperature $T$ and the measurement strength $\Gamma$ we have a two-dimensional phase diagram, as discussed in Sec.~\ref{sec:2Dphasediagram} below. However, the novel kinds of critical behavior occur at the critical temperature $T=T_c$, see \cite{NahumJacobsen2025,PutzGarrattNishimoriTrebstZhu2026}. In particular, the measurement problem in the paramagnetic phase is in the same universality class as at infinite temperature \cite{Iba1999} and is related to the Nishimori critical point.}

In all dimensions ${d>1}$ there is a stable {\em strong-measurement phase} for $\Gamma$ greater than some critical value $\Gamma_c$. This phase is described below. 
The nature of the phase that exists for  ${\Gamma\lesssim \Gamma_c}$ depends on the dimensionality. 
Both in ${d=2}$, and in ${d\lesssim 6}$, the phase at $\Gamma<\Gamma_c$ is a simple {\em weak-measurement phase} which flows to an RG fixed point at zero measurement strength. 
So, in both $d=2$ and for $d\lesssim 6$ the phase diagram has the simple topology
\be\label{eq:simplephaseidagram}
\begin{tikzpicture}
\tikzset{flow/.style = {thick, double = black,
      double distance = 0pt}}
\tikzset{fixedp/.style = {minimum width= 0.7*\minsize,inner sep=\innersep}}
  \node[fixedp] (o) at (-1.5,0) [] {};
  \node[fixedp] (u) at (2.5,0) [draw, circle, fill=\colU] {};
      \node[fixedp] (inf) at (6.5,0)[]{};
  \draw [very thick, CadetBlue,flow][-] (o.east) -- (u.west);
    \draw [very thick, BlueViolet,flow][->] (u.east) -- (inf.west);
  \node at (6.2,-0.4) {$\Gamma$};
  \node at (0.5,0.6) {{Weak measurement}};
\node at (4.5,0.573) {{Strong measurement}};
\end{tikzpicture} 
\ee
The blue dot is the critical point we will study.
(We discuss more general $d$ below.)

In the {\em strong-measurement phase}, the Edwards-Anderson correlator, averaged over measurement outcomes $M$,
tends to a nonzero value at large distances:
\be\label{eq:edwardsandserson}
\lim_{|x-y|\to\infty} \mathbb{E} \<S(x) S(y)\>_M^2  > 0.
\ee
($\mathbb{E}$ denotes the average over measurement outcomes. We have switched from labelling spins by  $i,j$ to using their spatial coordinates $x,y$.)
In contrast, the average value of the conditioned correlator, 
\begin{equation}\label{ESSM=SS}
    {\mathbb{E} \<S(x) S(y)\>_M=\<S(x) S(y)\>},
\end{equation}
is simply  the \textit{unconditioned} correlator, and tends to zero in the usual power-law fashion. 

Loosely speaking, Eq.~\ref{eq:edwardsandserson} means that in the strong-measurement  phase the information from measurements is sufficient to determine the relative orientation of asymptotically distant spins (with a success probability greater than~1/2).\footnote{Some thought shows that Eq.~\ref{eq:edwardsandserson} is also equal to the  covariance between $(S_{\text{true } i} S_{\text{true } j})$ and $(S_i S_j)$, where $\{S_\text{true}\}$ is the measured configuration, and $\{S_i\}$ is  sampled from the conditioned distribution \cite{ZdeborovaKrzakala2016} (see the discussion at the end of App.~\ref{a:Review_definitions}).}
 In terms of the heuristic cartoon in Fig.~\ref{isingcartoon},
this means using the noisy information in the right panel to guess whether two distant points lay in same-color or opposite-color domains for the configuration in the left panel.

In the weak-measurement phase, the Edwards-Anderson correlator instead decays to zero at large distances, indicating that (asymptotically) measurements give us no information about whether $S_i S_j$ is positive or negative.
Our main focus in this paper is the critical point $\Gamma_c$ between these two phases.
We use both a $d=6-\epsilon$ expansion, and numerical simulations in  $d=2$.

For completeness, let us comment on the regime ${0<\Gamma<\Gamma_c}$ in general $d$.
In both $d=6-\epsilon$ and in $d=2$ this is a  ``trivial'' weak-measurement phase  governed by the fixed point at zero measurement strength. 
For example, this means that $\mathbb{E}\<S(x)S(y)\>^2_M$  scales in the same way as $\<S(x)S(y)\>^2$
(modulo possible logarithmic corrections in $d=2$).
However, there is an intermediate range of dimensions $0<d<d_*$ (with $d_*$ between 4 and 6) where nontrivial fixed points appear and disappear within the ``weak measurement'' regime.
The  results in the following Sections support the second scenario for this, which is depicted in Fig.~4 of \cite{NahumJacobsen2025}.
Here we only note that we expect that, in $d=3$ and $d=4$, the phase at ${0<\Gamma<\Gamma_c}$ is  governed by a nontrivial fixed point, and has multiscaling for correlation functions (discussed below):
\be
\begin{tikzpicture}
\tikzset{flow/.style = {thick, double = black,
      double distance = 0pt}}
\tikzset{fixedp/.style = {minimum width= 0.7*\minsize,inner sep=\innersep}}
  \node[fixedp] (o) at (-1.5,0) [] {};
  \node[fixedp] (u) at (2.5,0) [draw, circle, fill=\colU] {};
      \node[fixedp] (inf) at (6.5,0)[]{};
  \draw [very thick, CadetBlue,flow][-] (o.east) -- (u.west);
    \draw [very thick,BlueViolet, flow][->] (u.east) -- (inf.west);
  \node at (6.2,-0.4) {$\Gamma$};
  \node at (0.5,0.6) {{Multifractal phase}};
\node at (4.5,0.573) {{Strong measurement}};
\end{tikzpicture} 
\label{eq:linearphaseidagram2}
\ee
However, despite the nontrivial evolution of the phase diagram with $d$,\footnote{In fact, according to this scenario there is a range of $d$ just above 4D where the regime $0<\Gamma<\Gamma_c$ splits into two distinct phases, one governed by the $\Gamma=0$ fixed point and one governed by a nontrivial fixed point. 
But it is possible that this range of $d$ does not include any integer values.}
we conjecture that the fixed point governing the transition into the strong-measurement phase  at $\Gamma_c$ (the blue dot)
evolves continuously (see Sec.~\ref{sec:RG}) from $6-\epsilon$ dimensions down to $d=1+\delta$ dimensions, albeit with a nonanalyticity in exponents at ${d=4}$.
 We expect 
this continuous evolution to connect the fixed point we study in ${6-\epsilon}$ dimensions to the one that we study numerically in two dimensions.

Our main focus in this paper will be the critical point at $\Gamma=\Gamma_c$.
At the critical point we expect the Edwards-Anderson correlator, and analogous higher moments of the conditioned two-point  function, to decay as power laws:
\be\label{eq:multiscaling}
\mathbb{E} \< S(x) S(y)\>_M^{l}
\sim |x-y|^{- {2 \Delta_{l}}}.
\ee
In the following Sections we will discuss these exponents in ${d=6-\epsilon}$ and $d=2$. The situation with a nontrivial $l$-dependence of the exponents is referred to in the disordered systems literature as multiscaling  or multifractality \cite{Ludwig1990,
Lewis1998,
OlsonYoung1999,
DavisCardy2000,
PalagyiChatelainBercheIgloi2000,
HoneckerPiccoPujol2001,
DeQueirozStinchcombe2003,
MerzChalker2002,
PiccoHoneckerPujol2006,
MonthusBercheChatelain2009,
MarinariMartinMayorParisiRicciTersenghiRuizLorenzo2024,
Castellani2005,
Duplantier2003}.

\subsection{Two-parameter phase diagram}
\label{sec:2Dphasediagram}

\begin{figure}
    \centering
    \includegraphics[width=0.98\linewidth]{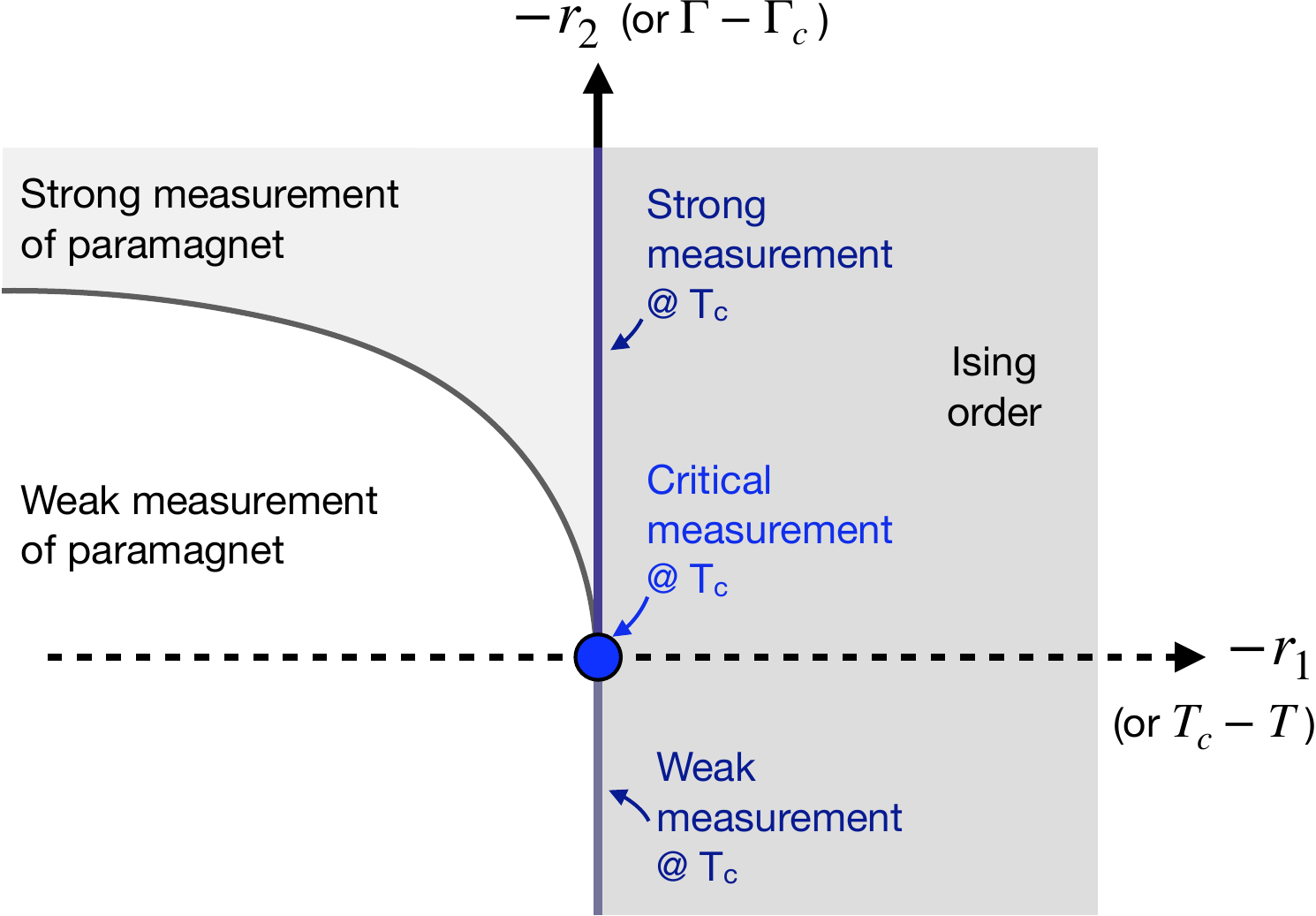}
\caption{Topology of the phase diagram for the field theory in the ${(r_1, r_2)}$ plane (Sec.~\ref{sec:lagrangians})
and also for the lattice model in the ${(T_c -T, \Gamma-\Gamma_c)}$ plane.
The vertical axis, 
where ${T=T_c}$, corresponds to the one-pameter phase diagram in (\ref{eq:simplephaseidagram}).
Our main focus will be the point at the origin, marked with a circle, 
where the measurement strength is also critical.
In the field theory, the diagonal line ${r_1=r_2}$ possesses an enlarged symmetry (\ref{eq:symmgpenhancedlattice}) if the cubic couplings also respect the enlarged symmetry.}
\label{fig:2dphasediag}
\end{figure}

Above we have described the phase diagram as a function of measurement strength $\Gamma$, in the case where the Ising model is fixed to its critical temperature ($T=T_c$).
For completeness, 
Fig.~\ref{fig:2dphasediag} puts this one-parameter phase diagram in the context of the \textit{two}-parameter phase diagram that arises when we allow both  $\Gamma$ and $T$ to vary \cite{PutzGarrattNishimoriTrebstZhu2026,NahumJacobsen2025}.
This phase diagram topology is valid both in 2D and in $6-\epsilon$ dimensions, 
and we have anticipated Sec.~\ref{Landau theory for the measurement transition} by labelling the axes with the 
field theory mass parameters $r_1$ and $r_2$ as well as  by  the microscopic parameters $T_c-T$ and $\Gamma-\Gamma_c$.

The one--parameter phase diagram  in Eq.~\ref{eq:simplephaseidagram}  
is the vertical axis of the present figure,
and the critical point 
${\protect\tikz[baseline=-2.5pt]
  \protect\node[minimum width=0,inner sep=0.2] (u) at (0,0) [draw, circle, fill=\colU] {\phantom{U}};}$
discussed immediately above 
is at the origin of the present figure (at ${T=T_c}$ and ${\Gamma=\Gamma_c}$).
In the context of the 
larger phase diagram, this point $(T_c,\Gamma_c$) 
is formally a  multicritical point.
However, the fact that the measurement process has no effect on thermodynamic quantities means that 
the restriction to the one-dimensional phase diagram of the previous Section is a natural one.\footnote{Since our idealized  measurements have no physical effect on the system (see e.g.\ Eq.~\ref{eq:edwardsandserson}), 
we can first identify $T_c$, then fix it once and for all, and then vary $\Gamma$ to find $\Gamma_c$.
This is unlike a standard  multicritical point, which we could only find by \textit{simultaneously} fine-tuning two parameters.  There is an analogy here with the standard phase diagram for an Ising magnet in the $(h,T)=\text{(magnetic field, temperature)}$ plane. The point $(0,T_c)$ is formally a multicritical point in this plane. 
But symmetry guarantees that varying $T$ cannot drive us off the $h=0$ axis, so it is   natural to think of $(0,T_c)$ as a critical point in a one-dimensional phase diagram parameterized by $T$.
In our problem, the $T=T_c$ axis is protected in the sense that varying $\Gamma$ cannot drive us off this line.}

To the right of the vertical axis (${T<T_c}$)
there is the thermodynamically ordered phase,\footnote{The thermodynamically ordered phase does not split into two phases on taking measurement into account.
The Edwards-Anderson correlator is trivially long-range ordered here, since the standard correlator is long-range ordered.}
 and to the left of the axis we have the thermodynamically disordered (paramagnetic) phase.
There is a measurement transition \textit{within} the paramagnetic phase which was mentioned briefly in Sec.~\ref{sec:setup} and discussed in  \cite{NahumJacobsen2025,PutzGarrattNishimoriTrebstZhu2026}.
This critical line, within the paramagnet, 
can be viewed as  flowing to infinite temperature $T$: as a result
it is in the same universality class as the critical point for inference in the infinite-temperature Ising model \cite{Iba1999},
which in turn can be mapped to  
the Nishimori multicritical point in the phase diagram of the Ising spin glass  \cite{Iba1999}. 
See Sec.~VIII of Ref.~\cite{NahumJacobsen2025} for a detailed review of the high-temperature inference problem.
The critical line for the transition inside the paramagnetic phase is  asymptotically vertical as it meets the ``multicritical'' point \cite{PutzGarrattNishimoriTrebstZhu2026}, as a result of the exponent 
values at the ``multicritical'' point \cite{LeDoussalHarris1989}
(Sec.~\ref{sec:critexponentsummary}).

To avoid confusion let us empasize that in the following we will usually use the terminology of Sec.~\ref{sec:phases}, 
so that the ``critical point'' will refer to the point  ${(T=T_c, \, \Gamma=\Gamma_c)}$ unless otherwise specified.

\subsection{Measurement protocols and enlarged symmetry}
\label{sec:latticeprotocols}

We now use the replica formalism to discuss a distinction between different measurement protocols at the level of symmetry. We begin with some review of the Ising measurement problem.
A brief review of the replica formalism used below is given in 
App.~\ref{a:Review_definitions} (see Sec.~II of  Ref.~\cite{NahumJacobsen2025} for more detail): here we  recall that the required replica limit is ${n\to 1}$.\footnote{This ensures that expectation values in the replica theory map to physical expectation values in which measurement outcomes are sampled with the correct probabilities, and \Eq{ESSM=SS} holds.  $n\to 1$ replica limits have also been used to study free energy landscapes
in glassy systems \cite{Monasson1995,FranzParisi1995,FranzParisi1997}.}

For Gaussian measurements with variance $\Delta^2$, the effective replica Hamiltonian 
contains a pairwise interaction between replicas of strength $\Gamma= 1/2\Delta^2$, and reads
 \be\label{eq:replicaHgaussian}
    \mathcal{H}_n= 
  -  \sum_{\<ij\>} \bigg[ 
    J \sum_{\alpha=1}^n S_{\alpha i} S_{\alpha j} 
    + \f{\Gamma}{n} \sum_{\substack{\alpha, \beta=1 \\ \alpha\neq \beta}}^n 
    (S_{\alpha i} S_{\alpha j}) (S_{\beta i} S_{\beta j})
    \bigg].
    \ee
The replica indices ${\alpha,\beta  = 1,\ldots, n}$ (Greek letters) appear as an additional subscript on the spin. For our purposes, we can replace  $\Gamma/n$ with $\Gamma$, since we are interested in the ${n\to 1}$ limit. 
The interactions appearing in the above Hamiltonian match those in the replica description for a magnet with quenched Gaussian randomness in the couplings,  except that here the relevant replica limit is ${n\to 1}$ rather than ${n\to 0}$.\footnote{It would not make sense to take $n\to 0$ for the correlators of Eq.~\ref{eq:replicaHgaussian}, because of the factor of $1/n$ in the second coupling. In contrast,  in the replica Hamiltonian for the disordered magnet, the corresponding coupling is proportional to the bond-disorder variance, and does not come with a factor of~$1/n$.}

The global symmetries of the replica Hamiltonian include $S_n$ permutations of the $n$ replicas, as well as an Ising-like $\mathbb{Z}_2$  symmetry for each replica, generating the group 
\be\label{eq:basicsymmetry}
\mathcal{G}_n \equiv 
(\mathbb{Z}_2)^n \rtimes S_n.
\ee
It is known in the random magnet context that there is a special coupling ratio $J/\Gamma=2$ with a \textit{larger} symmetry: at $n=0$ this is the  Nishimori line \cite{Nishimori1981,Nishimori2001,LeDoussalHarris1988,LeDoussalHarris1989,GruzbergReadLudwig2001}, in the replica language. This symmetry also plays a role here, as discussed below.

$\mathcal{H}_n$ can be used to express measurement-averaged correlation functions: for example
the correlator (\ref{eq:edwardsandserson}), which becomes long-ranged
 in the strong-measurement phase, 
can be written (for any $\alpha\neq\beta$) as 
\ba \notag
\mathbb{E} \<S_i S_j\>_M^2
& =\lim_{n\to 1} \< S_{\alpha i} S_{\beta i}  S_{\alpha j} S_{\beta j} \>
\\
& =
\lim_{n\to 1} \< (\Phi_i)_{\alpha\beta} (\Phi_j)_{\alpha\beta}\>.
\label{eq:eaorder}
\end{align}
That is, the strong-measurement phase is characterized by \textit{condensation} (ordering) of the overlap matrix $\Phi$,
\be\label{eq:EAmatrixlattice}
(\Phi_i)_{\alpha\beta} = S_{i\alpha}S_{i\beta} - \delta_{\alpha \beta}.
\ee
In the random-magnet context, i.e.\ for ${n\to 0}$, condensation of $\Phi$ would indicate a spin-glass phase and would not have a simple mean-field description.
For ${n\to 1}$, by contrast, there \textit{is} a simple ordered phase. 
The ordering pattern that we expect is 
\be\label{eq:orderedmoment}
\< \Phi_{\alpha\beta} (x) \>  = \chi_\alpha \chi_\beta  \Phi,
\ee
for arbitrary sign factors ${\chi_\alpha = \pm 1}$ and some value of the  ordered moment ${\Phi>0}$.\footnote{A given choice of signs, such as ${\vec \chi = (1, 1, \ldots, 1)}$, 
spontaneously breaks the global symmetry in Eq.~\ref{eq:basicsymmetry}  down to~${\mathbb{Z}_2\times S_n}$. For example, if we choose the $\chi_\alpha=1$ state, 
then permutations of the replicas, $S_\alpha(x)\rightarrow S_{\pi(\alpha)}(x)$, remain a symmetry ($\pi\in S_n$),
and a simultaneous $\mathbb{Z}_2$ spin flip on all $n$ replicas  remains a symmetry, since this leaves the expectation value of 
$\Phi_{\alpha\beta}$, which is even in the spins, invariant.
If we choose a different ordering pattern $\chi_\alpha$ then the $S_n$ operations come with sign changes.}
There are $2^{N-1}$ equivalent ordered states, defined by the \textit{relative}\footnote{Changing the sign of all of the $\chi_a$ simultaneously does not change the ordering pattern.} signs between different replicas.

$\mathcal{H}_n$ possesses the global  symmetry in Eq.~\ref{eq:basicsymmetry} for any values of the couplings. This symmetry is shared by any bond-energy measurement protocol, for example, those mentioned at the beginning of this Section.
As is well-known in the context of the random-magnet  \cite{Nishimori1981,LeDoussalHarris1988,LeDoussalHarris1989,GruzbergReadLudwig2001,GeorgesHanselLeDoussalMaillard1987}
 (and also of the Ashkin-Teller model for the case $n=2$ \cite{KohmotodenNijsKadanoff1981,Fan1972}),
 a Hamiltonian of the specific form (\ref{eq:replicaHgaussian})
acquires a larger symmetry when the ratio of the coupling for the quartic term to that of the quadratic term is $1/2$, i.e.\ when
\be
\Gamma = J/2
\ee
(we have put $n=1$ in the coupling).
Setting $J=J_c$, this gives ${\Gamma_c = J_c/2}$, which will turn out to be the critical measurement strength for Gaussian measurements of the critical Ising model.

Essentially, at ${\Gamma = J/2}$ there is --- in addition to the symmetry under permutation of replicas $S_\alpha$ --- also symmetry under operations of the form
\be\label{eq:newsymms}
(S_1,  \ldots, S_\alpha,  \ldots, S_n) 
\leftrightarrow 
(S_\alpha S_1, \ldots, S_\alpha, \ldots, S_\alpha S_n)
\ee
for any choice of $\alpha$
(we have omitted the site index $i$).
As a result, correlation functions at this 
value obey exact identities. 
For the random magnet, one of these identities was first obtained by Nishimori using non-replica methods \cite{Nishimori1981} and this was generalized in Refs.~\cite{ledoussalgeorges1988, LeDoussalHarris1988}.\footnote{To avoid possible confusion: as noted in Ref.~\cite{GruzbergReadLudwig2001}, the special feature of (\ref{eq:replicaHgaussian}) for $\Gamma/(nJ)=1/2$ is an enlarged \textit{global} symmetry. 
Eq.~\ref{eq:replicaHgaussian} does not have a {\em gauge} symmetry. See Sec.~\ref{sec:symmclar} for clarification.
However, the standard way to easily see the enlarged global symmetry is to rewrite Eq.~\ref{eq:replicaHgaussian} using a gauge-redundant parameterization, as we review below.}

Let us introduce the $(n+1)\times (n+1)$ matrix 
\be\label{latticesuperfield}
\Psi_i =
\lf
\begin{array}{cc}
    \Phi_i  & \,\,\,\, \vec S_i \\
         \vec S_i^T  & \,\,\,\, 0 
\end{array}\ri,
\ee
whose top-left block was defined in Eq.~\ref{eq:EAmatrixlattice} and where 
${\vec S_i = (S_{1i}, \ldots, S_{n i})^T}$. 
The  entries of $\Psi$ are 0 on the diagonal, and $\pm 1$ elsewhere. For ${\Gamma= nJ/2}$, we get 
\be\label{eq:Hnrewritten}
\mathcal{H}_n= 
- \f{J}{2}\sum_{\< i,j\>} \sum_{\substack{a, b=1 \\ a\neq b}}^{n+1} 
(\Psi_i)_{ab} (\Psi_j)_{ab}.
\ee
This Hamiltonian is invariant under  the enlarged global symmetry group
\be\label{eq:symmgpenhancedlattice}
\mathcal{G}_n^+ \equiv  
\mathbb{Z}_2^n\rtimes S_{n+1},
\ee
where $S_{n+1}$ permutations ${\pi\in S_{n+1}}$  act on $\Psi$ as 
${\Psi_{ab} \to \Psi_{\pi(a),\pi(b)}}$.
This formalizes the consequences of Eq.~\ref{eq:newsymms}, which corresponds to a permutation of the form $(\alpha,n+1)$ in cycle notation.

For completeness, let us recall an alternative standard rewriting (see e.g.\ \cite{GruzbergReadLudwig2001, GeorgesHanselLeDoussalMaillard1987})
via  an ${n+1}$ component spin vector $\widetilde{\vec S}$. 
The  overall sign of this spin vector is gauge-redundant, and the ``physical'' degrees of freedom are written as 
\ba\label{eq:gaugeredundantwriting}
S_{\alpha} = \widetilde S_\alpha \widetilde S_{n+1},
\,\,
\text{or equivalently}
\,\,
(\Psi_i)_{ab} = \widetilde{S}_{a,i} \widetilde{S}_{b,i} - \delta_{ab} .
\end{align}
In this formulation, the
Hamiltonian on the Nishimori line $J=2\Gamma$ takes the form
\be\label{eq:hamiltoniangaugeredundant}
\mathcal{H}_n =
-\Gamma \sum_{\<ij\>} \sum_{\substack{a,b=1 \\ a\neq b }}^{n+1} (\widetilde S_{ai} \widetilde S_{aj})  (\widetilde S_{bi} \widetilde S_{bj}) 
\ee
and the  enlarged permutation symmetry corresponds to permutations of the ${n+1}$ components of $\widetilde{\vec S}$. 
In the context of the Nishimori line, $\widetilde S_{n+1}$ is sometimes referred to as an ``extra replica''. However, note that $\widetilde{\vec S}$ is a gauge-redundant field, and is not  on the same footing as ${\vec S}$.
We will not need this gauge-redundant formalism in this paper.

Since Eq.~\ref{eq:newsymms} exchanges components between the spin field and the overlap field, it leads to useful identities between correlation functions that are well known in the context of the Nishimori line and which also hold for the present $n\to 1$ limit.
By writing the physical correlators in terms of replicas, we see that \cite{LeDoussalHarris1988,LeDoussalHarris1989}
\ba\label{18}
 \mathbb{E} \< S_i S_j\>_M &= \mathbb{E} \< S_i S_j\>_M^2, \, \, \text{and}
\\
 \mathbb{E} \< S_i S_j\>_M^{2k-1} &= \mathbb{E} \< S_i S_j\>_M^{2k} \, \, \text{for $k\in \mathbb{N}_+$}.
\end{align}
The form of these identities is the same here and for the standard Nishimori problem. 
(In the latter case we have an enlarged $S_{n+1}$ symmetry with ${n\to 0}$, while here we have  an enlarged $S_{n+1}$ symmetry with ${n\to 1}$.) 
However, there are also identities that are special to the measurement problem.\footnote{We note that while for convenience we use the replica formulation, the correlator identities can also be obtained without the use of a replica limit, by direct manipulation of e.g.\ the the probability of an overlap configuration.}

A key feature of the ${n\to 1}$ limit
is that any expectation value of the form ${\mathbb{E}\<\cdots\>_M}$ reduces to an expectation value $\<\cdots\>$ in the pure Ising model, without any measurements or disorder. In the presence of enlarged symmetry, this implies some \textit{exact} results for correlators.
In particular, \Eq{18} implies for the Edwards-Anderson correlator  ($r$ is the distance between $i$ and $j$)
\ba\label{eq:EAtrivialization2d}
\mathbb{E} \< S_i S_j\>_M^2 
&\sim r^{- 1/4}  &  & \text{for $d=2$},
\\
\mathbb{E} \< S_i S_j\>_M^2 
&\sim r^{- (d-2)}  &  & \text{for $d>4$}.
\label{eq:EAtrivialization}
\end{align}
The second line holds because the Ising correlator $\<S_iS_j\>$ is free-field-like in ${d>4}$.

The reduction of  correlators of the form $\mathbb{E}\<S_i S_j\ldots\>^2_M$ to correlators in the pure Ising model has a perhaps more intuitive interpretation,
noted already in the introduction.
Let  us temporarily use
$\Phi_i$  (without  replica indices) to denote   the overlap ${\Phi_i =S^{\text{true}}_i S_i}$ between the true measured configration $\{S^{\text{true}}_i\}_i$ and a configuration $\{S_i\}_i$ sampled from the conditioned distribution 
(Bayesian posterior) that 
results from measurement.
Then the  probability distribution of  $\{\Phi_i\}_i$ 
(after marginalizing over $S^\text{true}$, $S$, and the measurements) is
simply the  Boltzmann distribution of the pure Ising model.\footnote{Recall that, in the replica language, we are free to identify (say) the first replica with  $S^\text{true}$, and the second replica with $S$, so that correlators of $\Phi^{12}=S^1S^2$ are equivalent to correlators of $\Phi$ above.}
Loosely speaking, 
if we use the measurement data to ``guess'' a configuration,
the spatial pattern of errors in our guess 
itself looks like a pure Ising configuration.

In  contrast to Eq.~\ref{eq:EAtrivialization},
 higher moments of correlators, such as ${\mathbb{E}\<S_iS_j\>^4}$,
do not reduce to correlators in the pure Ising model, and
 involve new critical exponents even for ${4<d<6}$: we compute these multiscaling exponents in $6-\epsilon$ in Sec.~\ref{sec:RG} and in 2D in Sec.~\ref{sec:montecarlo}.

The   scaling of the Edwards-Anderson correlator in 
Eq.~\ref{eq:EAtrivialization} is nontrivial
--- by contrast,   its exponent would  be twice as large at ${\Gamma=0}$ (since the Edwards-Anderson correlator would just be the square of the spin correlator).  
Just as in the spin glass  problem, this nontrivial scaling helps fix the phase diagram \cite{LeDoussalHarris1988} --- it  strongly suggests\footnote{This is straightforward to see in $d=2$ and in high $d$, assuming the simple phase diagram structure in (\ref{eq:simplephaseidagram}).} that, for  Gaussian measurements, ${\Gamma_c = J_c/2}$ is the critical  point at the boundary of the strong-measurement phase.

So far we have discussed only Gaussian measurements. 
The enlarged  symmetry $\mathcal{G}_n^+$ in Eq.~\ref{eq:symmgpenhancedlattice}
can also hold for some other measurement protocols, as we mention briefly in Sec.~\ref{sec:enhancedsymmetrycondition} and describe in more detail in \cite{enhancedsymmunpublished}.
But, from the point of view of more general lattice models or measurement protocols,  the enlarged symmetry  represents a form of \textit{fine-tuning}.  More generic measurement protocols do not possess this symmetry for any value of the measurement strength.

For example, if the measurements are made for next-nearest neighbors $\langle\langle i,j\rangle\rangle$ instead of nearest neighbors, we clearly eliminate the symmetry between the two terms in Eq.~\ref{eq:replicaHgaussian}.
The protocol involving binary measurements with error probability $p_\text{err}$ described in Sec.~\ref{sec:setup} also breaks the replica symmetry, because in addition to the terms of the form ${(\Phi_i)_{\alpha\beta} (\Phi_j)_{\alpha\beta} = (S_i S_j)_\alpha  (S_i S_j)_\beta}$,
we obtain higher-order terms starting with 
$(S_i S_j)_\alpha  (S_i S_j)_\beta
(S_i S_j)_\gamma  (S_i S_j)_\delta$.
 We discuss lattice models without the enlarged symmetry in Sec.~\ref{sec:montecarlo}.

Since these  more general models do not 
possess the enlarged symmetry microscopically,
an important question is whether they acquire the larger symmetry in the IR. 
If they do, then a single universality class, with the enlarged symmetry (\ref{eq:symmgpenhancedlattice}) and obeying the above correlation function identities, could govern the transition into the strong-measurement phase
for generic measurement protocols of the critical Ising model.
Similar symmetry considerations are relevant to the analogous ${n\to 0}$ problem, i.e.\ to the multicritical point in the (temperature, disorder) plane of the random-bond Ising model \cite{GruzbergReadLudwig2001}.\footnote{This multicritical point is commonly studied in models with a Nishimori line, where enlarged symmetry holds microscopically. 
It is suspected \cite{GruzbergReadLudwig2001}  that even models without a Nishimori line show the same universality class for the multicritical point. 
This is true in $6-\epsilon$ dimensions \cite{LeDoussalHarris1989} but we are not aware of a direct numerical test in low dimensions (though there have been tests of universality between different models \textit{with} a Nishimori line \cite{DeQueiroz2006,DeQueiroz2009,AgrawalCugliandoloFaoroIoffePicco2024}). 
Such a test would be more challenging in the random magnet problem than in the measurement problem, because of the need to simultaneously tune two parameters to find the multicritical point. In the measurement problem, once the physical $T_c$ is known, then only the measurement strength $\Gamma$ needs to be tuned to criticality.}
In the following Sections
we will address this question for the measurement problem in $6-\epsilon$ dimensions and in two dimensions, showing that $\mathcal{G}_n^+$ does indeed emerge generically at the critical point.

Finally we recall that, if our measurement protocol is even more fine-tuned, it is possible for the replica Hamiltonian to have an $S_{2^n}$ global symmetry that is even larger than Eq.~\ref{eq:symmgpenhancedlattice} \cite{NahumJacobsen2025}.
This symmetry imposes the stronger identity 
\be\label{eq:strongeridentity}
{\mathbb{E} \<S_i S_j\>_M^k = \<S_i S_j\>}
\ee
for any $k$. This symmetry arises in a fairly simple binary measurement protocol in which 
bonds with  ${S_iS_j=-1}$  are always correctly reported (i.e.\ yield  $M_{ij}=-1$) but bonds with  ${S_iS_j=+1}$ are reported incorrectly with a critical error probability. 
However, it was argued that  this fine-tuning is infinitely unstable in the RG sense \cite{NahumJacobsen2025}
when we consider the (in principle) infinite-dimensional space of more general measurement protocols. Therefore, we will not consider this fine-tuned situation further in this paper.

\subsection{General criterion for enhanced replica symmetry}
\label{sec:enhancedsymmetrycondition}

In the previous Section we saw that, for Gaussian measurements, there is an exact microscopic enhancement of replica symmetry when $\Gamma=J/2$. 
It is natural to expect that 
replica symmetry enhancement can occur for other measurement protocols.
In  Ref.~\cite{enhancedsymmunpublished} we discuss the general criterion for this to occur: here we summarize some basic outcomes.

This criterion is an analog of the  ``Nishimori criterion'' which ensures identities for correlation functions
in the random-bond Ising model. 
In that case, the bond coupling $J_{ij}$ is random, and the  criterion constrains the bond-disorder distribution $p(J_{ij})$. 
In the present setting, we require a criterion for the  probability  $P(M_{ij}|\mathcal{E}_{ij})$ of the bond measurement outcome \textit{conditioned} on the true  value of the bond $\mathcal{E}_{ij}=S_iS_j$.
For example, in the Gaussian case this probability density had the form ${P(M|\mathcal{E})\propto e^{-\Gamma (M-\mathcal{E})^2}}$.

For simplicity, consider first the case where $M$ is a continuous variable  whose conditional probability density function has the special form 
\be\label{eq:symmetrycondit1}
P(M|\mathcal{E}) = f(M) e^{M \mathcal{E}}.
\ee
In this special case, one may  show that there is an exact, microscopic enlarged  replica symmetry if the indentity
\be\label{eq:symmetrycondit2}
P\big( M\big|-1\big) = P\big(-M-2J\big|-1\big)
\ee
is obeyed \cite{enhancedsymmunpublished}.
If we take $P(M|\mathcal{E})$ to be the probability of outcome $M$,
the equations above also guarantee symmetry in the case where the measurements are drawn from a discrete set of outcomes (where the set has the property that if  $M$ is a possible outcome, so is ${-M-2J}$).

At first sight (\ref{eq:symmetrycondit1})  only allows for probability distributions of a  special form. 
However, a key feature of the measurement problem, with no analog in the disordered systems problem, is that we can \textit{relabel} the measurement outcomes $M$ without changing the physics.
Given an arbitrary initial density 
$P_0(M|\mathcal{E})$, we can bring it into the form (\ref{eq:symmetrycondit1}) by the change of variable
\be\label{eq:measurementchangeofvariable}
M' = \f{1}{2} \ln \lf \frac{P_0(M|+1)}{P_0(M|-1)} \ri. 
\ee
Here $M'$ is the new name for the outcome previously denoted by $M$.\footnote{Two distinct $M$ values, say $M_1$ and $M_2$, can be mapped to the same $M'$. But in this case, the outcomes $M_1$ and $M_2$ give exactly the same information about the system, and it is not necessary to distinguish them.}
This allows (\ref{eq:symmetrycondit2}) to be extended to the general case.

To be more concrete, let us restrict for simplicity to the setting of discrete measurement outcomes.\footnote{We assume that all measurement outcomes give distinct values for the right-hand side of Eq.~\ref{eq:measurementchangeofvariable}, because outcomes with the same value of this quantity can simply be treated as the same outcome.} Then the general criterion  for the enlarged symmetry to be present (microscopically) is that 
there exists a bijection between measurement outcomes, ${M\mapsto \overline{M}}$, such that,
first, outcomes $M$ and $\overline M$ are equally likely when the bond is unsatisfied ($\mathcal{E}=-1$),
\be
P\big(\overline{M}\big|-1\big) = P\big({M}\big|-1\big),
\ee
and second,
\be
\frac{P\big(\overline{M}\big| +1\big)}{P\big(\overline{M}\big| -1\big)}
=
\exp(-4J)
\frac{P\big({M}\big|-1\big)}{P\big({M}\big|+1\big)}.
\ee
Remarkably, these relations show that microscopic enhanced symmetry can be achieved even for a   protocol with binary measurement outcomes, if we allow the  probability of incorrectly measuring a \textit{satisfied} bond ($\mathcal{E}=1$) to differ from the probability of correctly measuring an \textit{unsatisfied} bond ($\mathcal{E}=1$) \cite{enhancedsymmunpublished}.\footnote{In the notation of Sec.~IV~C of Ref.~\cite{NahumJacobsen2025}, this requires ${p_-=1/2}$, ${p_+=\f{1}{2}\lf
1+\sqrt{1-e^{-4J}}
\ri}$.}

\subsection{Clarification: global symmetries and gauge redundancies}\label{sec:symmclar}

This Section is not necessary for the subsequent development, 
but we summarize the symmetry properties of the replica theory in various cases of interest, in order to clarify some potentially confusing points. See also Ref.~\cite{GruzbergReadLudwig2001}, which  emphasized that the special property of the replica theory on the Nishimori line is an enlarged global symmetry, not an enlarged gauge symmetry.

Consider the replica Hamiltonian $\mathcal{H}_n$ in Eq.~\ref{eq:replicaHgaussian}. We ignore the factor of $1/n$ in the second coupling, since it tends to 1 in the case of interest to us, and would be absent if we were using replicas to describe the spin glass rather than the measurement problem.

The internal global symmetry of $\mathcal{H}_n$ is\footnote{The case $n=2$, where the spin $(S_{1i},S_{2i})$ takes four values, is the Ashkin-Teller model
(two Ising models coupled via their energy densities). The model at $J=2\Gamma$ is equivalent to the 4-state Potts model \cite{Fan1972,KohmotodenNijsKadanoff1981}, and its symmetry group $\mathcal{G}_2^+$ is equivalent to  the permutation group $S_4$ acting on the four spin values.}
\ba\label{eq:symmsummaryG}
\mathcal{G}_n&  \equiv  \mathbb{Z}_2^n\rtimes S_n ,
& & \text{for $J/\Gamma$ generic},
\\
\label{eq:symmsummaryGplus}
\mathcal{G}_n^+&  \equiv  \mathbb{Z}_2^n\rtimes S_{n+1} ,
& & \text{for $J/\Gamma=2$},
\end{align}
as discussed above.
Note that $\mathcal{G}_n^+$ is not the same thing as $\mathcal{G}_{n+1}$. 
Nevertheless, there is a relation between theories with $n$ and $n+1$ replicas \cite{GeorgesHanselLeDoussalMaillard1987}, as discussed below. 

The two cases above are the ones relevant to this paper, but 
another special case is where $J=0$. 
In this case, the $\mathbb{Z}_2$ symmetry operation which flips all of the replicas simultaneously, $\vec{S}_{ i}\rightarrow -\vec{S}_{ i}$, 
is promoted to a local symmetry (i.e.\ 
one that can be performed  independently at each site $i$). 
In a loose notation we may then write the internal symmetry group as
\ba\label{eq:symmsummaryJ0}
& (\mathbb{Z}_2^\text{local} \times \mathbb{Z}_2^{n-1})\rtimes S_n ,
& & \text{for $J=0$}.
\end{align}
We refer to a ``local symmetry'', rather than a gauge redundancy, 
because we view  $\vec S$ as our physical degree of freedom: $\vec S$ and $-\vec S$ are physically distinct states.
(In the $n\to 1$ context, the case $J=0$ describes measurement in the infinite temperature state; local symmetries appear generically when we measure ``singlet'' degrees of freedom in paramagnets  \cite{NahumJacobsen2025}.)

Finally, let us check that these statements about symmetry are compatible with the known relation between theories at different $n$ \cite{GeorgesHanselLeDoussalMaillard1987, GruzbergReadLudwig2001}.
To review this relation: 
We are free to rewrite the spin in terms of the   gauge-redundant variable ${(\widetilde S_1, \ldots, \widetilde S_{n+1})}$ as in Eq.~\ref{eq:gaugeredundantwriting}.
On the Nishimori line ${J=2\Gamma}$, the resulting expression (\ref{eq:hamiltoniangaugeredundant}) for $\mathcal{H}_n$  resembles the Hamiltonian for ${J=0}$, 
except that the spin is now replaced with 
$\widetilde{\vec S}$, which has $n+1$ components instead of $n$. This maps the partition function of the $n$-replica system at ${J=2\Gamma}$ to that of the ${n+1}$-replica system at $J=0$ (up to a constant).

The physical symmetry of the theory at $J=2\Gamma$ is $\mathcal{G}_n^+$, as noted above.
However, when we rewrite the Hamiltonian using $\widetilde S$, we introduce a gauge redundancy, 
so that the rewritten Hamiltonian 
is now invariant under transformations from ${\mathbb{Z}_2^\text{gauge} \times \mathcal{G}_n^+}$, i.e.\ from
\be
(\mathbb{Z}_2^\text{gauge} \times \mathbb{Z}_2^{n}) \rtimes S_{n+1}.
\ee
We see that this is indeed formally the same symmetry group as we have for the $J=0$ problem (Eq.~\ref{eq:symmsummaryJ0}) with $n+1$ replicas, the only difference being in the interpretation of the first $\mathbb{Z}_2$ factor (which is now a gauge redundancy rather than a local symmetry that acts on physical variables).

We now return to the main development and to field theory descriptions of the measurement problem.

\section{Landau theory for the measurement transition}
\label{Landau theory for the measurement transition}

In this Section we describe the field theory that is useful for studying  the transition into the strong-measurement phase at $\Gamma_c$ in ${6-\epsilon}$.
We conjecture that this fixed point can be continued down to ${d=4}$ without encountering any nonanalyticities.
As explained below, we in fact expect that the fixed point can be followed all the way down to $d=2$, but with a nonanalyticity in the critical exponents at ${d=4}$.

As discussed in Sec.~\ref{sec:setup}, we expect that the nature of the adjacent \textit{phase} at  ${\Gamma<\Gamma_c}$ changes at a critical value of ${d=6-\epsilon}$, becoming multifractal at some value of $\epsilon$ between 0 and 2.
Some aspects of the phase diagram in intermediate dimensions
have not previously been resolved, and our results for the fixed point will clarify this.

\subsection{Lagrangians}
\label{sec:lagrangians}

The universal behavior of the unmeasured ensemble is captured by the usual $\phi^4$ Lagrangian $\mathcal{L}_*(\phi)$, 
where  $\phi$ is a continuum analog of the lattice spin $S$.
When the measurements are weak (small $\Gamma$) their effect is   captured by a replica theory of the form
\be\label{eq:Lphionly}
\mathcal{L}_n = 
\sum_{\alpha =1}^n \mathcal{L}_*(\phi_\alpha) 
 -   \Gamma \sum_{\alpha \neq\beta} \phi_\alpha^2 \phi_\beta^2.
\ee
where 
$ \phi_\alpha^2$ 
is the leading term when we expand the measured lattice observable in terms of continuum operators.

A basic conclusion from the quartic Lagrangian above is that \textit{weak} measurements are irrelevant for ${d>4}$. 
However, even for ${d>4}$ there is a strong-measurement phase at large enough $\Gamma$. In order to study the transition into this phase it is useful to switch to a different field-theory representation \cite{NahumJacobsen2025}, which allows an epsilon expansion around $d=6$ dimensions.\footnote{The quartic Lagrangian can  be used to show that, just above four dimensions, there is an unstable RG fixed point at ${\Gamma=O(d-4)}$.
At first glance, we may think that this fixed point was the same as the one which we will study in a $6-\epsilon$ expansion. However, as we discuss later, we do not expect that this is the case. The likely RG flow topology near four dimensions is shown in the right-hand panel of Fig.~4 in \cite{NahumJacobsen2025}.}

In the alternative formulation the replica overlap,
which on the lattice reads 
\be\label{eq:EAmatrix}
{\Phi_{\alpha\beta} = S_\alpha S_\beta - \delta_{\alpha\beta} \< S_\alpha^2\>}\qquad (\text{on lattice}),
\ee
is promoted to an additional \textit{independent} continuum field $\Phi_{\alpha\beta}$.  It is symmetric, $\Phi_{\a\b}= \Phi_{\b\a}$, and has zeros on the diagonal, $\Phi_{\a\a}=0$.
This is standard  in the spin glass \cite{OerdingCornellBray1997,ChenLubensky1977, LeDoussalHarris1988} (and in other contexts, e.g.~\cite{FeiGiombiKlebanov2014}).
One heuristic way to motivate the introduction of the continuum field $\Phi_{\alpha\beta}$ is through a Hubbard-Stratonovich decoupling of the $\Gamma$ term in Eq.~\ref{eq:Lphionly}. 
 Instead we simply postulate the continuum theory on grounds of symmetry and matching of phases.

As discussed around Eq.~\ref{eq:eaorder},
the Edwards-Anderson correlator (\ref{eq:edwardsandserson}) becomes long-ranged
 in the strong-measurement phase.
In the field theory this is the two-point function of $\Phi_{\alpha\beta}$.
That is, the strong-measurement phase is characterized by the condensation of the $\Phi$ field.

Writing the leading symmetry-allowed terms for $\phi$ and $\Phi$ gives the Lagrangian already stated in the introduction:
\begin{eqnarray}
\notag
\mathcal{L}_n
& = & 
 \frac1{2} \sum_{\alpha=1}^n\! \Big[ (\nabla \phi_{\alpha})^2  {+} r_1 \phi_\alpha^2 \Big]
{+}
 \frac1{4} \sum_{\alpha,\beta=1}^n \!\Big[ (\nabla \Phi_{\alpha\beta})^2 {+} r_2   \Phi_{\alpha\beta}^2  \Big] 
\\
   & +&    \frac{\lambda_1}{2} \sum_{\alpha,\beta=1}^n \phi_{\alpha}  \phi_{ \beta } \Phi_{ \alpha \beta } 
+ 
 \frac{\lambda_2}{6}  \sum_{\alpha,\beta,\gamma=1}^n \Phi_{\alpha \beta}\Phi_{ \beta\gamma }\Phi_{\gamma \alpha }
 \label{eq:introducecubicL}
\end{eqnarray}
Recall that the diagonal elements of the symmetric matrix $\Phi_{\alpha\beta}$ vanish; the combinatorial factors compensate for the multiple occurences of equivalent terms (e.g.\ due to the identification $\Phi_{\a\b} \equiv \Phi_{\b \a}$). 

In the remainder of this Section we assume that ${d>4}$. 
Therefore we  may neglect quartic terms such as $\sum_\alpha \phi_\alpha^4$  in $\mathcal{L}_n$, which are  irrelevant for $d>4$.\footnote{The quartic term $\phi_\alpha^4$  is dangerously irrelevant above 4D (as  $\phi^4$ is in standard $\phi^4$ theory) and must be taken into account for some purposes, such as computing the ordered Ising moment below the critical temperature \cite{CardyBook}.
However it will not play any role in the RG flows above 4D.} 
The Lagrangian above has previously been applied, in the limit ${n\to 0}$, to disordered magnets near the Nishimori line \cite{LeDoussalHarris1988,LeDoussalHarris1989,ChenLubensky1977}.

In the spirit of Landau-Ginsburg theory, we may imagine that the values of the couplings ${(r_1,r_2,\lambda_1,\lambda_2)}$ in the \textit{effective} Lagrangian are all functions of the microscopic temperature $T$ and of the microscopic measurement strength  $\Gamma$. The transition we are interested in is formally a multicritical point $(T_c, \Gamma_c)$ in the two-dimensional parameter space $(T,\Gamma)$.
Near this point we expect\footnote{On the left hand side we have as usual absorbed a shift of the critical value of $r_2$ that arises from  fluctuations.}  (recall that we have absorbed the inverse temperature into $J$, so 
${J_c-J \propto T-T_c}$):
\ba\label{eq:massidentification1}
r_1 & \simeq - c_1 (J - J_c),
\\ \label{eq:massidentification2}
r_2 & \simeq- c_2 (\Gamma- \Gamma_c) - c_3 (J-J_c),
\end{align}
where $c_1, c_2$, and (naively) $c_3$ are all positive constants. 
Eq.~\ref{eq:massidentification1} follows straightforwardly
from the relation between the replica theory and the original ``unmeasured'' Ising model, recalled in  Sec.~\ref{sec:onereplicareduction} immediately below.
Eq.~\ref{eq:massidentification2} follows from identifying the phase in which $\Phi_{\alpha\beta}$ has a negative squared mass with a condensed phase for $\Phi_{\a \b}$, i.e.\ with the strong-measurement phase (see the discussion around Eqs.~\ref{eq:eaorder},~\ref{eq:orderedmoment}).
Increasing $\Gamma$ or  $J$ both make it easier to infer long-distance correlations, hence the signs in Eq.~\ref{eq:massidentification2}.\footnote{For the model with Gaussian measurements in Eq.~\ref{eq:replicaHgaussian}
we have ${2c_1=c_2+2c_3}$, so that the   enhanced  symmetry line $J=2\Gamma$ of the lattice model maps to 
the line $r_1=r_2$ where the field theory can have the enhanced symmetry (Sec.~\ref{sec:enhancedsymmfieldtheory}).}

To avoid confusion, we now set ${J=J_c}$, 
so that we are studying a  \textit{one-dimensional} phase diagram for $\Gamma$, and as in the discussion at the beginning of this Section we use the word ``strong-measurement phase'' to indicate the range ${\Gamma \in (\Gamma_c, \infty]}$.
We will comment on the two-dimensional parameter space later.

After this restriction to the thermal critical point, 
\ba\label{eq:massidentification12}
r_1 & =0,
\\ \label{eq:massidentification22}
r_2 & \simeq - c_2 (\Gamma- \Gamma_c).
\end{align}
The bare values of the couplings $\lambda_1$ and $\lambda_2$ are generically nonvanishing at the critical point of interest.
Mean-field arguments suggest they should both be negative.\footnote{We expect that ${\lambda_1<0}$ so that $\Phi_{12}$ is positively correlated with $\phi_1 \phi_2$ . This sign can also be motivated using the  Hubbard-Stratonovich transformation mentioned above. If ${\lambda_2<0}$ 
then when ${r_2<0}$ there are saddle-point solutions with a positive ordered moment ${\Phi = r_2/\lambda_2}$. In the field theory, changing $\Phi_{\a\b} \to - \Phi_{\a\b}$ changes the signs of both $\lambda_1$ and~$\lambda_2$.}

When $r_2$ is positive, the naive ``mean field''  guess is that we should flow all the way to ${r_2=\infty}$. Assuming that this is the case sufficiently close to ${d=6}$, then in this range of dimensions the phase diagram as a function of $\Gamma$ has only two phases: 
the strong-measurement phase where  $\Phi$ is condensed, 
and a phase where $\Phi$ is massive and 
can be integrated out, leaving only irrelevant quartic interactions between the $\phi_\alpha$.
The latter is the ``trivial'' weak-measurement phase where measurements are irrelevant.
This justifies the picture in Sec.~\ref{sec:setup} for the structure of the phase diagram in sufficiently high dimensions.

Next we discuss a simple but important property of the ${n\to 1}$ limit, where various correlators and RG equations simplify.

\subsection{Single-replica  sector: free field reduction}
\label{sec:onereplicareduction}

The squared mass $r_1=r_1(T)$ of $\phi$ 
has a protected role: it  is set by the distance from the critical temperature in the microscopic Ising model,
and is completely unaffected by the measurement strength: we have
${r_1(T)\sim (T-T_c)}$ as in the usual Ising Landau-Ginsburg theory.
To see this, we recall a basic general feature of the  replica formalism for measurements.\footnote{See Sec. ~III~C of \cite{NahumJacobsen2025} for further discussion.}  
If we set ${n=1}$ directly in the Lagrangian, so that the terms with multiple replica indices vanish, the resulting Lagrangian, 
\ba\label{eq:freeL}
\mathcal{L_\text{free}}
= & 
 \frac1{2} (\nabla \phi)^2 
+\frac{r_1}{2 }  \phi^2,
\end{align}
describes the standard Ising model without
conditioning on measurement.
That is, $r_1$ is a property of the unmeasured Ising model, so  depends only on $T$, in the standard way.

Correlators that  involve a single replica index, such as  ${\lim_{n\to 1} \< \phi_1(x) \phi_1(y)\>  = \<\phi(x)\phi(y)\>}$, reduce to correlators in the  free theory (\ref{eq:freeL}). 
Therefore, when $n\to 1$, the scaling 
dimension of the field $\phi_\alpha$ 
is protected, 
remaining equal to its free-field value ${\Delta_{1} = (d-2)/2}$
independently of the values of the couplings $r_2$, $\lambda_1$ and $\lambda_2$.

\subsection{Enhanced symmetry line}
\label{sec:enhancedsymmfieldtheory}

To avoid clutter let us now  restrict to the massless theory, so that we are concerned only with the couplings $(\lambda_1, \lambda_2)$. In this space, the line  
\be
\lambda_1=\lambda_2
\ee
has an enhanced symmetry.
As done for the lattice in \Eq{latticesuperfield}, 
on this line it is convenient to package the fields into an 
enlarged ${(n+1)\times (n+1)}$ matrix. In our field-theory notation it reads
\be\label{superfield}
\Psi =
\lf
\begin{array}{cc}
     \Phi  & \phi \\
         \phi^T  & 0 
\end{array}\ri
\ee
and the Lagrangian becomes 
\be\label{eq:Lsymmetricline}
\mathcal{L}_n^\text{symm} = \f{1}{4} \sum_{a,b=1}^{n+1} (\nabla \Psi_{ab})^2 + 
\f{\lambda}{6} \sum_{a,b,c=1}^{n+1}  \Psi_{ab}\Psi_{bc}\Psi_{ca},
\ee
where ${\lambda = \lambda_1=\lambda_2}$.

We may now permute the $n+1$ index values for $a$ without changing the action. 
As a result, the action on the symmetric line has the enlarged ${\mathcal{G}_n^+ = \mathbb{Z}_2^n\rtimes S_{n+1}}$ symmetry that was discussed on the lattice around Eq.~\ref{eq:symmgpenhancedlattice},
rather than just the  
${\mathcal{G}_n = \mathbb{Z}_2^n\rtimes S_{n}}$
symmetry that is present at generic points in the $(\lambda_1, \lambda_2)$ plane.

In the following we will find that in the massless theory the RG for ${n=1}$ drives the theory to the symmetric line
(as also happens at $n=0$ \cite{LeDoussalHarris1989}), 
so that the enlarged symmetry emerges in the IR.

Remarkably, the replica limit ${n\to 1}$ implies that the field $\Phi$ has a vanishing anomalous dimension on the symmetric line, despite the nontrivial cubic interaction.
This vanishing follows from the free-field reduction in Sec.~\ref{sec:onereplicareduction} (which fixes the scaling dimension of $\phi$) together with the enlarged symmetry, which forces all components of $\Psi$ to have the same dimension.

The rewriting in Eq.~\ref{eq:Lsymmetricline} also reveals a relation between the theories for $n$ and $n+1$ replicas.
(This is reminiscent of the
relation between lattice models in Sec.~\ref{sec:symmclar}, though not equivalent.)
Above we obtained the Lagrangian (\ref{eq:Lsymmetricline}) from the theory $\mathcal{L}_n$ for $n$ replicas, by imposing 
\be
(\lambda_1, \lambda_2) = (\lambda, \lambda)
\qquad
\text{for $\mathcal{L}_n$}.
\ee
We obtain a Lagrangian of the same form
from the theory $\mathcal{L}_{n+1}$ with ${n+1}$ replicas,
by first imposing 
\be
(\lambda_1, \lambda_2) = (0, \lambda) \qquad \text{for $\mathcal{L}_{n+1}$},
\ee
and then neglecting the decoupled $\phi$ field.
This relationship between $\mathcal{L}_n$ and $\mathcal{L}_{n+1}$ provides a check on the RG equations for $\lambda_1$, $\lambda_2$.

\subsection{Comparison between  $n\to 1$ and $n\to 0$ limits}\label{sec:nvalues}

The Lagrangian (\ref{eq:introducecubicL}) was studied long ago in the ${n\to 0}$ limit, in the context of disordered magnets \cite{ChenLubensky1977,LeDoussalHarris1988,LeDoussalHarris1989}. There, $\phi$ is the magnetization and $\Phi$ the Edwards-Anderson spin-glass order parameter. The point where both are massless is the Nishimori multicritical point, which, in high enough dimensions, is a meeting point of the ferromagnetic phase, the spin glass phase, and the paramagnetic phase.\footnote{Slightly confusingly, the Nishimori fixed point is also related to Bayesian inference/measurement of the \textit{infinite} temperature Ising model \cite{Iba1999,ZdeborovaKrzakala2016}. 
In the present language, we can see this by starting with $\mathcal{L}_n$ for $n\to 1$ which describes the measurement problem.
In order to describe measurement of the Ising model at high temperature, we turn on a large mass for the field $\phi$, eliminating it from the IR theory. We are left with a theory for $\Phi$  only, at $n\to 1$. By the mapping in the previous Section, this is equivalent, after a change of variables to $\mathcal{L}_{n}$ at ${n\to 0}$, with coupling constants on the symmetric line.}

Here we are interested in the ${n\to 1}$ model which describes measurements of the critical Ising model. 
The replica trick allows perturbation theory to be done for both problems simultaneously.
The one-loop results in \cite{LeDoussalHarris1988,LeDoussalHarris1989,ChenLubensky1977} and the present two-loop results can be applied to both problems, inserting the appropriate value of $n$. 

However, there are significant  differences between ${n=0}$ and ${n=1}$.
As we will see, the structure of the RG flows is different near $d=6$, 
which is the reason why a 2-loop calculation is required to determine whether or not a fixed point exists for $n=1$.
The reduction discussed in  Sec.~\ref{sec:onereplicareduction} also enforces constraints on the phase diagram and scaling dimensions that are specific to $n=1$.

The phase that one obtains when one attempts to condense $\Phi$ is also very different in the two different replica limits. Consider for simplicity the regime of large positive $r_1$ and large negative $r_2$.
In the ${n\to 1}$ limit this phase has a simple desription in terms of the order parameter in Eq.~\ref{eq:eaorder}. 
By contrast, in the disordered magnet  (${n\to 0}$) this is the spin-glass phase, which does not have a simple description of this form. 

Some of the above differences between $n=0$ and $n=1$ are relevant to a wider range of replica Lagrangians.
Another structural difference between $n=0$ and $n=1$ is shown by the c-theorem obtained for a  class of weakly measured 2D systems in 
Ref.~\cite{PatilLudwig2025}.\footnote{The conventional c-theorem \cite{Mussardo2010} does not apply in either replica limit.}  
A positivity property for a correlation function at $n=1$ implies that the  effective central charge decreases under RG for a large class of RG flows induced by weak measurement, whereas the effective central charge typically increases in the analogous flows for disordered systems~\cite{PatilLudwig2025}.

\section{Renormalization in $6-\epsilon$ dimensions}
\label{sec:RG}

We have argued that the massless cubic theory $\mathcal{L}_n$
describes the phase transition of   the critical Ising model at the critical measurement strength. We now describe its RG flows in ${6-\epsilon}$ dimensions, in the plane ${(\lambda_1, \lambda_2)}$ of the two cubic couplings.

Recall from  Sec.~\ref{sec:enhancedsymmfieldtheory} that the line ${\lambda_1=\lambda_2}$ has an enhanced ${\mathcal{G}_n^+=\mathbb{Z}_2^n\rtimes S_{n+1}}$ global replica symmetry.  Lattice-measurement protocols that  have this exact, \textit{microscopic} ${\mathcal{G}_n^+}$ 
symmetry (at their transition point $\Gamma_c$) correspond to points on this line, once we drop irrelevant higher couplings. Lattice-measurement protocols that do not have the ${\mathcal{G}_n^+}$ symmetry correspond to points away from this line, with ${\lambda_1 \neq \lambda_2}$.

The basic questions are whether nontrivial fixed points exist within the epsilon expansion, and what their 
their stability and  symmetry are.\footnote{A Landau theory for the quantum measurement phase transition gives an example where the RG flows run  to strong coupling below 6 dimensions, instead of showing a perturbatively-accessible fixed point \cite{NahumWiese2023}.}
In order to describe the \textit{generic} version of the phase transition, the fixed point should be stable in both directions of the ${(\lambda_1, \lambda_2)}$ plane.\footnote{Recall that we have already tuned the two masses to zero, corresponding to tuning the Ising model to its critical temperature, and  the measurement strength to its critical value.} 
Below we find that there is a stable fixed point in $d=6-\epsilon$ dimensions, and that it lies on the higher-symmetry line ${\lambda_1=\lambda_2}$. This implies that the enlarged symmetry   emerges without fine-tuning at least close to the upper-critical dimensions.

What happens as we descend in $d$?  As we discuss in the following paragraph, the simplest hypothesis is that we can follow the $6-\epsilon$ fixed point down to low dimensions, with the exponents evolving continuously, but with a nonanalyticity in $d=4$.
This  nonanalyticity arises from a quartic coupling $\lambda_\mathrm{WF}$ in the replica theory,
which becomes the coupling of the usual ${\phi^4}$ term of the Ising model when we set $n=1$.\footnote{In more detail: The symmetries of the replica theory allow several quartic couplings, and $\lambda_{\rm WF}$ is the linear combination which sets the coupling of $\phi^4$ when ${n=1}$.
Consider the $\mathcal{G}_n^+$-symmetric theory for simplicity. The allowed quartic terms are
${g_1 \Psi_{ab} \Psi_{ab} \Psi_{cd} \Psi_{cd} / 4}
 +
 {g_2 \Psi_{ab} \Psi_{ab} \Psi_{ab} \Psi_{ab} / 2}
 +
 {g_3 \Psi_{ab} \Psi_{bc} \Psi_{cd} \Psi_{da} / 2}
 +
 {g_4 \Psi_{ab} \Psi_{ab} \Psi_{ac} \Psi_{ac} / 2}$. We may check that this reduces to $\lambda_{\rm WF}\phi^4$ at ${n=1}$, with $\lambda_{\rm WF}= {g_1 + g_2 +g_3 + g_4}$.}
 
In general, we must distinguish coupling constants 
like $\lambda_{\rm WF}$, 
which multiply operators that remain nonzero at $n=1$, 
from those
(like the cubic couplings $\lambda_{1,2}$)
which multiply operators that vanish at $n=1$.
The beta functions of the former are  independent of the latter at ${n=1}$
(see Sec.~IIIC of \cite{NahumJacobsen2025} for further discussion).
Here the beta function for $\lambda_{\mathrm{WF}}$ 
is independent of the cubic couplings and is just the usual Wilson-Fisher beta function for the Ising model.
Therefore the (stable) fixed-point value of $\lambda_{\mathrm{WF}}$ becomes nonzero below four dimensions: ${\lambda_{\mathrm{WF}}\propto \operatorname{max} \{0, \, (4-d) + \ldots \}}$.
This nonanalyticity will infect all exponents, because $\lambda_{\mathrm{WF}}$ can appear in the beta functions for all other couplings.\footnote{It may be possible to eliminate this nonanalyticity by measuring a tricritical (or tetracritical, etc.) Ising model; this remains to be explored.}

Similar RG flows arise in the $Q$-state Potts model in the limit of ${Q\to 2}$ \cite{WieseJacobsen2024}.
Though the cubic replica theory is nontrivial in $6-\epsilon$ dimensions
--- as reflected in the geometrical properties of Fortuin Kasteleyn clusters --- 
the single-replica sector reduces to the standard Ising model, and is agnostic to the cubic couplings. Loop models  show a similar effect~\cite{NahumChalkerSernaOrtunoSomoza2013}.

Ref.~\cite{NahumJacobsen2025} gave two speculative scenarios for the evolution with $d$ of the RG flows for the monitored critical Ising model. The picture  above is consistent with the second scenario, but not with the first.\footnote{The assumption that the $6-\epsilon$ fixed point with enlarged symmetry remains stable down to $d=4$ is sufficient to rule out the first scenario in Fig.~4 of \cite{NahumJacobsen2025}. In that scenario, the $6-\epsilon$ fixed point would be identified with a weak-coupling fixed point that exists in $4+\varepsilon$ dimensions. But the latter fixed point manifestly does not have the enlarged symmetry, giving a contradiction.} The second scenario also agrees with the naive expectation that  the transition we are studying, which is driven by changing the sign of the mass$^2$ for $\Phi$, lies on the boundary of the $\Phi$-ordered (strong measurement) phase, for any value of $d$.

\subsection{RG flows}

The RG equations $\partial_\ell \lambda_i = \beta_i(\lambda_1, \lambda_2)$ (where $\ell$ is the logarithm of a lengthscale)  give the IR flow of the coupling constants $\lambda_i$. They are obtained via dimensional regularization and minimal subtraction in a massless scheme.   We refer to App.~\ref{Summary of the RG procedure} for details of our procedure.

\begin{figure}[t]
\fig{0.9}{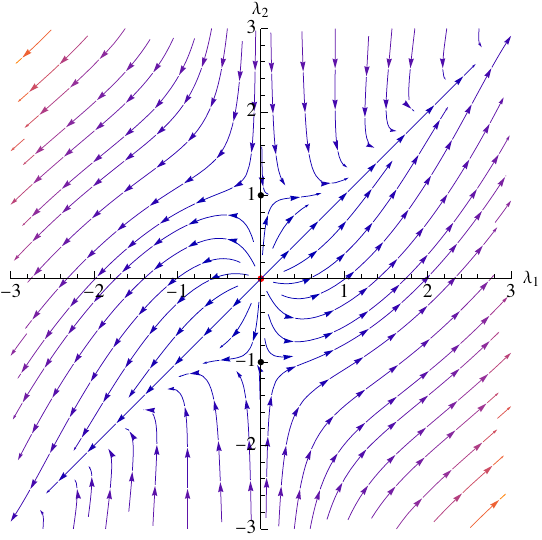}
\caption{Flow diagram at 1-loop order for $n=1$, $\epsilon=1$.}
\label{flow-n=1}
\end{figure}

After setting ${n=1}$ (equations for general $n$ are given  below) we have
\begin{align}\label{beta1-n=1}
&\!\!\!\beta_1(\lambda_1,\lambda_2) = \frac{\lambda _1 \epsilon }{2}+\frac{1}{6}    \left(5 \lambda _1^3{-}6 \lambda
   _2 \lambda _1^2{+}\lambda _2^2 \lambda _1\right) \nn\\
&{+}\frac{1}{108}   \left(21
   \lambda _1^5{-}69 \lambda _2 \lambda _1^4{-}152 \lambda _2^2 \lambda _1^3{-}15
   \lambda _2^3 \lambda _1^2{-}\lambda _2^4 \lambda _1\right)  + ... \\
&\!\!\!\beta_2(\lambda_1,\lambda_2) = \frac{\lambda _2 \epsilon }{2}{+}\frac{1}{2}    \left(2 \lambda _1^3{-}\lambda _2
   \lambda _1^2{-}\lambda _2^3\right)\nn\\
   &{+}\frac{1}{36}   \left(27 \lambda
   _1^5{+}69 \lambda _2 \lambda _1^4{-}3 \lambda _2^2 \lambda _1^3{+}10 \lambda _2^3
   \lambda _1^2-175 \lambda _2^5\right) 
    + ...  \label{beta2-n=1}
\end{align}
In the spin-glass problem, one is interested in ${n = 0}$: in that case, the one-loop RG equations show a nontrivial fixed point \cite{LeDoussalHarris1988,LeDoussalHarris1989,ChenLubensky1977}.
However,  at ${n=1}$, the one-loop RG flows, which are shown in Fig.~\ref{flow-n=1}, are attracted to the diagonal $\lambda_1 = \lambda_2$, but do not possess a nontrivial fixed point.   
This is  seen from the flows on the higher-symmetry line 
$\lambda_1 = \lambda_2 = \lambda$, which are
\bea\label{beta-iso-n=1}
\beta_1(\lambda,\lambda) &=& \beta_2(\lambda,\lambda)  = \frac{\lambda  \epsilon }{2}-2  \lambda ^5+\ldots,
\eea 
\begin{figure}[t]
\fig{0.9}{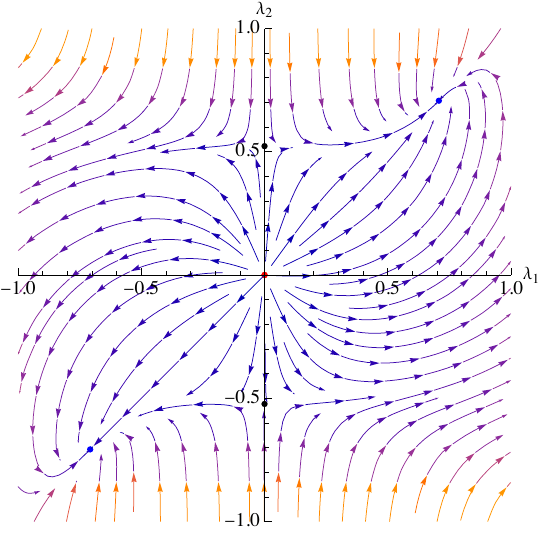}
\caption{Flow at 2-loop order for  $n=1$, $\epsilon=1$. Compared to Fig.~\ref{flow-n=1}, a non-trivial fixed point has appeared at $\lambda_1 = \lambda_2 = \lambda^*$, see \Eq{lambda*}.}
\label{flow-n=2}
\end{figure}%
There is no 1-loop term $\sim \lambda^3$, so the flow topology is left undetermined at one-loop order.
However we see that the two-loop term   prevents the flow from running to strong coupling. 
Fig.~\ref{flow-n=2} shows the two-loop flows; there is indeed a stable fixed point (or rather two equivalent such fixed points, related by changing the signs of both couplings, which amounts  to reversing the sign convention for $\Phi$). The stable fixed point (for positive $\lambda_1=\lambda_2=\lambda$) is at 
\be\label{lambda*}
\lambda^* =  \sqrt[4]{\frac{\epsilon}4}.
\ee 
We describe the properties of the fixed point in the next subsection.

The form of the enlarged symmetry discussed in Sec.~\ref{sec:enhancedsymmfieldtheory} implies a relation between the beta functions $\beta^n_i$ for different $n$ 
which is a check on our results.
For ${\lambda_1=0}$   the field $\phi$ decouples, and the resulting theory for $\Phi$ resembles  the theory for the {\em superfield} $\Psi$   on the symmetric line 
(see Eqs.~\ref{superfield}-\ref{eq:Lsymmetricline}), but with one  replica fewer. 
As a result, our $\beta$-functions need to satisfy
\be\label{expeceted-relations}
\beta_2^{n}(0,\lambda) = \beta_1^{n-1}(\lambda,\lambda)=\beta_2^{n-1}(\lambda,\lambda)
\ee
There is also a correspondence between the fixed point on the vertical axis for $n=m+1$ replicas and the fixed point on the symmetric line for $n=m$.

The beta functions for general $n$ read 
\begin{align}\label{beta1-all-n}
&\beta_1(\lambda_1,\lambda_2) = \frac{\lambda _1
   \epsilon }{2} \nn\\
&+\frac{1}{6} \left(7 \lambda_1^3{-}12 \lambda _2 \lambda _1^2{+}2 \lambda _2^2 \lambda _1{-}2 \lambda _1^3 n{+}6    \lambda _2 \lambda _1^2 n{-}\lambda _2^2 \lambda _1 n\right)
 \nn\\
 & + \frac{1}{108} \Big[{-}59 \lambda _1^5{-}228 \lambda _2 \lambda _1^4{-}174 \lambda _2^2
   \lambda _1^3{-}60 \lambda _2^3 \lambda _1^2{-}4 \lambda _2^4 \lambda _1\nn\\
&\qquad\quad~~ {+}11 \lambda_1^5 n^2{-}45 \lambda _2 \lambda _1^4 n^2{+}65 \lambda _2^2 \lambda _1^3 n^2
{-}15\lambda _2^3 \lambda _1^2 n^2
\nn\\
& \qquad\quad ~~ {-}\lambda _2^4 \lambda _1 n^2{+}69 \lambda _1^5 n{+}204
   \lambda _2 \lambda _1^4 n{-}43 \lambda _2^2 \lambda _1^3 n\nn\\
& \qquad\quad ~~   {+}60 \lambda _2^3
   \lambda _1^2 n{+}4 \lambda _2^4 \lambda _1 n\Big]   ,
\\
\label{beta2-all-n}
&\beta_2(\lambda_1,\lambda_2) = \frac{\lambda _2   \epsilon }{2}  + \frac{1}{2} \left(2 \lambda _1^3{-}\lambda _2 \lambda _1^2{-}2 \lambda _2^3{+}\lambda
   _2^3 n\right)\nn\\
   &{+}  \frac{1}{36} \Big[48 \lambda _1^5{+}31 \lambda _2 \lambda _1^4{-}6 \lambda _2^2
   \lambda _1^3{+}20 \lambda _2^3 \lambda _1^2{-}268 \lambda _2^5{+}5 \lambda _2^5
   n^2\nn\\
&\qquad~~ {-}21 \lambda _1^5 n{+}38 \lambda _2 \lambda _1^4 n{+}3 \lambda _2^2 \lambda _1^3
   n{-}10 \lambda _2^3 \lambda _1^2 n{+}88 \lambda _2^5 n\Big].  \nn
  \\
 \end{align} 
On the symmetric line, they reduce to  
\begin{align}
\label{beta-isotropic}
\beta_1(\lambda,\lambda) &= \beta_2(\lambda,\lambda)= \frac{\lambda  \epsilon }{2}+\frac{1}{2}   \lambda ^3 (n-1) \nn\\
&+ 
    \lambda ^5 \Big[ \frac{5}{36} (n-1)^2+3 (n-1)-2 \Big] + ...
\end{align}
For $n<1$ the fixed point is at 
\be \label{lambda*-n<1}
(\lambda^*)^2 =  { \frac{\epsilon }{1-n}+\frac{ 175-n (5 n+98) }{18 (n-1)^3} \epsilon^2 + \ca O (\epsilon ^3) }
\ee
For the check mentioned above we set $\lambda_1=0$, in which case the $\beta$-function for $\lambda_2$ must reduce to the one of  \Eq{beta-isotropic} after changing $n\to n-1$
\begin{align}
\beta_2(0,\lambda) & = \frac{\lambda  \epsilon }{2}+\frac{1}{2}    \lambda ^3 (n-2) \nn\\
&+ 
   \lambda ^5 \left[ \frac{5}{36} (n-2)^2+3 (n-2)-2\right] +...
\end{align}
We see that the expected relations \eqK{expeceted-relations} are obeyed. The results above for general $n$ are also applicable to the spin glass problem and to the standard Nishimori multicritical point, by taking $n\to 0$.

For measurements, $n=1$, and Eq.~\ref{beta-isotropic} reduces to Eq.~\ref{beta-iso-n=1} with a fixed point at $\lambda^*$, see Eq.~\ref{lambda*}.

\subsection{Critical exponents and multiscaling}
\label{sec:critexponentsummary}

We now discuss critical exponents.
We give a summary of the results for the measurement problem ($n=1$) here, and defer more detailed results for general $n$, some of which are also relevant to the spin glass multicritical point,
to the following Section.
While the  discussion is framed in terms of the $\epsilon$ expansion, many basic features are relevant to low dimensions as well.

First, let us consider correlation length exponents.  The critical point we are discussing is formally a multicritical point:  it arises when we tune the temperature to the critical temperature $T_c$
and the measurement strength to the 
critical value $\Gamma_c$.
In the field theory this means tuning both $r_1$ and $r_2$ to zero.
These two mass terms yield two RG-relevant scaling variables,
and therefore two correlation length exponents, 
which govern the lengthscales that emerge when we traverse the critical point along various lines in parameter space.

One of the correlation length exponents is the standard value for a thermal perturbation in the Ising model (above~4D):
\be\label{eq:nuthermal}
\f{1}{\nu_T} = 2.
\ee
This simple value is special to $n=1$
(Sec.~\ref{sec:onereplicareduction}); 
for ${n\neq 1}$ there are nontrivial corrections (Sec.~\ref{sec:massterms}).
The other correlation length exponent is nontrivial at ${n=1}$:
\be\label{eq:nuM}
\frac{1}{\nu_M} = 2 - \sqrt{\epsilon} + \ca O(\epsilon).
\ee
The scaling variable associated to (\ref{eq:nuthermal}) is simply $r_1$, and that associated to (\ref{eq:nuM}) is ${r_1-r_2}$.\footnote{These linear combinations correspond to \textit{left} eigenvectors of the matrix in the RG equation, see Sec.~\ref{sec:massterms}.}
In terms of these scaling variables the mass terms take the form
\ba
\delta \mathcal{L}
&= 
\f{r_1}{4}
\sum_{a,b=1}^{n+1}  \Psi_{ab}^2
+ 
\f{r_2-r_1}{4} 
\sum_{\alpha,\beta=1}^n \Phi_\alpha^2.
\label{eq:massrewritten}
\end{align}
Note that the first scaling variable preserves the enlarged $\mathcal{G}_n^+$ symmetry.
The special property of the second scaling variable in Eq.~\ref{eq:massrewritten} is that it has no effect on single-replica correlators.

As noted around Eq.~\ref{eq:massidentification1}, $r_1(T)$ is independent of the microscopic measurement rate $\Gamma$, and varies only with the microscopic temperature~$T$.
For the single-parameter phase diagram depicted in (\ref{eq:simplephaseidagram}), 
 the  temperature is held fixed at  $T_c$, 
and only the measurement strength $\Gamma$ is varied (Eq.~\ref{eq:massidentification22}).
This phase diagram therefore corresponds to the line  ${r_1=0}$ in the $r_1$, $r_2$ plane, as shown in Fig.~\ref{fig:2dphasediag}.

Therefore
 $\nu_M$ in  (\ref{eq:nuM}) 
sets the scaling of the correlation length 
for the measurement phase transition in  (\ref{eq:simplephaseidagram}).
For example, the Edwards-Anderson correlator will cross over to its asymptotic form 
at a lengthscale 
\be
{\xi_\text{EA}\sim |\Gamma-\Gamma_c|^{-\nu_M}},
\ee
in either of the two adjacent phases.

Note that the values in Eqs.~\ref{eq:nuthermal},~\ref{eq:nuM} imply that the ``thermal'' perturbation $r_1$ is more RG-relevant than the perturbation  induced by varying to the measurement strength. 
This ordering dictates the geometry of the phase boundaries  \cite{LeDoussalHarris1989} 
that was shown in Fig.~\ref{fig:2dphasediag}, 
for the phase diagram in the two-dimensional $(r_1, r_2)$ plane
[corresponding to the two-dimensional $(J,\Gamma)$ plane].
The phase boundary for the ``paramagnet measurement transition'' is asymptotically parallel to the thermal phase boundary at $r_1=0$ [i.e.\ at $J=J_c$].

Next we consider the leading irrelevant perturbations to the action.
Once we tune to the critical point of a given microscopic model,
we  still in general have ``corrections to scaling'' of order $L^{-\omega_i}$ (on lengthscale $L$) 
that are due to irrelevant perturbations with RG eigenvalues $-\omega_i$.
Here the leading such perturbations are combinations of ${\delta\lambda_i=\lambda_i-\lambda_*}$, and as discussed in Sec.~\ref{sec:correctionscaling}
they have the values
\begin{eqnarray}
\label{omega1*}
    \omega_1^{n=1} &=&  2 \epsilon +\ca O(\epsilon^{3/2}), \\
    \omega_2^{n=1} &=&\frac{2
   \sqrt{\epsilon }}{3}+\ca O(\epsilon).
   \label{omega2*}
\end{eqnarray}
The  smaller exponent, $\omega_1$, is associated with  perturbations that preserve the enlarged symmetry.

Finally  we consider the moments 
\be\label{eq:momentgeneral}
{\mathbb{E}\<S(x)S(y)\>_M^l}
\sim |x-y|^{-2 \Delta_l}
\ee
of the physical correlation function in the microscopic model.

As discussed in
Sec.~\ref{sec:enhancedsymmfieldtheory}, the anomalous dimension of the field $\phi_\alpha$
is forced to vanish at $n=1$.
The emergent symmetry of the fixed point means that the anomalous dimension of $\Phi_{\alpha\beta}$ also vanishes,
so that the Edwards-Anderson correlator 
has the simple scaling already noted in
Eq.~\ref{eq:EAtrivialization}.
However, composite operators can have nontrivial dimensions.

Using replicas,
\be\label{eq:momentgeneral}
{\mathbb{E}\<S(x)S(y)\>_M^l}
=
\lim_{n\to 1} \< \lf S_1 S_2\cdots S_l \ri (x)
\lf S_1 S_2\cdots S_l \ri (y) \>.
\ee
Therefore we wish to map the lattice operator 
$S_1 S_2\cdots S_l$ to a continuum operator.
The leading operators with the same symmetry 
properties are 
\ba
\mathcal{O}_{2k-1} & = [\phi_1(x)\Phi_{2,3}(x) ... \Phi_{2k-2,2k-1}(x)]_{S},
\\
\mathcal{O}_{2k} & = [\Phi_{1,2}(x) ... \Phi_{2k-1,2k}(x)]_S,
\end{align}
for $l=2k-1$  and $l = 2k$ respectively.
Here, the ${[\cdots]_{S}}$ indicates that we symmetrize over all possible ways of assigning the indices $1,2, \ldots, l$ to the operators inside the brackets:\footnote{This symmetrization ensures that  
$\mathcal{O}_l$ is invariant under permutations of the first $l$ replica indices, just as the lattice operator $S_1 S_2 \cdots S_l$ is. 
Note that $\mathcal{O}_l$ also has the same quantum numbers under $\mathbb{Z}_2^n$ as $S_1S_2\cdots S_l$.} for example,
\be
\mathcal{O}_3 = \phi_1 \Phi_{23} + \phi_2 \Phi_{13} + \phi_3  \Phi_{12}
\ee
(the normalization is arbitrary).

The multiscaling exponents $\Delta_l$ are the scaling dimensions of the $\mathcal{O}_l$.
The emergent ${\mathcal{G}_n^+ = \mathbb{Z}_2^n\rtimes S_{n+1}}$ symmetry at the fixed point  unites $\mathcal{O}_{2k-1}$ and $\mathcal{O}_{2k}$ into the same multiplet, ensuring that $\Delta_{2k-1}=\Delta_{2k}$ as expected.

Let us write
\ba\label{eq:mfexpts}
\Delta_{2k-1} = \Delta_{2k} 
= k \Delta^\text{Ising} + \delta \Delta_{2k},
\end{align}
where
\be
\Delta_\text{Ising} = \f{d-2}{2}
\ee
is scaling exponent for the unconditioned Ising correlator (given here by free field theory, since $d>4$).
We find in Sec.~\ref{sec:multifraceps} that the spectrum of multiscaling exponents is given to leading order in $\epsilon$ by 
\be
\delta \Delta_{2k}  = 
-k (k-1)\sqrt{\epsilon} + \ldots.
\ee
This confirms the existence of multiscaling within the epsilon expansion.\footnote{Above six dimensions,  the exponents take free field values. However the cubic interactions are dangerously irrelevant there, since they appear in the leading scaling of some correlators such as 
${{\mathbb{E} \<S(x) S(y)\>_M \<S(x) S(z)\>_M}\sim
{\< \Phi_{12}(x) \phi_1(y) \phi_2(z)\>}}$ and 
${{\mathbb{E}\hspace{-0.2mm} \< S(x)S(y)\>_M\hspace{-0.9mm} 
 \< S(y)S(z)\>_M \hspace{-0.9mm} 
\< S(z)S(x)\>_M} \hspace{-0.9mm} \sim\hspace{-1mm} 
{\< \Phi_{31}(x)\Phi_{12}(y)\Phi_{23}(x)\>}}$. The right-hand sides of these expressions would   vanish if we set the cubic couplings exactly to zero. Perturbation theory gives e.g.\  ${\< \Phi_{31}(x)\Phi_{12}(y)\Phi_{23}(z)\>} {\sim \lambda_2 R^{-2(d-3)}}$ for separations of order $R$.}
We have commented on   the relation between the $6-\epsilon$ fixed point and the low-dimensional fixed point  at the beginning of Sec.~\ref{sec:RG}: the   simplest conjecture is that the critical exponents  evolve continuously from $6$ dimensions down to $1+\delta$ dimensions, but with a nonanalyticity at $d=4$.

\subsection{Exponents for general $n$  and for the spin glass}

In the following subections we give more information about the RG flows, including results for other $n$ values and results away from the symmetric fixed point.

\subsubsection{Correction-to-scaling exponents $\omega$}
\label{sec:correctionscaling}

To assess the  stability of the fixed point
in the $(\lambda_1, \lambda_2)$ plane, we set  ${\lambda_i=\lambda^*+\delta \lambda_i}$. The $\beta$-function becomes 
\begin{equation}
\beta_i(\lambda_1,\lambda_2)  = \beta_i(\lambda^*,\lambda^*)  -  \sum_j \Omega_{ij} \delta \lambda_j + \ca O(\delta \lambda^2),
\end{equation}
with the stability matrix 
\begin{equation}
 \Omega_{ij}:= -   \frac{\partial }{\partial \lambda_j} \beta_i(\lambda_1,\lambda_2).
\end{equation}
Its  eigenvalues $\omega_i$  read, for general $n$,
\begin{eqnarray}
    \omega_1 &=& -\frac{\epsilon }{2} -\frac{3}{2} \lambda ^2
   (n{-}1)-\frac{5}{36} \lambda ^4 \left(5 n^2{+}98
   n{-}175\right)+... \\
   \omega_2 &=& -\frac{\epsilon }{2}-\frac{1}{6} \lambda ^2 (5
   n{-}13)- \frac{1}{108} \lambda ^4 \left(39 n^2{+}1048
   n{-}3011\right)\nn\\
   &&+ ...
\end{eqnarray}
Following standard conventions, 
these are  (minus) the RG eigenvalues of the leading irrelevant scaling 
operators at the critical point.
The eigenvectors are 
 $(1,1)$  for $\omega_1$ and 
${(2-n,3)}$ for $\omega_2$.

Assuming   a fixed point, we see that at leading order in $\lambda$,   $\omega_1$ changes from attractive for $n<1$ to repulsive for $n>1$. 
The (slightly lengthy) result for $n<1$ is given in \Eqs{a:SRomega1}-\eqK{a:SRomega2} of App.~\ref{app:RGfunctionsandexponents}.
For spin glasses ($n=0$) they read
\begin{eqnarray}
    \omega_1^{n=0} &=& \epsilon +\frac{175 \epsilon
   ^2}{18} +\ca O(\epsilon^{3}) \\
    \omega_2^{n=0} &=& \frac{5 \epsilon
   }{3}+\frac{184 \epsilon ^2}{27} +\ca O(\epsilon^{3})
\end{eqnarray}
The result for $n=1$ was given in Eqs.~\ref{omega1*},~\ref{omega1*}. 
Since, at $n=1$,
we know   $\lambda^*$ only to leading order
(see Eq.~\ref{lambda*}) we do not have the subleading terms for $\omega_i^{n=1}$; the latter  require the 3-loop results for the $\beta$-functions.

\subsubsection{Anomalous field dimensions $\eta$}
\label{s:Anomalous field dimensions: eta}

Each of the two fields $\phi$ and $\Phi$ renormalizes, and 
for general $n$ aquires an anomalous dimension $\eta$. 
The RG functions are 
\bea
\eta_1(\lambda_1,\lambda_2) &=&  \frac{ \lambda _1^2}{3}     (n{-}1) -\frac{ \lambda _1^2 (n{-}1)}{108}  \nn\\
& & \times
   \Big[\lambda _1^2 (11 n{-}24){-}24 \lambda _2 \lambda _1
   (n{-}2)+11 \lambda _2^2 (n{-}2)\Big]\nn\\
   && +...  \\
\eta_2(\lambda_1,\lambda_2) &=&    \frac{\lambda _1^2}{3}+\frac{\lambda _2^2}{3} 
   (n{-}2) +\frac{1}{54}  \Big[\lambda _1^4 (23{-}11 n)\nn\\
   && {+}24 \lambda _2
   \lambda _1^3 (n{-}2)-11 \lambda _2^2 \lambda _1^2 (n{-}2)+\lambda _2^4
   (n{-}2)^2\Big] \nn\\
   &&+...
\eea 
At the isotropic fixed point
\be
\eta_1(\lambda,\lambda) = \eta_2(\lambda,\lambda) = \frac{  \lambda ^2}{3}   (n-1) +\frac{\lambda ^4}{54}     (n-1)^2+... 
\ee
Thus there are no perturbtative corrections to either $\eta_1$ or $\eta_2$ at $n=1$. This is expected.  First, $\eta_1$ is the anomalous dimension of the Ising spin, which vanishes in dimensions $d>4$ and so must be zero in our $\epsilon$-expansion. The enhanced symmetry then implies that the same result holds for the field $\Phi$. 

For $n<1$,
\begin{equation}
    \eta_1 = \eta_2 =-\frac{\epsilon }{3}+\frac{(88-2 n (n+25)) \epsilon ^2}{27
   (n-1)^2}+\ca O(\epsilon ^3).
\end{equation}
Note that the limit $n\to 1$ is highly singular. 

For the Nishimori multicritical point in  spin glasses (${n=0}$), 
\begin{equation}
    \eta_1 = \eta _2 = -\frac{\epsilon }{3}+\frac{88 \epsilon
   ^2}{27}+\ca O (\epsilon ^3).
\end{equation}
The one-loop term is in agreement with Refs.~\cite{LeDoussalHarris1989,ChenLubensky1977}.

\subsubsection{Mass insertions: $\nu$}\label{sec:massterms}

In Sec.~\ref{sec:correctionscaling} we considered moving away from the fixed point by  changing $\lambda_i \to \lambda^*+\delta \lambda_i$.
Now we consider perturbating by a (squared) mass $r_1$ for $\phi$ or $r_2$ for $\Phi$. 
(The interpretation of the masses in terms of microscopic couplings was given in Eq.~\ref{eq:massidentification1}). 
The squared masses $r_1$ and $r_2$ mix under RG. To leading order one has to evaluate 
\begin{equation}
\mathcal M_{ij}(\lambda_1,\lambda_2) :=  \frac {\delta r_i^ {\rm eff} }{\delta r_j} = \left( \begin{array}{cc}
    (n{-}1)  \lambda_1^2     &   (n{-}1) \lambda_1^2  \\
     2 \lambda_1^2    &  2  (n{-}2) \lambda_2^2
    \end{array}\right)\triagdiag .
\end{equation}
To get the contribution for the anomalous mass dimensions, 
one varies the RG scale, and subtracts the contributions  $\eta_i$  from the fields, 
\begin{equation}
\delta \gamma_{ij}(\lambda_1,\lambda_2):=   -\mu \frac{\partial}{\partial  \mu }  \mathbb M_{ij}(\lambda_1,\lambda_2)    -  \eta_i \delta_{ij},
\end{equation}
To leading order 
\begin{equation}
    \delta \gamma_{ij} = \left( \begin{array}{cc}
    \frac23 (n-1)  \lambda_1^2     &   (n-1) \lambda_1^2  \\
     2 \lambda_1^2  ~  &~\frac{1}{3} \left[5 \lambda _2^2
   (n-2)-\lambda _1^2\right]
    \end{array}\right) . 
\end{equation}
Eigenvalues as well as right and left eigenvectors at 2-loop order are 
\begin{align}
\label{79a}
\delta \gamma_1 &= \frac{5}{3} 
   (1{-}n)\lambda ^2 -\frac{49}{27} 
   (1{-}n)^2 \lambda ^4, & \\  
v_1^{\rm r}     &= \left({1 \atop 1 }\right), \qquad   v_1^{\rm l}  =\left({1 , \frac{n-1}2  }\right) \\
   \label{80a}
\delta \gamma_2 &= \frac{2}{3}
   \lambda ^2 (4{-}n) {-}\frac{269 {-}115
   n{+}8 n^2}{54} \lambda^4 , \\
    v_2^{\rm r} &= \left({\frac{1-n}2 \atop 1 }\right), \qquad    v_2^{\rm l} = \left( 1, -1 \right) .
\end{align}
The left eigenvectors dictate the linear combinations of $r_1$ and $r_2$ that form scaling variables, while the right eigenvectors set the linear combinations of $\f{1}{2}\sum_\alpha \phi_\alpha^2$ and $\f{1}{4}\sum_{\alpha\beta}\Phi_{\alpha\beta}^2$ that form scaling operators.

The eigenvalues at the fixed point give the two inverse correlation length exponents. 
Their values at ${n=1}$, 
 ${1/\nu_T=2}$
 and 
 $ 1/\nu_M={2 - \sqrt{\epsilon} + \ca O(\epsilon)}$,
 were given in
Sec.~\ref{sec:critexponentsummary}.
The RG eigenvalue $1/\nu_T$ corresponds to a perturbation that preserves the enlarged symmetry of the fixed point,
while $1/\nu_M$ corresponds to a perturbation that breaks it.
The vanishing of the correction to $1/\nu_T$ at this value of $n$ is  required by the the single-replica reduction in Sec.~\ref{sec:onereplicareduction}. Just as for $\omega$ in \Eq{omega2*}, to improve on  the $O(\sqrt{\epsilon})$ result for $\nu_M$  we would need to go to 3-loop order for $\lambda^*$. 
We  defer these higher-loop corrections to \cite{DavletbaevaKompanietsWiese2026}, where results at 5-loop order will be given.

For $n<1$, where the  flow has a   fixed point already at 1-loop order, given in \Eq{lambda*-n<1}, we obtain 
\ba
\f{1}{\nu_{\rm s}} & = 2 - \f{5\epsilon}{3}+\frac{\left(973 -686
   n+ 73 n^2\right) \epsilon ^2}{54
   (n-1)^2}, \\
\f{1}{\nu_{\rm a}} & = 2  - \f{2(4{-}n)\epsilon}{3(1{-}n)}
 + \frac{\left(1669 {-}1518
   n {+}279 n^2 {+} 2 n^3\right) \epsilon ^2}{54
   (1-n)^3}.
\end{align}
Here $1/\nu_{\rm s}$ is the RG eigenvalue of the perturbation that preserves the $\mathcal{G}_n^+$ symmetry, and $1/\nu_{\rm a}$ is that of the perturbation that breaks it.
The 1-loop terms agree with \cite{ChenLubensky1977,LeDoussalHarris1989}.
For the spin glass problem  we find
\begin{align}
  \frac{ 1}{\nu_{\rm s}} &= 2-\frac{5\epsilon}3 + \frac{973 \epsilon ^2}{54}+ \ca O(\epsilon^3), \\
  \frac1{\nu_{\rm a}} &=2-\frac{8\epsilon}3+\frac{1669 \epsilon ^2}{54}+ \ca O(\epsilon^3).
\end{align}
Higher-loop results are deferred to~\cite{DavletbaevaKompanietsWiese2026}.

\subsubsection{Multifractal  exponents $\Delta_l$}
\label{sec:multifraceps}

In Sec.~\ref{sec:critexponentsummary} we noted that the multiscaling exponents $\Delta_l$ were the scaling dimensions of the operators 
\ba
\mathcal{O}_{1} & = \phi_1,
& 
\mathcal{O}_{2} & = \Phi_{12},
\\
\mathcal{O}_{3} & = \phi_1 \Phi_{23} + \ldots,
& 
\mathcal{O}_{4} & = \Phi_{12}\Phi_{34} + \ldots,
\end{align}
where the ellipses represent symmetrization over the possible index orderings.
These operators are continuum avatars of the lattice operators ${\mathcal{O}_l^\text{latt} = S_1S_2\cdots S_l}$.

In general, renormalization of products of $\phi$ and $\Phi$ operators is tedious due to operator mixing, but the task simplifies considerably when all indices are distinct, as above. 
Consider what two $\Phi$ operators or one $\Phi$ and one $\phi$ generate at 1-loop order:
\bea
\Phi_{1,2} \Phi_{3,4} &\to&  2 \lambda_2^2 \left[ \Phi_{1,4} \Phi_{2,3} + \Phi_{1,3} \Phi_{2,4} \right] \triagdiag \\
 \phi_{3}  \Phi_{1,2}&\to&  \lambda_1 (\lambda_1 + \lambda_2) \left[ \phi_{2} \Phi_{1,3} + \phi_{1} \Phi_{2,3} \right]  \triagdiag
\eea
After symmetrization, and at the fixed point, both expressions 
 yield the same constant.
In general, for $k$ $\Phi$ operators, or ${(k-1)}$  $\Phi$ operators and one $\phi$ operator, this becomes
\bea
\ca O_{2k}  &\to& 2 k (k-1) \lambda^2 \ca O_{2k} \triagdiag \\
\ca O_{2k-1}  &\to& 2 k (k-1) \lambda^2 \ca O_{2k-1} \triagdiag
\eea
As a consequence, writing, for integer $k$
\begin{eqnarray}\label{eq:Deltageneral1}
\Delta_{2k} &=& { k\f{d-2+\eta}{2}} +  \delta \Delta_{2k} \\
\Delta_{2k-1} &=&  { k\f{d-2+\eta}{2}} +  \delta \Delta_{2k-1}
  \label{eq:Deltageneral2}
\end{eqnarray}
we have 
\be\label{eq:anomdelta}
\delta \Delta_{2k-1}  =
 \delta \Delta_{2k} =  - 2 k (k-1)\lambda^2.
\ee
At $n=1$, where $\eta=0$, the non-trivial fixed point \eqK{lambda*} yields the result in  Eq.~\ref{eq:mfexpts}.

The result in Eq.~\ref{eq:anomdelta} also gives the multifractal exponents for the spin glass multicritical point to $O(\epsilon)$,
after inserting the appropriate $n=0$ fixed-point coupling value and the $n=0$ value $\eta=-\epsilon/3$ in Eqs.~\ref{eq:Deltageneral1},~\ref{eq:Deltageneral2}.

According to the one-loop expression, 
$\Delta_{2k}/(2k)$ is linear in $k$.
 For a fixed $\epsilon$, the one-loop approximation must become poor at large $k$, because otherwise Eq.~\ref{eq:anomdelta} would predict negative exponents for large $k$ (cf. Ref.~\cite{BrezinHikami1996}).
Perturbation theory predicts more generally that
\begin{equation}
    \frac{\Delta_{2k} }{2k} = \eta + \ca A_1 (k-1) + \ca A_2 (k-1) (k-2) + ...
\end{equation}
The leading term $\mathcal{A}_1$ appears at 1-loop order, and comes with a combinatorial factor of ${( {2k \atop 2})/(2k)\sim (k-1)}$. At next order, we can connect up to four $\Phi$ fields, leading to the next two terms in the expansion, and so on. As a result, $\ca A_1$ receives corrections at all orders, while $\ca A_{4\ell}$ starts at $\ell$-loop order.   
We leave a more systematic exploration of these observables for future work \cite{DavletbaevaKompanietsWiese2026}.

\section{Simulation results in two dimensions}
\label{sec:montecarlo}

In 2D, fixing the Ising model to be at its critical temperature, and varying measurement strength, we have a direct transition between the weak-measurement phase and the strong-measurement phase (Eq.~\ref{eq:simplephaseidagram}) \cite{NahumJacobsen2025,PutzGarrattNishimoriTrebstZhu2026}.

This transition was studied numerically in Ref.~\cite{PutzGarrattNishimoriTrebstZhu2026} (for binary measurements) using tensor network techniques to evaluate the Ising partition function in each realization of $M$. 
Here we employ direct Monte Carlo sampling. This limits us to smaller system sizes. 
Nevertheless, we will provide new information, both  about exact and emergent symmetries in 2D, and about the multiscaling  of correlations.

Our aims are (1) 
to demonstrate the existence of a critical fixed point with enlarged ${\mathbb{Z}_2\rtimes S_{n+1}}$ symmetry; 
(2)   to characterize the multiscaling properties of this critical point; 
and  (3)  to explore the whether  this universality  class remains stable when we change the model in a way that reduces the \textit{microscopic} replica symmetry (at the critical point) from  ${\mathbb{Z}_2\rtimes S_{n+1}}$ to the more generic ${\mathbb{Z}_2\rtimes S_{n}}$. 
We will start with the Gaussian measurement protocol, where the enlarged symmetry is present microscopically at the critical point, and then look at other protocols. 

\subsection{Simulation method}\label{sec:simmethod}

 Recall that quantities such as ${\mathbb{E} \<\cdots\>_M^l}$ or
${\mathbb{E} \ln \<\cdots\>_M}$ involve both the average $\mathbb{E}$ over measurement realizations ${M=\{M_{ij}\}}$
and the conditioned ``thermal'' average
$\<\cdots\>_M$.

The most direct sampling approach is as follows. 
First, to generate a measurement realization, we produce a thermalized configuration $\{S^\text{ref}_i\}$ of the Ising model using Metropolis dynamics, and then pick $M_{ij}$, for each bond $\<ij\>$, using the distribution $P(M_{ij}|S^\text{ref}_i S^\text{ref}_j)$ that defines our measurement protocol. 
Then, $\<\cdots\>_M$ can be sampled by performing a long simulation using the ``effective'' Hamiltonian $\mathcal{H}_M$ that defines the conditioned ensemble $P(S|M)$ (this was shown in 
Eq.~\ref{eq:HeffSM} for the Gaussian case\footnote{More generally, $\mathcal{H}_M[S] = \mathcal{H}[S] - \sum_{\<ij\> }\ln P(M_{ij}|S_iS_j)$.}).

An alternative method which is preferable for quantities of the form ${\mathbb{E} \<\cdots\>_M^l}$  is to use ``real'' replicas \cite{MarinariMartinMayorParisiRicciTersenghiRuizLorenzo2024}.
We write 
\be\label{eq:realreplicaexpl}
\mathbb{E} \<\mathcal{O}\>_M^l
=
\mathbb{E} \<\mathcal{O}_1\mathcal{O}_2\cdots \mathcal{O}_l\>_M,
\ee
where now the thermal average involves  $l$  
samples which are equilibrated independently using $\mathcal{H}_M$  for the same $M$.
Therefore the average can be taken by running $l$ (or more generally $N\geq l$) 
samples simultaneously in the Monte Carlo, with the same $\mathcal{H}_M$.
Below we will use up to $N=6$ real replicas.

All simulations are for 2D Ising on the square lattice with periodic boundary conditions.
The temperature is fixed to its (known)
critical value throughout this Section,
and we study the measurement transition as a function of measurement strength $\Gamma$.
Details of the simulations, such as the warm-up times and lengths of runs, are given in App.~\ref{a:Sim_details}.

\subsection{Observables}
\label{sec:numericalobservables}

The simplest order parameter for the transition is the replica overlap $\Phi_{\alpha\beta}$ (Eq.~\ref{eq:EAmatrixlattice}).
For the purposes of this Section we  denote the spatially averaged order parameter,  expressed using two ``real'' replicas $S_{1}$ and $S_{2}$, by $\Phi$:
\be
\Phi=\frac{1}{|\Lambda|}\sum_{x\in \Lambda}S_{1x}S_{2x},
\ee
where ${\Lambda = [1,L]^2}$ is the spatial lattice.

 Using $\Phi$, we may define a Binder cumulant, which is a useful tool for locating the critical measurement strength $\Gamma_c$ in cases where we do not know it analytically.
 Slightly modifying the standard definition of the Binder cumulant also gives a useful diagnostic for enlarged symmetry.

We define the modified\footnote{Recall that the standard Binder cumulant for an order parameter $\Phi$ is ${U=1 -{\<\Phi^4\>}/{(3\langle\Phi^2\rangle^2)}}$.} Binder parameter as 
\be\label{eq:bindermod}
A(L,\Gamma)=\frac{\<\Phi^4\>}{\<\Phi^2\>^2}\frac{\<S^2\>^2}{\<S^4\>}-1,
\ee
where
\be
S=\frac{1}{|\Lambda|}\sum_{x\in \Lambda} S_x.
\ee
As in the conventional Binder cumulant, the ratios are chosen so that 
the $L$-dependences of  numerator and denominator cancel out at the critical point.\footnote{Note that $\<\Phi^k\>\sim L^{-k\Delta_2}$ and $\<S^k\>\sim L^{-k\Delta_1}$, where $\Delta_2$ and $\Delta_1$ are the scaling dimensions of the local operators $\Phi_{\alpha\beta}(x)$ and $S_\alpha(x)$ respectively.} This gives a simple scaling form
(the temperature does not appear as a separate argument, because here we are fixing it to the critical value)
\be\label{eq:bindermodscaling}
A(L,\Gamma) = \hat A \lf L^{1/\nu_M} (\Gamma-\Gamma_c)\ri.
\ee
Therefore we expect that $A$ takes a universal value $\hat A(0)$ at the measurement critical point ${\Gamma=\Gamma_c}$.
A standard method for accurately locating a continuous phase transition is then to plot  $A$ against the control parameter (here $\Gamma$) for different values of $L$, and look for the value of $\Gamma$ where the curves cross.

The inclusion of the moments of $S$ in Eq.~\ref{eq:bindermod} means that 
$A$ \textit{vanishes} whenever the enhanced $S_{n+1}$ symmetry is present (either as a microscopic symmetry or an emergent one). 
Therefore  the value of $A$ at the critical point 
gives a test for $S_{n+1}$ symmetry in the models where 
it is not present microscopically.

As another test for symmetry, we will show the quantity
\be\label{eq:Rquantity}
R(L, \Gamma)={\< \Phi^2\>}/{\<S^2\>}.
\ee
Let $R_\infty = \lim_{L\to \infty} R(L, \Gamma_c)$ be its asymptotic value at $\Gamma_c$.
In the Gaussian protocol, the microscopically exact $\mathcal{G}_n^+$ symmetry implies that $R_\infty =1$.
If $\mathcal{G}_n^+$  emerges only in the IR, 
then the equality of the scaling dimensions
of $\Phi$ and $S$ should imply that 
$R_\infty$ is a positive but nonuniversal constant.\footnote{The numerical value of $R_\infty$ is nonuniversal because in general it depends on the nonuniversal normalization of the lattice operators.}
If $\mathcal{G}_n^+$  does not emerge,
then we would expect that $\Phi$ and $S$ have distinct scaling dimensions so that $R_\infty$  will either vanish or diverge.

We will also examine correlation functions at the critical point. 
To minimize finite-size effects, we extract the scaling dimensions for moments ${l=1,2,...,6}$ 
from the correlation function
at the maximum separation possible for a given $L\times L$ lattice:

\begin{align} \label{eq:latticecorrelator}
G_l(L)  \equiv
\frac{1}{|\Lambda'|}
\sum_{\mathbf x\in \Lambda'}
\mathbb E \left[ \<S_{\mathbf x}S_{\mathbf x'}\>^l_M \right], 
 \\
\text{with} \hspace{0.1 cm}
\mathbf{x'} =\mathbf{x}+\left(\f{L}{2},\f{L}{2}\right)
\,\,\, (\operatorname{mod} L).
\end{align}
We have  averaged ${\bf x}$  over the left-half of the system, ${\Lambda'=[1,\frac{L}{2}]\times [1,L]}$, $L$ being taken even. 

In the simulation, we have  $N=6$ real replicas which are evolved independently (see discussion around Eq.~\ref{eq:realreplicaexpl}): then 
\begin{equation}\notag
    \mathbb E [\<S_{\mathbf x}S_{\mathbf x'}\>^l_M]=\binom{N}{l}^{-1}
    \hspace{-3mm}
    \sum_{\{a_1, \ldots, a_l\}}
    \hspace{-1mm}
    \<\prod_{i=1}^l S^{(a_i)}_{\mathbf x}S^{(a_i)}_{\mathbf x'}\>_\text{MC},
\end{equation}
where the sum implements an average over choices of a subset  $\{a_1,a_2,\ldots, a_l\}$ containing exactly $l$ of the $N=6$ real replicas. 
The bracket ${\<..\>_\text{MC}}$ is now the full  Monte-Carlo average of this replica quantity
 over both different $M$ realizations and over different Monte-Carlo timesteps for a given $M$.

We also define the \textit{typical} magnitude  of the correlator, $G_0$, 
as ${G_0:=\exp(\mathcal E_{\text{typ}})}$ where
\begin{align}\label{eq:typdefn}
    \mathcal E_{\text{typ}}=\frac{1}{|\Lambda'|}\sum_{\mathbf x\in \Lambda'} \mathbb E \left[ \ln |\<S_{\mathbf x}S_{\mathbf x'}\>_M| \right].
\end{align}
This is averaged using data for a single ``real'' replica.

\begin{figure}[t]
\fig{0.85}{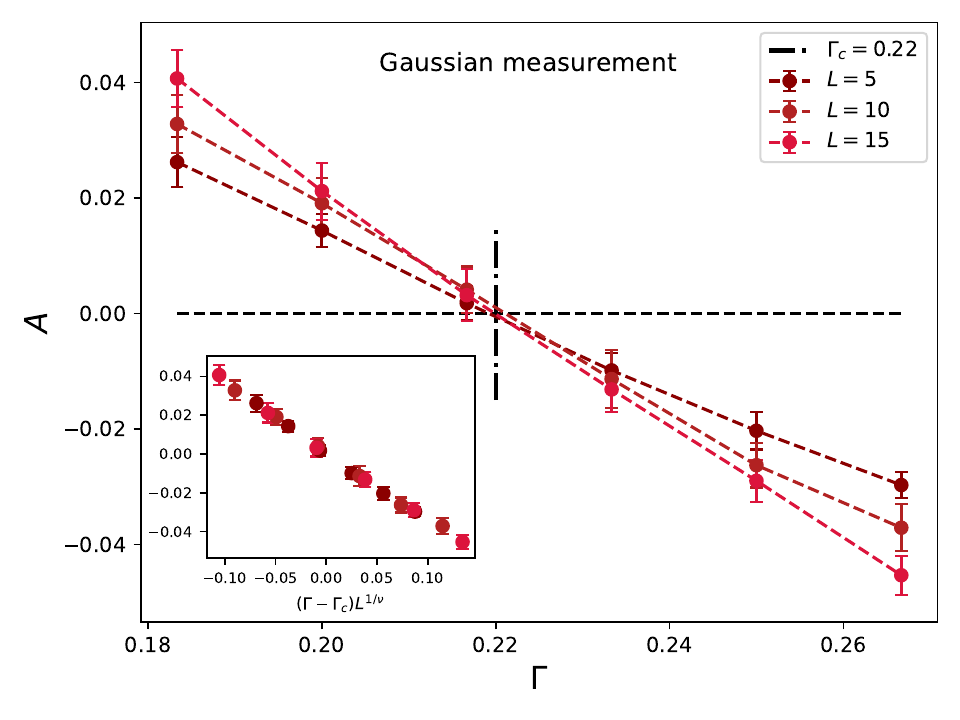}
\fig{0.85}{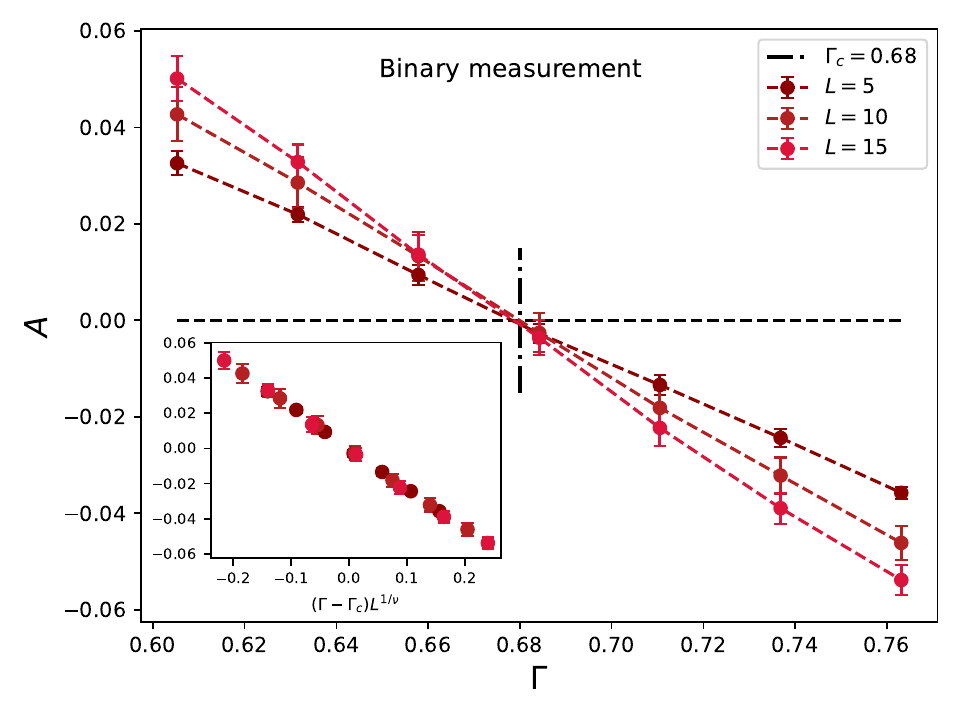}
\fig{0.85}{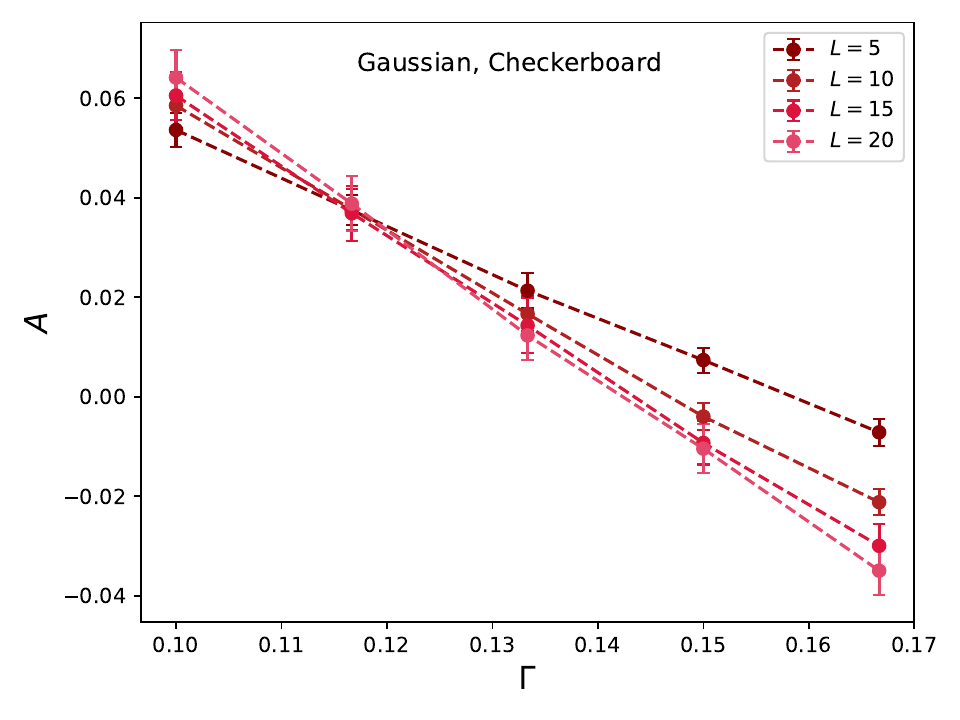}
\caption{Top: Modified Binder cumulant for the Gaussian measurement protocol. The crossing is at $\Gamma_c\approx 0.22$ (cf. Eq.~\ref{eq:GammacGaussian}). Middle: For the binary measurement protocol. The crossing is at $\Gamma_c \approx 0.68$. Bottom: For the Gaussian protocol on a checkerboard lattice, where only the diagonal bonds are measured. In the first two cases, we get a good scaling collapse with exponent $1/\nu_M \approx 0.39$. (The data for the checkerboard case is consistent with a wide range of $\nu_M$ values so we do not show a collapse.)}
\label{fig: Binders}
\end{figure}

\begin{figure}[t]
    \fig{0.85}{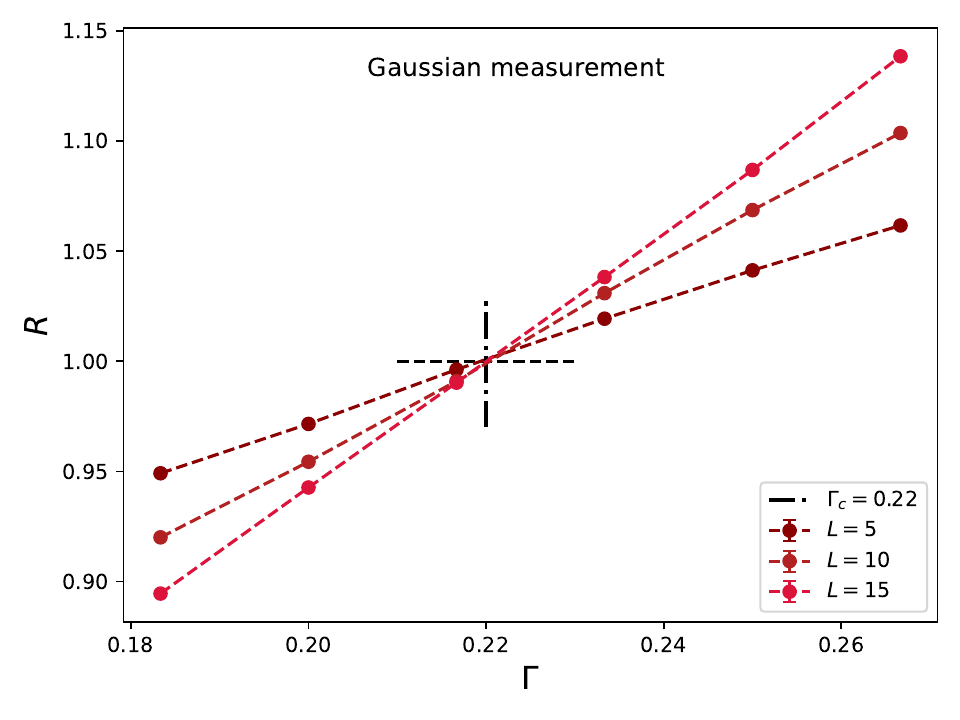}
    \fig{0.85}{R_Binary_meas.pdf}
    \fig{0.85}{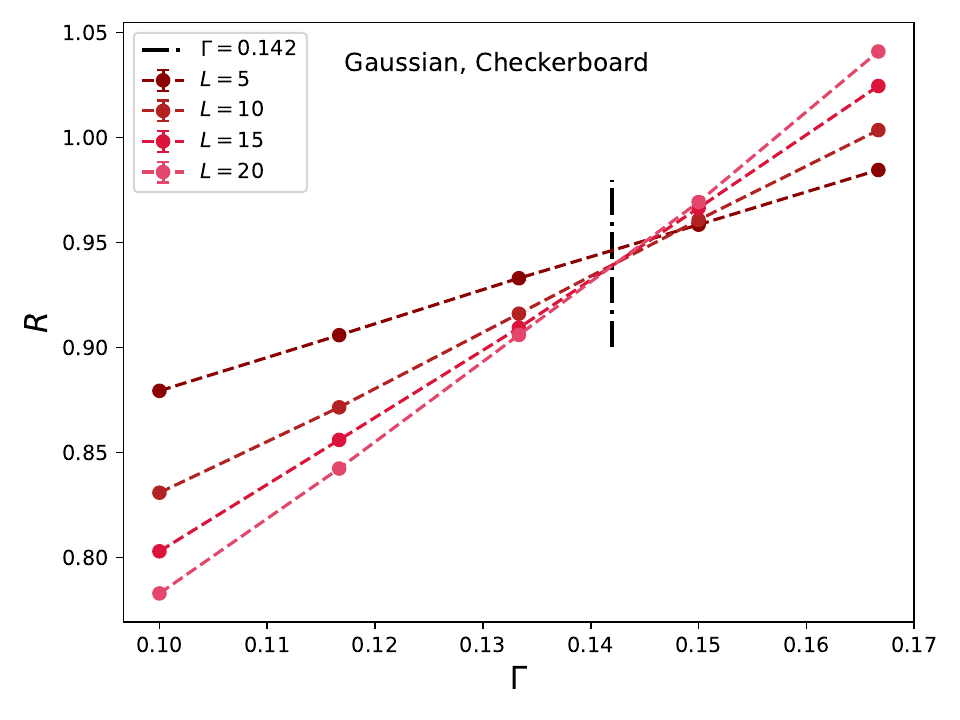}
    \caption{The quantity $R$ shown for Top: Gaussian measurement, Middle: Binary measurement and Bottom: Gaussian measurement in the checkerboard lattice. The error bars are smaller than the markers. The horizontal dashed line in the first two panels indicates $R=1$.
    (In the final panel, the vertical dashed line indicates the $\Gamma$ value for which correlators are plotted in Fig.~\ref{fig:Checkboard_moments}.)}
    \label{fig:R_quantity}
\end{figure}

\subsection{Gaussian measurement protocol}
\label{sec:gaussiansim}

In the Gaussian protocol  the measurements have Gaussian errors of variance $\Delta^2 = 1/2\Gamma$ (Sec.~\ref{sec:latticeprotocols}):
\ba
M_{ij} \sim \mathcal{N} \lf S^\text{ref}_i S^\text{ref}_j , \, \f{1}{\sqrt{2\Gamma}}\ri,
\end{align}
where $\{S^\text{ref}_i\}$ is the measured configuration.
This protocol was illustrated schematically in 
Fig.~\ref{fig:Snapshot}.

The thermodynamic critical point of the square-lattice Ising model is fixed by Kramers-Wannier duality to be at \cite{KardarBook}  
\be
J_c  = \f{\ln (1+\sqrt{2})}{2}
\ee
(where $e^{-2J_c}=\tanh J_c$). 
Our simulations are performed at this $J$ value, and we vary only the measurement strength $\Gamma$.
We identify the critical point of the measurement process as
\be\label{eq:GammacGaussian}
\Gamma_c = \f{\ln (1+\sqrt{2})}{4},
\ee
by using the criterion $\Gamma=J/2$ for the enlarged symmetry. 
The exact knowledge of  $\Gamma_c$ is very useful for simulations.

The modified Binder cumulant for the Gaussian protocol is shown in the Top panel of Fig.~\ref{fig: Binders}. The crossings of the curves around
 ${\Gamma_c \simeq 0.22}$
are consistent with the expected $\Gamma_c$ in Eq.~\ref{eq:GammacGaussian},
and the critical value of $A$ is also  consistent with  zero, the value expected from the exact symmetry of this protocol at $\Gamma_c$.

The inset shows  a scaling collapse using Eq.~\ref{eq:bindermodscaling}, which yields an estimate of the (inverse) correlation length exponent 
\be
\f{1}{\nu_M} \approx 0.39,
\ee
close to the value reported in Ref.~\cite{PutzGarrattNishimoriTrebstZhu2026}.
Note that this value is strikingly small compared to standard critical points.
This indicates that, while changing $\Gamma$ is a relevant perturbation, it is a relatively weakly relevant one.

The data for $R$ (Eq.~\ref{eq:Rquantity}) also shows a  good crossing at the expected value of $\Gamma_c$ (Fig.~\ref{fig:R_quantity}, first panel).

\begin{figure}[t]
\fig{1.0}{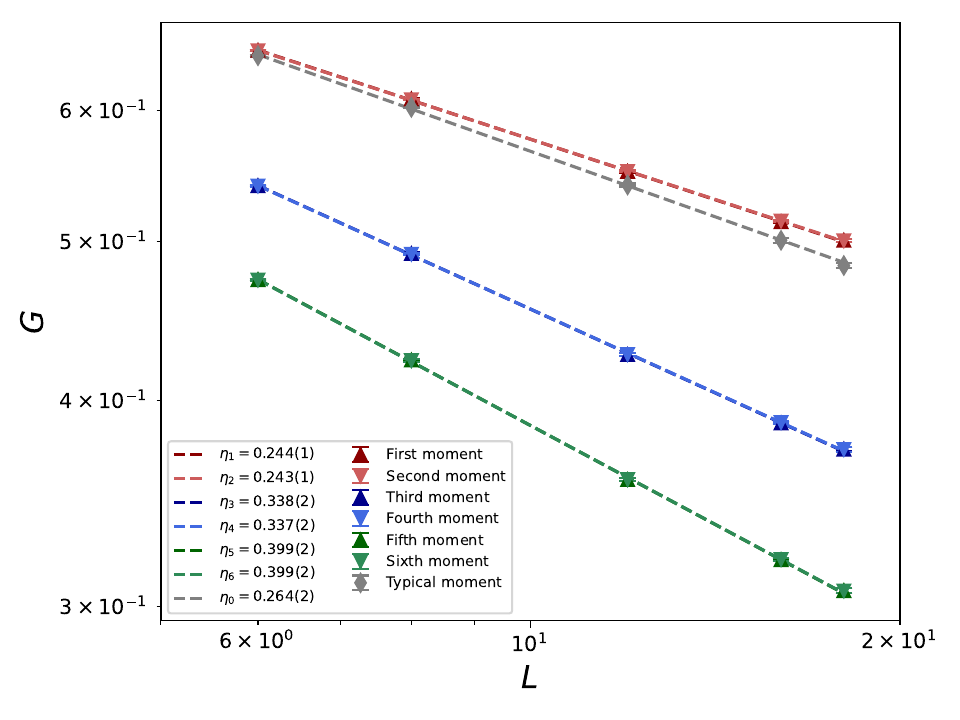}
\caption{Log-log plot of the first six moments of the correlator and of the typical value (Eq.~\ref{eq:typdefn}),
at the critical point of the Gaussian measurement protocol (where the enlarged symmetry holds exactly).
The fits giving the exponents $\eta_{k}$, $k=1,...,6$ for the moments and  $\eta_0$ for the typical are shown.
Note that some of the lines are almost on top of each other.}
\label{fig:Moments}
\end{figure}

Finally, 
the correlation functions at the critical point,
shown in Fig.~\ref{fig:Moments},
 give the most striking signature of enlarged symmetry.
We write 
\be
G_{l}(L)\sim L^{-\eta_l},
\ee
where ${\eta_l = 2\Delta_l}$ in the notation of Sec.~\ref{sec:RG}.
Fitting the correlator for different system sizes gives us the scaling dimensions with $\eta_0$ being that for  typical value (Eq.~\ref{eq:typdefn}). The results are 
\begin{align}\label{eq:GslopesGauss}
    \eta_1&=0.244(1), & \eta_2&=0.243(1), \nonumber\\
    \eta_3&=0.338(2), &  \eta_4&=0.337(2), \nonumber \\
\eta_5&=0.399(2), & \eta_6&=0.399 (2) ,\nonumber \\
\end{align}
consistent with the identities $\eta_{2k-1}=\eta_{2k}$ expected from enlarged symmetry,
and 
\ba
    \eta_{0}=0.264(2).
\end{align}
The error bars shown are purely the statistical errors in the fits.

It should be borne in mind that the system sizes range only over a factor of three, so finite size effects could be large. In an attempt to quantify these, we have looked at the variation with $L$ of effective exponents $\eta_l(L)$, defined by fitting only three consecutive system sizes: see Fig.~\ref{fig: Exponents_variation} in App.~\ref{a:Sim_details}.

Recall that ${\eta_1=\eta_2 = 1/4}$ is exact as a result of symmetry together with the single-replica reduction property (Eq.~\ref{eq:EAtrivialization2d}). 
It is intriguing that some of the other exponents are close to simple rational values. In view of the small sizes, this may be a coincidence, but it deserves further investigation.

Note that the identities 
${\eta_{2k-1}=\eta_{2k}}$
between exponents 
make it clear that both positive and negative values of the correlator are important. 
This is a qualitative difference between the critical point discussed here and 
the multifractal phase that 
exists for ${\Gamma<\Gamma_c}$ in dimensions just above two (Eq.~\ref{eq:linearphaseidagram2}).\footnote{That multifractal phase can be accessed in a binary measurement protocol with very small measurement strength, such that the couplings in the effective Hamiltonian $\mathcal{H}_M$ are all ferromagnetic, and the conditioned correlators are all positive (by the Griffiths-Kelly-Sherman inequality).}
There is a universal sense in which the former fixed point involves ``frustration'' while the latter does not.

The data above clearly show that there is a measurement critical point in the Gaussian protocol, and 
the exponents above characterize the corresponding universality class. 
The next question is whether this universality class remains generic in models without microscopic enlarged symmetry.
The  $6-\epsilon$ expansion 
in Sec.~\ref{sec:RG}
suggests this is plausible.
We now give evidence that it is the case in 2D.

\subsection{{More general measurement protocols in 2D: \\ Test of symmetry emergence}}
\label{sec:moregeneralprotocols2d}

To address the question of whether the above universality class is generic, we look at other protocols whose replica Hamiltonian does not have the full ${\mathcal{G}_n^+=\mathbb{Z}_2^n\rtimes S_{n+1}}$ symmetry at the lattice level, but instead only ${\mathcal{G}_n=\mathbb{Z}_2^n\rtimes S_{n}}$.
We give evidence that, at the measurement transition, full $\mathcal{G}_n^+$ emerges in the infrared.

First we consider the binary protocol \cite{PutzGarrattNishimoriTrebstZhu2026} (Sec.~\ref{sec:setup}), with  $M_{ij}\in \{\pm 1\}$ for the measured bond energies, with conditional probabilities 
\be
P(M_{ij}|S_iS_j)=\mathcal Ne^{-\Gamma M_{ij} SiS_j }
\ee
where $\mathcal N=(e^\Gamma+e^{-\Gamma})^{-1}$ is the normalization constant.

The modified Binder cumulant $A(\Gamma)$ is shown in Fig.~\ref{fig: Binders} (second panel), and has a crossing at ${\Gamma_c\approx 0.68}$. 
For this protocol we again find the correlation length exponent ${1/\nu_M \approx 0.39}$, roughly consistent with  \cite{PutzGarrattNishimoriTrebstZhu2026}, 
which fitted a different quantity.
In addition, we find that   ${A(\Gamma_c)}$ is very close to zero near the crossing, consistent with an enhanced symmetry.

At first glance, this looks like extremely strong support for the idea 
that $\mathcal{G}_n^+=\mathbb{Z}_2\rtimes S_{n+1}$ emerges generically for the measurement transition of critical Ising.
However, it is necessary to be more careful. 
At lowest nontrivial order in $\Gamma$ (order $\Gamma^2$), the lattice replica Hamiltonian for the binary protocol matches that for the Gaussian protocol (App.~B, \cite{NahumJacobsen2025}). 
The four-replica terms that prevent the enlarged symmetry appear only at order $\Gamma^4$, with a relatively small coefficient. 
It could be that we see approximate symmetry at intermediate sizes only because the microscopic replica Hamiltonian happens to be \textit{almost} $\mathcal{G}_n^+$--symmetric at $\Gamma_c$. 

Indeed, the quantity $R$ (\ref{eq:Rquantity}) shows a crossing at a value extremely close to 1 (Fig.~\ref{fig:R_quantity}, second panel). 
As noted below Eq.~\ref{eq:Rquantity}, this suggests that 
the symmetry is (approximately) present even at the microscopic level.
In this situation,  very large sizes might be needed in order to determine whether the symmetry becomes more or less accurate at large scales, i.e.\ whether the residual $\mathcal{G}_n^+$-breaking terms are RG-relevant or irrelevant.

Therefore, we formulate a measurement protocol in which we definitely do not have approximate $\mathcal{G}_n^+$ symmetry at the lattice scale.
To do this we  make Gaussian measurements of next-nearest-neighbors instead of nearest neighbors.  The replica Hamiltonian then resembles Eq.~\ref{eq:replicaHgaussian}, except that the $\Gamma$ term now runs over next-nearest neighbors (diagonal pairs) instead of lattice bonds.
This microscopic Hamiltonian is far from having enlarged symmetry.

The modified Binder cumulant for this ``checkerboard'' protocol is shown in Fig.~\ref{fig: Binders}, third panel, and the quantity $R$ is shown in Fig~\ref{fig:R_quantity} (third panel). 
At these sizes, $A$ is of order $0.04$ in the region where the curves have crossings. 
The data does not allow a precise determination of $\Gamma_c$, but it is clear that the critical measurement strength is smaller than for the nearest-neighbor protocol, as we might expect (longer-range measurements are more constraining).
The quantity $R$ crosses at a somewhat larger value of $\Gamma$, but with the crossings not being very well defined (moving to the left).

The data for $A$ show that
${\langle X^4\rangle/\langle X^2\rangle^2}$ is within ${\sim 4\%}$ of the value of ${\langle S^4\rangle/\langle S^2\rangle^2}$ in the critical region. 
Given the range of $L$ considered,
we interpret this small value as modest evidence that the enlarged symmetry will become exact as ${L\to \infty}$.
In this scenario,  both sets of curves will 
show a crossing at some $\Gamma_c$ in the limit of large $L$,
with $R$ taking a nonuniversal constant value at the crossing, and $A$ taking the value~0.
To confirm this scenario using these observables alone would require larger sizes. 

\begin{figure}[t]
\fig{1.0}{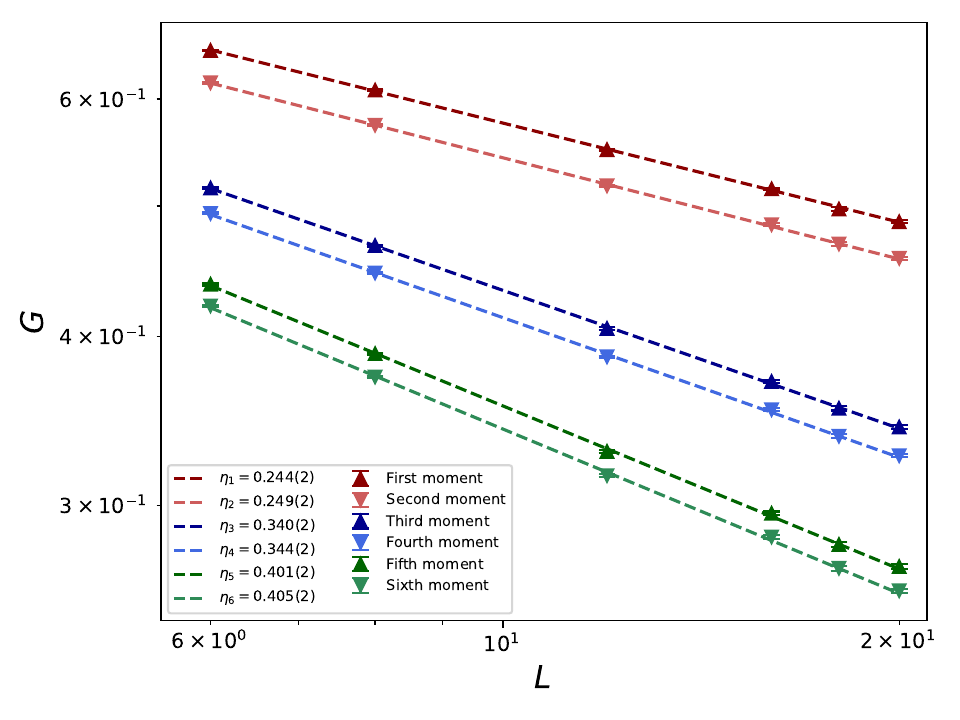}
\caption{Log-log plot of the first six moments of the correlator for the checkerboard lattice, at ${\Gamma=0.142}$. The fits giving exponents ${\eta_k,k=1,2..,6}$ are shown. 
(Typical moment is omitted  to avoid clutter.)
Note that the lines for the $2k$th and $(2k-1)$th moment are close to being parallel, but are not on top of each other, consistent with the emergence of enlarged symmetry in the IR.}
\label{fig:Checkboard_moments}
\end{figure}

However, looking at the correlation functions gives a clearer indication. Fig.~\ref{fig:Checkboard_moments} shows the correlators at ${\Gamma=0.142}$ (indicated by the dashed vertical line in the lower panel of Fig.~\ref{fig:R_quantity}).
Note first the remarkable pairwise consistency of the slopes, consistent with the  identities ${\eta_{2k-1}=\eta_{2k}}$ that are expected if the flow is to the $\mathcal{G}_n^+$--symmetric universality class. 
Further, the fitted values of the slopes (shown in the legend of the figure) are in good agreement with the values in Eq.~\ref{eq:GslopesGauss}. 
(For the typical correlator, not shown in the figure, we find $\eta_{0}=0.268(2)$.)

We conclude that generic bond measurement protocols (for critical Ising) show a measurement critical point governed by the same  $\mathcal{G}_n^+$--symmetric conformal field theory that governs the Gaussian measurement problem.

\section{Long-range models: a setting where the expansion parameter $\epsilon$ can be varied continuously}\label{sec:longrange}

At the level of dimensional analysis for the couplings, lattice models or field theories with 
with long-range, power-law decaying interactions are similar to higher-dimensional short-range models, though the  RG equations are in general different.

In this Section we consider low-dimensional models with power-law spin interactions, and with long-range {\em measurements} (defined below). Fixing the spatial dimension --- for example to ${d=1}$ --- we may perform an $\epsilon$-expansion by varying the exponents for the decay of interactions and measurements. We show that much of the resulting phenomenology is similar to that in the ${6-\epsilon}$ expansion.  Since we now operate in fixed $d$, a continuously varying  $\epsilon$ is accessible in Monte Carlo simulations.

There are various ways to define long-range measurement protocols. Here we consider a conceptually simple (if perhaps impracticable) 
one in which we make measurements $S_i S_j$ 
not only for nearest-neighbor bonds but for all pairs $\{i,j\}$ of spins, but with the measurement strength ${\Gamma_{ij} = 1/2\Delta_{ij}^2}$ decreasing as a power law with the distance. 

The lattice formalism for Gaussian measurements reviewed in Sec.~\ref{sec:latticeprotocols} 
carries over directly. If the physical spin  Hamiltonian has long-range interactions $J_{ij}$, then\footnote{We have set $n=1$ in the second coefficient; the restriction $i<j$ indicates that we sum over each distinct pair once. The dimension $d$  is arbitrary.} 
 \begin{equation}\label{eq:replicaHgaussianLR}
    \mathcal{H}_n= -
    \sum_{i<j} \Bigg[ 
   \sum_{\alpha=1}^n 
    J_{ij}
   S_{\alpha i} S_{\alpha j} 
    + \sum_{\substack{\alpha, \beta=1 \\ \alpha\neq \beta}}^n  \Gamma_{ij} \,
   (\Phi_i)_{\alpha\beta}
    (\Phi_j)_{\alpha\beta}
    \Bigg].
    \end{equation}
As in Sec.~\ref{sec:latticeprotocols}, there is an enlarged symmetry if we pick ${\Gamma_{ij} = \f{1}{2} J_{ij}}$,; in this sense the Gaussian protocol is fine-tuned. However, the form above is sufficient to motivate the continuum theory which we study here. 

For the case where  ${J_{ij}\sim r_{ij}^{-\sigma_J}}$,  ${\Gamma_{ij}\sim
r_{ij}^{-\sigma_M}}$,
we take the cubic Lagrangian defined by  Eq.~\ref{eq:introducecubicL}, and  replace the $k^2$-terms  with the momentum dependences obtained by Fourier-transforming $J_{ij}$ and $\Gamma_{ij}$,
\begin{align}
\mathcal{S}_n & =
\f{1}{2}
 \sum_{\alpha} 
\int_k |k|^{\alpha_J} \phi_\alpha (k) \phi_\alpha (-k)
\\
& + 
\f{1}{4}
\sum_{\alpha\beta}
\int_k |k|^{\alpha_M} \Phi_{\alpha\beta}(k) \Phi_{\alpha\beta}(-k) + \text{interaction terms}.
\notag
\end{align}
with ${\alpha_{J,M} = \sigma_{J,M} - d}$.
\begin{figure}
\fig{0.95}{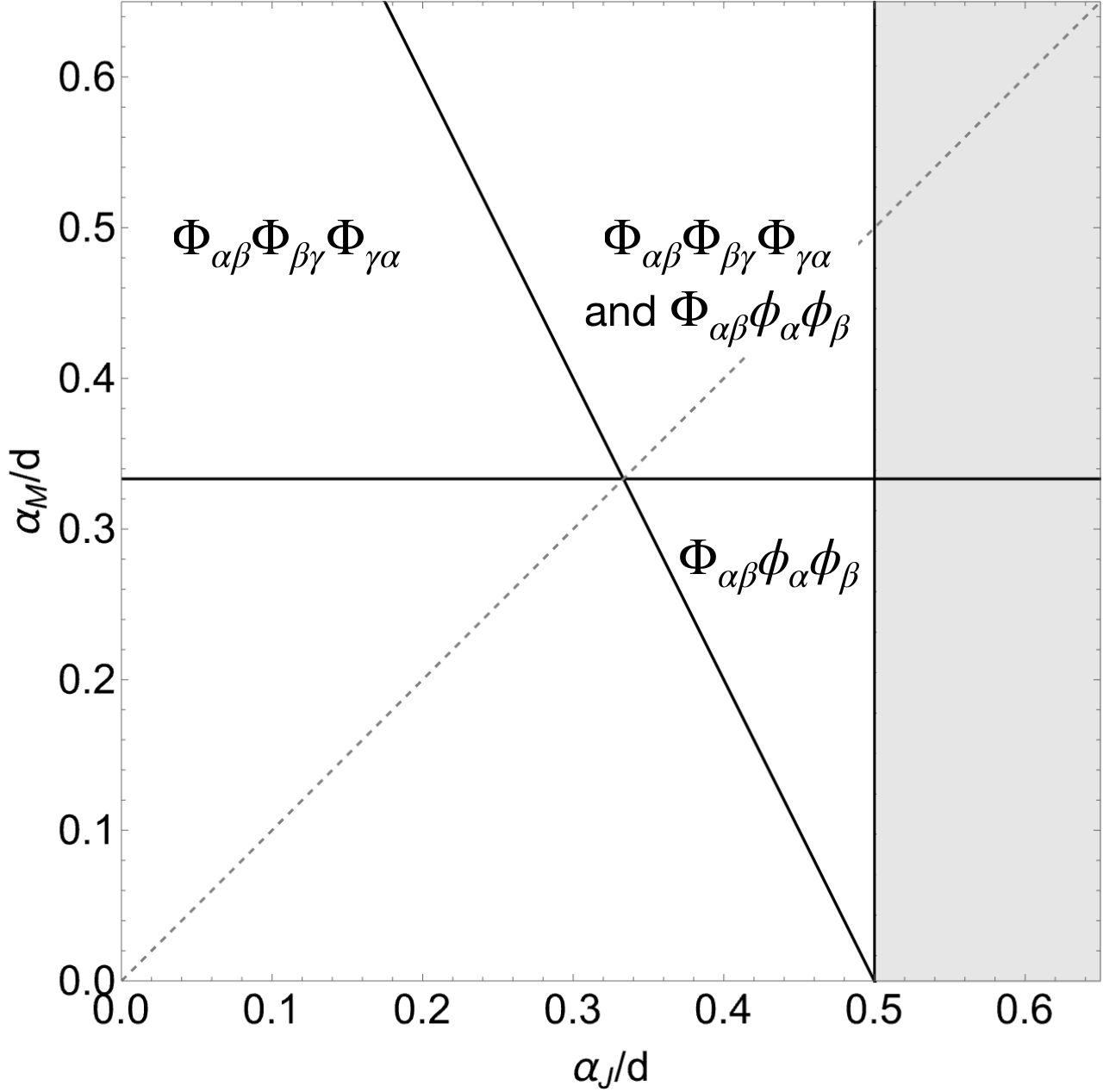}
\caption{Regions of the $(\alpha_J/d, \alpha_M/d)$ plane are labelled according to which cubic terms are relevant. 
The region that touches the origin is the ``mean-field'' region where all interactions are irrelevant.
In the shaded region a quartic operator becomes relevant (see Sec.~\ref{sec:RG}), invalidating the naive $\epsilon$ expansion (unless its coupling is tuned away).
The dashed line indicates $\alpha_M = \alpha_J$, in which case there is the possibility of an enlarged symmetry relating $\phi$ and~$\Phi$.}
\label{fig:alpharegions}
\end{figure}
We restrict to the range
\begin{equation}
0 < \alpha_J, \alpha _M < 2,
\end{equation}
since if a given $\alpha$ is greater than 2 the  corresponding $|k|^\alpha$ term is subleading compared to the usual local derivative term (and since $\alpha<0$ gives a superextensive action).

\begin{figure}
    \centering
    \includegraphics[width=0.9\linewidth]{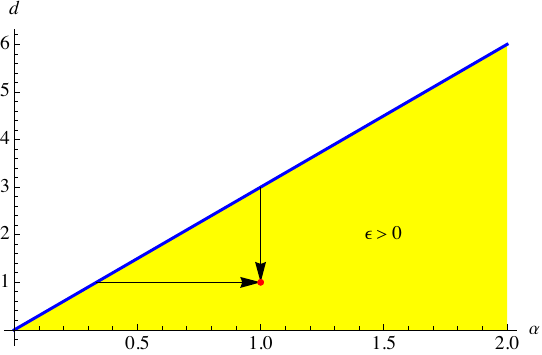}
\caption{The range where $\epsilon(\alpha,d)>0$ (in yellow) and two possible expansion paths to reach $\alpha= d=1$.}
\label{measuremets-relevant2}
\end{figure}

Depending on our choices for $\alpha_J$ and $\alpha_M$, both, one, or neither of the two cubic interactions is relevant at tree level. This is illustrated in Fig.~\ref{fig:alpharegions}, which shows which cubic operators are relevant in the various regions of the $(\alpha_J/d, \alpha_M/d)$ plane. 
In the shaded region a  quartic operator  becomes relevant, which invalidates the naive epsilon expansion, as explained in Sec.~\ref{sec:RG} --- 
although we are of course also free to consider measurement of a tricritical Ising model where this term is tuned away (or even perhaps measurement of a noninteracting field).
The effective values of ``$\epsilon$''
for the two coupling constants are 
\begin{align}
 \epsilon_1 & 
 =  2 \alpha_J + \alpha_M - d, 
\\ 
\epsilon_2& 
= 3 \alpha_M - d,
\end{align} 
i.e.\ the RG eigenvalues of $\lambda_1$ and $\lambda_2$ at the free fixed point are $\epsilon_1/2$ and $\epsilon_2/2$ respectively.

Our results for the beta functions in  App.~\ref{s:LR-beta} cover both the case where the two cubic operators are (weakly) relevant,  and the cases where only one of them is (weakly) relevant. 
For simplicity, here we  only show explicit results on the line ${ {\alpha_M=\alpha_J:=\alpha}}$, i.e. 
\be\label{epsilon-LR}
\epsilon_1 = \epsilon_2 = \epsilon := 3 \alpha-d,
\ee
where both operators have the same tree level dimension, and where there is the possibility of an enlarged symmetry uniting the two fields $\phi$  and $\Phi$.
We comment on the more general case below.

At ${n=1}$, the $\beta$-functions look very much like those for the short-range model near its upper critical dimension. 
Again, the diagonal   ${\lambda_1=\lambda_2=\lambda}$ is attractive. On this diagonal the $\beta$-functions read
\begin{eqnarray}
\label{beta-LR-iso1}
\beta_1(\lambda,\lambda) = \beta_2(\lambda,\lambda) 
&=& \frac{\lambda  \epsilon }{2}- \frac{\Gamma  (\frac{d}{6} )^3 \Gamma  (\frac{d}{2} )}{     \Gamma  (\frac{d}{3} )^3}    \lambda^5,
\end{eqnarray}
and yield a stable fixed point at 
\begin{equation}
    \lambda^* = \sqrt[4]{ \frac{\epsilon  \Gamma (\frac{d}{3})^3}{2
   \Gamma (\frac{d}{6})^3 \Gamma
   \left(\frac{d}{2}\right)}}.
\end{equation}

Above, we imagine that the spatial dimension $d$ is fixed and we vary $\epsilon$ by varying $\alpha$.
In fact, to access a given point of interest, it is possible to expand around any point on the critical line ${\epsilon(\alpha,d)=0}$. 
This is illustrated in Fig.~\ref{measuremets-relevant2}, which shows two different ways to access the point $(1,1)$.\footnote{In fact this particular point lies in the regime where a quartic term would invalidate the $\epsilon$ expansion. 
However, we can remove this term by considering the measured system to be free in the IR rather than interacting, as mentioned above.}
 One can even use the freedom to choose the expansion point to optimize the result (the principle of minimal sensitivity, see e.g.\ \cite{WieseDavid1995,DavidWiese1996,WieseDavid1997}). 
If we fix $\alpha$ and expand in $d$, the beta functions read
\begin{eqnarray}
\label{beta-LR-iso2}
\beta_1(\lambda,\lambda) = \beta_2(\lambda,\lambda) 
&=& \frac{\lambda  \epsilon }{2}- \frac{\Gamma  (\frac{\alpha }{2} )^3 \Gamma  (\frac{3
   \alpha }{2} )}{ \Gamma (\alpha )^3}  \lambda ^5 .
\end{eqnarray}
They have a stable fixed point at 
\begin{equation}
    \lambda^* = \sqrt[4]{\frac{\epsilon  \Gamma (\alpha )^3}{2 \Gamma
   (\frac{\alpha }{2})^3 \Gamma (\frac{3
   \alpha }{2})}}.
\end{equation}

We recover the  results for the short-range model when we set $\alpha\to 2$.
Note however that this property is special to the case $n=1$.
At this value of $n$, there is no wavefunction renormalization in either model.
At $n\neq 1$, there is wavefunction renormalization in the short-range model but not in the long-range model (a well-known property of long-range models \cite{HohenbergHalperin1977}).

To get  $\omega$ and $\nu$   at leading order, one 
simply uses the results for the short-range model with the definition of $\epsilon$ in \Eq{epsilon-LR} and the appropriate fixed-point value. 
We remind the reader that to improve the precision for  the measurement problem, we  need to go to 3-loop order. 

Our results also apply to a long-range version of the Nishimori multicritical point, by taking ${n\to 0}$ instead of ${n\to 1}$. We refer the interested reader to appendix \ref{a:LR}, and to a forthcoming paper \cite{DavletbaevaKompanietsWiese2026}.

The discussion above is for ${\epsilon_1=\epsilon_2}$, where the fields $\phi$ and $\Phi$ have the same tree-level scaling dimension. Let us finally discuss other regimes, for the case $n=1$.

\textit{Case where only $\Phi_{\alpha\beta}\Phi_{\beta\gamma}\Phi_{\gamma\alpha}$ is relevant:}
When ${\epsilon_1<0<\epsilon_2}$, $\lambda_1$ is irrelevant  while $\lambda_2$ is not (the region marked $\Phi_{\alpha\beta}\Phi_{\beta\gamma}\Phi_{\gamma\alpha}$ in Fig.~\ref{fig:alpharegions}). 
We may then set the coupling $\lambda_1$ of $\phi \phi \Phi$ to zero. 
As a result,  $\phi$ and $\Phi$ fields  decouple, and the $\phi$ field becomes a long-range free field.
The $\Phi$ field, on the other hand, is governed by a fixed point that is equivalent to that for long-range measurements of an \textit{infinite temperature} Ising model! (This fixed point is equivalent to the Nishimori fixed point, as a result of the relations between different $n$ values mentioned in Sec.~\ref{sec:enhancedsymmfieldtheory}.)

As a result, the moments $\mathbb{E}\<S_iS_j\>^{2k}$, which map to correlators involving only the $\Phi$ field, take the same exponents as for (long-range) measurement of an infinite-temperature Ising model, but 
moments $\mathbb{E}\<S_iS_j\>^{2k-1}$,
which map to correlators involving both $\Phi$ and $\phi$, reveal that the measured model was critical.

\textit{Case where only $\Phi_{\alpha\beta}\phi_\alpha\phi_\beta$ is relevant:}
In the opposite case when  $\epsilon_2<0<\epsilon_1 $,  $\Phi\Phi\Phi$ is irrelevant but $\phi \phi \Phi$ is relevant, and we can set the coupling $\lambda_2$  of $\Phi \Phi \Phi$ to zero. The  remaining beta function is $\beta_1=\epsilon_1 \lambda_1 + \lambda_1^3 + \ldots$, there is no perturbative fixed point and the  flow goes to strong coupling.

\begin{figure}
    \centering
    \includegraphics[width=\linewidth]{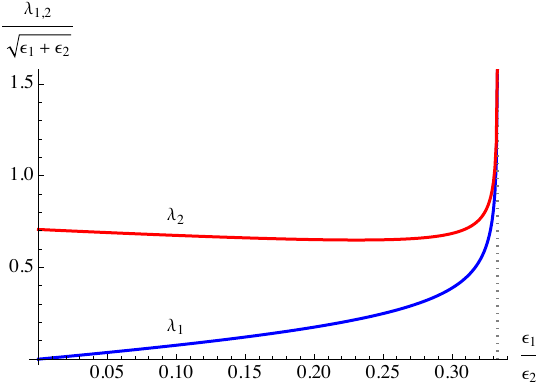}
    \caption{The  coordinates of the fixed point for ${0< \epsilon_1/\epsilon_2<1/3}$, when it is visible at one-loop order.}
    \label{f:LR-fixed-point}
\end{figure}

\textit{Case where both $\Phi_{\alpha\beta}\Phi_{\beta\gamma}\Phi_{\gamma\alpha}$ and  $\Phi_{\alpha\beta}\phi_\alpha\phi_\beta$ are relevant:}
We have already discussed the case $\epsilon_1=\epsilon_2$. 
Interestingly, it turns out that 
when ${\epsilon_1<\epsilon_2/3}$,
a stable fixed point is already present at one-loop.
We parameterize
\begin{equation}
    \epsilon_1 = \sqrt a \epsilon, \quad \epsilon_2 = \frac{\epsilon}{\sqrt a},
\end{equation}
and search for a fixed point as a function of $a=\epsilon_1/\epsilon_2$ at fixed $\epsilon=1$ (At 1-loop order changing $\epsilon$ only rescales the solution.) As figure \ref{f:LR-fixed-point} shows, there is a non-trivial 1-loop fixed point for  
\be
0 < \frac {\epsilon_1}{\epsilon_2} < \frac13. 
\ee
As $ {\epsilon_1}/{\epsilon_2}\to 0$ the fixed point tends to the $\lambda_1=0$ axis: 
in the limit 
$ {\epsilon_1}/{\epsilon_2}\to 0$ we are left with a copy of the Nishimori fixed point, together with a decoupled free field.
For $ {\epsilon_1}/{\epsilon_2}\to 1/3$, both couplings diverge; as their ratio goes to 1, they are finally stopped by the 2-loop term.  We expect that for $ {\epsilon_1}/{\epsilon_2}\geq 1/3$, the fixed point is visible only at 2-loop order.

\section{Measurement of $\mathbb{Z}_2$ gauge theory}
\label{sec:gaugemeasurement}

We have discussed how an enlarged replica symmetry can arise for measurement of Ising models, implying correlation function identities and constraints on the phase diagram.
Before ending this paper, we  note  analogous symmetry enhancements that are possible for measurement of a classical gauge theory ($\mathbb{Z}_2$ or ``Ising'' gauge theory).   For concreteness, we focus on the 3D case.

While the  standard expression (\ref{eq:Hgaugeinit}) below for the gauge theory Hamiltonian may appear formal, the model has numerous physical applications. For example, depending on the parameters, measurement of 3D $\mathbb{Z}_2$ gauge theory may be related   to measurement of membranes 
\cite{huse1991sponge}; 
or to  error correction, during time evolution, for a toric code quantum memory \cite{dennis2002topological,weinstein2024computational}.
These applications are usually most easily understood by thinking of the gauge-theory degrees of freedom in geometric terms, as membranes or flux lines  \cite{huse1991sponge,NahumJacobsen2025}.\footnote{In certain regimes, the measurement of flux lines in the 3D gauge theory gives a phase transition equivalent to that for measurement of anyons in an error correction protocol for the 2D toric code \cite{dennis2002topological}: in this interpretation the third coordinate is physical time, and the ``flux lines'' are analogous to worldlines of anyon excitations.}

There are also connections between the transitions discussed below and  the question of how easy it is to perform real space RG for gauge theories \cite{NahumJacobsen2025}.
 Via the mappings mentioned in the introduction, some of these inference phase transitions can be repurposed as phase transitions related to 
 classical ergodicity breaking (due to the imposition of constraints),  or  
 as phase transitions related to 
 ``strong-to-weak'' symmetry breaking \cite{lee2023quantum,chen2024symmetry,lessa2025strong,wang2026strong,hauser2026strong}.\footnote{More precisely, the ``weak'' symmetry becomes meaningful only if the classical probability distribution is embedded in a quantum density matrix}

Consider the  partition function $Z = e^{-\mathcal{H}}$ for $\mathbb{Z}_2$ lattice gauge theory on a cubic lattice \cite{wegner1971duality,fradkin1979phase,kogut1979introduction,tupitsyn2010topological},  with
\be\label{eq:Hgaugeinit}
\mathcal{H} = - J \sum_{\<ij\>} \sigma_{ij} S_i S_j - K \sum_{\<ijkl\>} \sigma_{ij}\sigma_{jk}\sigma_{kl} \sigma_{li}.
\ee
The Higgs field ${S_i =\pm 1}$ is defined on sites, and the gauge field ${\sigma_{ij}=\pm 1}$ is defined on links $\<ij\>$.
The $K$ term is summed over square plaquettes, denoted $\<ijkl\>$.
There is a gauge redundancy under ${S_i \rightarrow S_i \chi_i}$, ${\sigma_{ij} \rightarrow \sigma_{ij}\chi_i \chi_j}$ for ${\chi_i= \pm 1}$ (and we are free to choose the gauge ${S_i =1}$ $\forall i$).

For a brief review of the thermodynamic phase diagram of this model in three dimensions, see the early parts of \cite{somoza2021self}.
This phase diagram contains a  ``deconfined'' phase (at sufficiently large $K$ and sufficiently small $J$).
Generically the transition out of this phase is either  a \textit{Higgs} transition (loosely speaking, driven by increasing $J$ so as to condense  the Higgs field $S$) or a \textit{confinement} transition (driven by decreasing $K$ so that 
gauge flux proliferates). 
Both of these thermodynamic phase transitions are in the  ``Ising$^*$'' universality class. (The local scaling operators correspond to the scaling operators of the Ising CFT that are even under Ising symmetry.)
Either of them takes the system to the thermodynamically trivial phase, which we will refer to as the confined phase.
There is also a multicritical point where the Higgs and confinement transition lines meet \cite{vidal2009low,somoza2021self}.
Below we will consider measurement of the various kinds of critical state.

Each type of local term in $\mathcal{H}$ takes the values $\pm 1$, and we can consider Gaussian measurements of each type of term in  analogy to the  bond energy measurements in the Ising model.
We allow separate measurement strengths $\Gamma_J$ and $\Gamma_K$ for the link observables and plaquette observables respectively.
This leads to the following replica Hamiltonian   [where the site sums and site indices have been left implicit to avoid clutter, and we write e.g.\ $(\sigma SS)_\alpha$ as a shorthand for $\sigma_{\alpha\,ij} S_{\alpha i} S_{\alpha j}$]:
\ba\label{eq:replicaHgaussianGauge} \notag
    \mathcal{H}_n=  
  &- \bigg[ 
    J \sum_{\alpha=1}^n (\sigma S S)_\alpha 
    + {\Gamma_J} \sum_{\substack{\alpha, \beta=1 \\ \alpha\neq \beta}}^n 
(\sigma S S)_\alpha (\sigma S S)_\beta
    \bigg]
    \\
      & - \bigg[ 
    K \sum_{\alpha=1}^n (\sigma \sigma \sigma \sigma)_\alpha 
    + {\Gamma_K} \sum_{\substack{\alpha, \beta=1 \\ \alpha\neq \beta}}^n 
(\sigma \sigma \sigma \sigma)_\alpha (\sigma \sigma \sigma \sigma)_\beta
    \bigg].
    \end{align}
We have set $n=1$ in the coefficients.

This model has a  rich phase diagram which we will not map out in full here.
Ref.~\cite{NahumJacobsen2025}
discussed various measurement transitions  that can take place when the model is in a massive phase: either in the deconfined thermodynamic phase
(most simply studied by setting $J=0$ and $K=\infty$)
or when it is in the 
confined thermodymamic phase
(e.g.\ at $J=K=0$).
These transitions are either Higgs transitions of inter-replica \textit{composite}  Higgs fields $S^a S^b$ or confinement transitions of inter-replica composite gauge fields $\sigma^a \sigma^b$.

Here we instead consider measurement  phase transitions that can take place when the original thermodynamic ensemble is \textit{critical}, and note that some of them  have an enlarged symmetry analogous   to that discussed in the earlier parts of the paper. 

The enlarged replica symmetry is present microscopically whenever the measurement strengths obey ${\Gamma_J = J/2}$ and ${\Gamma_K=K/2}$.
Just as in the Ising case (\ref{latticesuperfield}), we define  the ${(n+1)\times (n+1)}$ matrix 
\ba
\Psi_i =
\lf
\begin{array}{cc}
   \vec{S}_i \vec{S}_i^T - \mathbbm{1}  & \,\,\,\, \vec S_i \\
         \vec S_i^T  & \,\,\,\, 0 
\end{array}\ri
\end{align}
at each site (vectors are in replica space) and on each bond we define
\ba
\Omega_{ij} =
\lf
\begin{array}{cc}
   \vec{\sigma}_{ij} \vec{\sigma}_{ij}^T - \mathbbm{1}  & \,\,\,\, \vec{\sigma}_{ij} \\
         \vec{\sigma}_{ij}^T  & \,\,\,\, 0 
\end{array}\ri.
\end{align}
Then if  ${\Gamma_J = J/2}$ and ${\Gamma_K=K/2}$ 
 we may write 
\ba\label{eq:replicaHgaussianGaugeSymm} \notag
    \mathcal{H}_n=&   
 - \f{J}{2}\sum_{\< i,j\>} \sum_{\substack{a, b=1 \\ a\neq b}}^{n+1} 
(\Omega_{ij})_{ab} (\Psi_i)_{ab} (\Psi_j)_{ab}
\\
& - \f{K}{2} \sum_{\< ijkl\>}
\sum_{\substack{a, b=1 \\ a\neq b}}^{n+1} 
(\Omega_{ij})_{ab} (\Omega_{jk})_{ab} (\Omega_{kl})_{ab} (\Omega_{li})_{ab},
    \end{align}
and similar arguments to the Ising case demonstrate the presence of an $S_{n+1}$ replica symmetry.
As for Ising, the symmetry can also emerge in the IR when it is not present microscopically and we will see examples of this below.

The initial gauge theory in Eq.~\ref{eq:Hgaugeinit} has various thermodynamic phase transitions: most importantly, the Higgs transition and the confinement transition (each of these is a line in the phase diagram \cite{tupitsyn2010topological}). 
It is interesting to consider the effectiveness of  measurements
for these two kinds of  critical ensembles.

The Higgs and  confinement transitions are \textit{thermodynamically} equivalent: 
their partition functions and free energies are mapped onto each other by a duality transformation. In a transfer matrix formulation of the partition function, this duality involves a change of basis.

An interesting feature here is that these two kinds of critical state are no longer ``equivalent''
when we consider their measurement phase diagrams. This is because measurement of a classical spin system comes with a preferred choice of transfer matrix basis \cite{NahumJacobsen2025}, i.e.\ the one  in which  observables are diagonal, and this means that duality transformations may not extend to the measured systems.\footnote{Relatedly, 
in classical systems it is necessary to  differentiate  symmetries that would be equivalent in a quantum context, according to the relation between the symmetry generators and the ``classical'' basis \cite{NahumJacobsen2025}.}

\subsection{Measuring membranes at the Higgs transition} \label{sec:higgs}

First consider the case where the system lies at the  thermodynamic Higgs transition.
The Higgs transition is a line in the $(J,K)$ plane: 
for simplicity we fix the point  ${(J,K)=(J_c, \infty)}$ on this line.\footnote{Here $J_c$ is by definition the critical value of $J$ for the case $K=\infty$. This is also the critical coupling strength for 3D Ising on the cubic lattice \cite{wegner1971duality,fradkin1979phase,kogut1979introduction}. See \cite{tupitsyn2010topological,somoza2021self} for numerical phase diagrams for general parameter values.}
We probe this system with  measurements of $(\sigma SS)$, i.e.\ we take $\Gamma_J>0$.
(The value of $\Gamma_K$ is unimportant, as will be apparent  below, but we may take ${\Gamma_K=0}$.)

The system can be visualized as an ensemble of closed ``membranes'',
made up of occupied plaquettes of the dual cubic lattice.
A plaquette of the dual lattice is occupied if $(\sigma S S)=-1$ for the corresponding link of the original lattice.
We make  measurements of the local membrane  density
(plaquette occupation number).

This problem is directly related to the Ising measurement problem. The replica Hamiltonian can be taken to be 
\ba \label{eq:HgaugeJGammaJ}
    \mathcal{H}_n=  
  &- \bigg[ 
    J_c \sum_{\alpha=1}^n (\sigma S S)_\alpha 
    + {\Gamma_J} \sum_{\substack{\alpha, \beta=1 \\ \alpha\neq \beta}}^n 
(\sigma S S)_\alpha (\sigma S S)_\beta
    \bigg],
    \end{align}
together with the constraint 
$(\sigma\sigma\sigma\sigma)=1$ on every plaquette that arises from the limit $K\to\infty$.\footnote{In three dimensions, the universal behavior is stable to nonzero $1/K$.}
Since ${(\sigma\sigma\sigma\sigma)=1}$, the value of $\Gamma_K$ is now moot.

This constraint means that  \cite{kogut1979introduction}, 
for the purpose of understanding the RG flows, 
we can simply choose the gauge ${\sigma_{ij}=1}$ on every link.\footnote{This is sufficient in considering the bulk phase diagram and RG flows but neglects e.g.\ boundary condition effects.}
The Hamiltonian then resembles that for the measured Ising model discussed in Sec.~\ref{eq:replicaHgaussian}.

Therefore the RG flows are precisely as in the Ising case, and the phase diagram has the  topology already shown in Eq.~\ref{eq:linearphaseidagram2}.
There is a ``multifractal'' weak-measurement phase, and a phase transition into a strong-measurement phase.
Enlarged replica symmetry is present at the transition between strong and weak measurement.

The membrane picture makes the  corresponce with the Ising measurement problem intuitive, because the membranes can be put in correspondence with Ising domain walls.
In the gauge theory, 
there is no local observable corresponding to   $S_i$, since $S_i$ is not gauge-invariant, but the  strong-measurement phase may be defined using line operators \cite{NahumJacobsen2025}.

[In formal terms, the unmonitored system at $J_c$ is described by a $\mathbb{Z}_2$ orbifold of the Ising CFT,\footnote{That is, since we gauge the Ising theory with a ``flat'' gauge field [with zero flux, ${(\sigma\sigma\sigma\sigma)=1}$], all $\mathbb{Z}_2$--odd operators are eliminated from the CFT spectrum.} sometimes denoted Ising$^*$ \cite{lammert1993topology,senthil2002microscopic,misguich2008quantum,grover2010quantum,moessner2001ising,slagle2014quantum,serna2024worldsheet}, 
and the measurement critical point at ${J=J_c}$, ${\Gamma_J=J_c}$ is described by a $\mathbb{Z}_2^n$ orbifold of the ${\mathcal{G}_n^+=\mathbb{Z}_2^n\rtimes S_{n+1}}$---symmetric fixed point described in the earlier parts of this paper.]

\subsection{Measuring flux lines at the confinement transition} 
\label{sec:confinementtr}

Next, consider the case where the original ensemble lies at the confinement transition.
Again, the confinement transition is a line in the $(J,K)$ plane: for simplicity we fix the point
 ${J=0}$, ${K=K_c}$ on this line.

 It is convenient to think of this thermodynamic ensemble as a critical ensemble of unoriented \textit{flux lines}
 that live on the links of the dual lattice.
 A link of the dual lattice  hosts a flux line if ${(\sigma\sigma\sigma\sigma)=-1}$ for the corresponding plaquette of the original lattice. Varying $K$ in the original ensemble controls the bare line tension  of flux lines. At large $K$ long flux lines are suppressed, while at small $K$ they proliferate. 

(In fact there is a distinction between the gauge theory and a simple flux line ensemble, if we are being more precise; the two ensembles are related to each other by a gauging procedure.\footnote{At $J=0$, the gauge theory is closely related to an ensemble of flux lines. 
However the flux line configuration satisfies global parity constraints, as a consequence of the definition of the flux  in terms of $\sigma$, and the $\sigma$ configuration also contains additional global information beyond the the flux line configuration (the three products $\prod \sigma$ along paths winding the spatial torus, if the BCs are periodic). 
The unconstrained ensemble of flux lines, with no additional degrees of freedom, corresponds instead to a closely-related theory in which  $\sigma$ is itself gauged by a higher-form gauge field $\mu$ that lives on plaquettes of the lattice. 
The resulting partition function is then equivalent to the partition function of the 3D Ising model, with the flux lines mapping to the lines that appear in a high-temperature expansion of the Ising model. 
(That is, the theory with $\sigma$ and $\mu$ may be used to formalize the idea of  measuring  high-temperature graphs of 3D Ising.) 
This gauging does not change the RG flows (because, in the relevant limit, the higher-form gauge field does not have any local excitations).} However, this subtlety  can be neglected for the discussion of the RG flows below.)

Measurements of the local flux line density, i.e.\ ${\Gamma_K>0}$, give
\ba\label{eq:replicaHgaussianGauge2} 
    \mathcal{H}_n=  - \bigg[ 
    K_c \sum_{\alpha=1}^n (\sigma \sigma \sigma \sigma)_\alpha 
    + {\Gamma_K} \sum_{\substack{\alpha, \beta=1 \\ \alpha\neq \beta}}^n 
(\sigma \sigma \sigma \sigma)_\alpha (\sigma \sigma \sigma \sigma)_\beta
    \bigg].
    \end{align} 

As already noted, the 
initial  ensemble lies at an Ising$^*$ criticial point, described in the continuum by an orbifold of the Ising CFT.
In the scaling limit, the leading CFT operator contributing to the   measured observable $(\sigma\sigma\sigma\sigma)$ is the operator corresponding to the Ising energy operator.

This operator identification in the continuum theory is sufficient to determine the RG flows  starting from an \textit{infinitesimally weak}
measurement strength: 
these
map onto the flows  for the critical 3D Ising model at weak measurement strength. 
Recall that, in 3D, weak measurement is  RG-relevant, and leads to a ``multifractal'' phase, governed by an RG fixed point that is stable to small variations of $\Gamma_K$.
(Shown in the phase diagram in Eq.~\ref{eq:linearphaseidagram2}.)
The gauge theory at $K=K_c$ and small enough $\Gamma_K$ therefore flows to an orbifold of this multifractal fixed point.


Therefore the RG flows for weak measurement strength map onto those in the 
Higgs case
(Sec.~\ref{sec:higgs}).
However, this does not hold for larger $\Gamma_K$.

At large $\Gamma_K$ we expect a stable phase  where the flux-line configurations from different replicas lock together.
(Physically, a pair of  these replicas could represent the true configuration and a sample from the posterior measure.)
However, this phase is distinct from the strong-measurement phase encountered immediately above in Sec.~\ref{sec:higgs}.

This can be seen formally by inspecting the partition functions.
In Sec.~\ref{sec:higgs}, the limit ${\Gamma_J\to\infty}$ completely trivialized the ``inter-replica'' fluctuations.
By contrast, here the limit ${\Gamma_K\to\infty}$ does not quite  have  this effect:  there remain global Wilson-line degrees of freedom that are not quenched by the limit ${\Gamma_K\to\infty}$.\footnote{The values of Wilson loops ${\prod_{\<ij\>\in C} (\sigma_{ij})_\alpha (\sigma_{ij})_\beta}$, where the path $C$ wraps the torus.}
Therefore the present strong-measurement phase does not map on to the strong-measurement phase in Sec.~\ref{sec:higgs}.

Further, we  expect that there is  an unstable fixed point separating the multifractal phase 
from the strong-measurement phase, 
(i.e.\ separating the small-$\Gamma_K$ and large-$\Gamma_K$ phases of Eq.~\ref{eq:replicaHgaussianGauge2}),
and we expect that this unstable fixed point 
\textit{also}
differs from the unstable fixed point described in Sec.~\ref{sec:higgs}.
Nevertheless, \textit{both} kinds of unstable fixed point have enlarged symmetry.
We discuss the phase transition of (\ref{eq:replicaHgaussianGauge2}) below.

As is clear from the preceding, there is no general duality between the present measurement problem
and the one in Sec.~\ref{sec:higgs}.
The replica partition functions  for (\ref{eq:replicaHgaussianGauge2}) and  (\ref{eq:HgaugeJGammaJ}) 
are not dual to each other in general. However,  they \textit{do} permit a duality if we expand only to lowest order in $\Gamma_K$ on each plaquette.
This is consistent with the discussion above of the RG flows for infinitesimal measurement strength.

Let us turn to the unstable fixed point separating the weak and strong-measurement phases. 
We expect this transition to take place at ${\Gamma_K = K_c/2}$, where the model acquires enlarged symmetry and has the replica Hamiltonian
\begin{align}
\label{eq:confcriticalGamma}
\mathcal{H}_n=&   
 - \f{K_c}{2} \sum_{\< ijkl\>}
\sum_{\substack{a, b=1 \\ a\neq b}}^{n+1} 
(\Omega_{ij})_{ab} (\Omega_{jk})_{ab} (\Omega_{kl})_{ab} (\Omega_{li})_{ab}.
\end{align}
If we introduce an additional redundancy in the parameterization by writing ${\sigma_\alpha = \tilde \sigma_\alpha \tilde \sigma_{n+1}}$, this becomes 
\begin{align}
\label{eq:Hgaugeredundant}
\mathcal{H}_n & = - K_c \sum_{\substack{a, b=1 \\ a \neq b}}^{n+1} (\widetilde\sigma \widetilde\sigma\widetilde\sigma\widetilde\sigma)_a
 (\widetilde\sigma \widetilde\sigma\widetilde\sigma\widetilde\sigma)_b
 & 
&  \text{[when $2\Gamma_K=K=K_c$]}.
\end{align}

Let us make a connection with a known problem.
So far here we have taken ${K=K_c}$.
The (thermodynamically trivial) ensemble with $K=0$ also shows a  phase transition as a function of $\Gamma_K$ between a weak-measurement phase and a strong-measurement phase (the weak-measurement phase is trivial in the ${K=0}$ case).
This ${K=0}$ transition  is relevant to error correction in the toric code, because if one  of the three dimensions is interpreted as time, there is an analogy to worldlines of anyon excitations; this phase transition was discussed in Ref.~\cite{dennis2002topological}.
The $K=0$ model is described by the replica Hamiltonian  \cite{NahumJacobsen2025} 
\ba \label{eq:HgaugeKzero}
    \mathcal{H}_n & =  - 
{\Gamma_K} \sum_{\substack{\alpha, \beta=1 \\ \alpha\neq \beta}}^n 
(\sigma \sigma \sigma \sigma)_\alpha (\sigma \sigma\sigma\sigma)_\beta & & [\text{when $K=0$}].
    \end{align}
The formal similarity between this Hamiltonian and the previous one (but differing in the number of fields and in their physical interpretation) is in complete analogy to the structures we saw in the measured Ising model.

We expect that Eq.~\ref{eq:confcriticalGamma} or Eq.~\ref{eq:Hgaugeredundant}, 
for $n\to 1$, describes another measurement critical point with enlarged replica symmetry, whose properties it might be interesting to study.
[It is possible to make a duality mapping from the gauge theories in (\ref{eq:Hgaugeredundant},\ref{eq:HgaugeKzero}) to ``spin'' models (see Sec.~6 of Ref.~\cite{takeda2004self} and App.~H of Ref.~\cite{NahumJacobsen2025}) but it is not clear whether this representation allows a useful field theory approach.]

\subsection{Lack of duality between the two measured ensembles; \\
Ising spins versus Ising high-temperature graphs} 

We noted above that, although the partition functions for the (unmeasured) Higgs and confinement transitions permit an exact duality, this duality does not extend to the measurement problems
at finite measurement strength. 

In Sec.~\ref{sec:higgs} we set $K=\infty$, while in Sec.~\ref{sec:confinementtr} we set $J=0$. In both of these limits the gauge theory partition function is closely related to the 3D Ising partition function.
The distinction between the two measurement problems above 
translates to a distinction between two measurement problems which arise from the Ising partition function.

The measurement problem  discussed in the bulk of this paper concerns measurement of Ising bond energies. 
Another problem arises if we reexpress the 3D Ising partition function in terms of high-temperature graphs, giving a kind of ``loop model'' \cite{winter2008geometric}, and then measure the density of bonds in the high-temperature graphs.
These two problems are related (by orbifolding) to the problems in Secs.~\ref{sec:higgs} and \ref{sec:confinementtr} respectively.




\subsection{Thermodynamic multicritical point}

Finally, the thermodynamic phase diagram of the gauge theory contains a  multicritical point $( J_{\rm MCP}, K_{\rm MCP})$ with a self-duality symmetry \cite{tupitsyn2010topological,somoza2021self,vidal2009low}. 
We do not discuss the phase diagram in the ${(\Gamma_J, \Gamma_K)}$ plane here, except to note that it contains a point ${(\Gamma_J, \Gamma_K)=( J_{\rm MCP}, K_{\rm MCP})}$ that we expect to yield another critical measurement problem with  enlarged replica symmetry.

\section{Outlook}

The primary aim of  this paper was to characterize the critical point, at the threshold measurement strength, for imaging of critical Ising configurations.
We gave results from an $\epsilon$--expansion 
and from numerical simulations in 2D, 
which indicated that the character of the critical point is in many ways similar in high dimensions and in 2D
(and even in 1D for certain long-range models).
As a result of an unexpected symmetry, 
which  emerges without fine-tuning near the upper critical dimension and probably also in 2D, 
some of the results for critical exponents are exact, despite the fact that the conditioned ensemble is nontrivially interacting. 
The analysis also clarified the structure of the phase diagram in general~$d$.
The enlarged symmetry is a structure that is also relevant to other inference problems 
(see e.g.\ Sec.~\ref{sec:gaugemeasurement}), 
and it relates expectation values of ``overlaps'' to expectation values in the original Gibbs state.

We now mention some miscellaneous open questions (starting with more heuristic questions about the inference problem) and possible extensions of the calculations above.

\textit{Phase transitions for particular algorithms.}
The results in this paper are for the ideal, unbiased estimates ${\<\bullet\>_M}$ of observables in the measured configuration $S^\text{ref}$.
For example, the conditioned correlator 
 $\<S_i S_j\>_M$ is an unbiased estimate of $S_i^\text{ref} S_j^\text{ref}$.
Our simulations corresponded to direct Monte Carlo sampling of  $\<S_i S_j\>_M$,
which becomes slow if we are close to the critical measurement strength (in 2D, tensor networks are an alternative method for computing unbiased estimates \cite{PutzGarrattNishimoriTrebstZhu2026}).
But in practical inference problems it is often more efficient to use faster algorithms that estimate $S_i^\text{ref} S_j^\text{ref}$ 
\textit{without} averaging over $P(S|M)$ exactly \cite{Mussardo2010}.

For example, minimal weight matching of defects, as is familiar in the context of the toric code \cite{DennisKitaevLandahlPreskill2002, WangHarringtonPreskill2003}, corresponds here to a simple algorithm for repairing the  corrupted image of the domain walls of a  2D Ising configuration (Fig.~\ref{isingcartoon}, Right). 
In higher dimensions we can define even simpler algorithms for repairing domain walls \cite{SernaSomozaNahum2024}.
These give non-ideal estimates of 
${S_i^\text{ref} S_j^\text{ref}}$.

It may be interesting to study phase transitions in the success or failure of specific algorithms, 
and to ask how these algorithms should be adapted to exploit the  criticality of the underlying measured system.

\textit{Error correction for critical states.} Relatedly, do the inference problems we have considered here serve as toy models for  practical error correction tasks? 
Two of the basic features of the present problems are that: 

\noindent
(a)  we are interested in decoding only the large scale features of the underlying ``message'', and
not the microscopic state of every single bit
(this is also the case if we must determine the  value of a logical operator in the toric code from  a readout of the qubits \cite{SemeghiniLevineKeeslingEbadiWangBluvsteinVerresenPichlerKalinowskiSamajdarothers2021,eckstein2025learning,
yang2026decoding,
yan2026non,
weinstein2026intrinsic}, or 
if we must determine the homology class of an error from the syndrome measurements \cite{DennisKitaevLandahlPreskill2002}); and 

\noindent
(b) the ``message'' contains long-range, critical correlations.

\textit{Nonoptimal inference.}
We have restricted in this paper to an optimal inference setting \cite{ZdeborovaKrzakala2016},
where the measurer knows the true prior distribution $e^{-\mathcal{H}[S]}/Z$ and the the true $P(M|S)$.
It may be interesting to explore nonoptimal inference for the critical Ising model in which the conditioned ensemble is defined using an ``incorrect'' $\mathcal{H}$ or $P(M|S)$.
One application is to the partial quenches  mentioned in Sec.~\ref{sec:intro}, in which the conditioned-ensemble formalism describes not an inference problem but instead the thermodynamics of a system after the freezing of some of the degrees of freedom.
If we (say) simply quench the values of some of the bonds 
${S_i S_j}$,  but without changing the parameters in $\mathcal{H}$ (including the temperature), 
then this maps to  optimal inference.
But if, more realistically, the parameters of 
$\mathcal{H}$ are left slightly changed after the quench, this maps to a nonoptimal inference problem.
The effect of weak perturbations away from optimality will be controlled by the scaling dimensions of perturbations that break $\mathcal{G}_n$ symmetry.

Turning to the concrete methods in this paper:

\textit{Extensions of the models.} In terms of simulations, larger sizes in 2D would allow more precise
exponent estimates. 
It would also be interesting to find more efficient Monte Carlo schemes for conditioned ensembles.
More efficient numerical methods would allow us to check the  conjectured phase diagram in three dimensions, and to study models with more complex symmetries than the Ising model.\footnote{In our simulations we briefly examined the three-state Potts model with bond energy measurements. For this model a straightforward Metropolis update becomes non-ergodic at large measurement strength, so another approach is needed (e.g.\ tensor networks \cite{PutzGarrattNishimoriTrebstZhu2026}).}

The field-theory approach (which includes both the critical field and an Edwards-Anderson order parameter) could be applied to other measured systems with other symmetries.
Analogs could be applied to measured quantum critical ground states \cite{GarrattWeinsteinAltman2023} via the imaginary time path integral  (with ``spin'' fields living in the bulk and overlap fields living only on the temporal boundary), and to monitored dynamical systems, either classical  \cite{NahumJacobsen2025, GopalakrishnanMcCullochVasseur2026} or quantum \cite{AgrawalZabaloChenWilsonPotterPixleyGopalakrishnanVasseur2022}.

In Sec.~\ref{sec:longrange} we studied a simple model of long-range measurement in which all spin pairs are measured, with low precision for distant pairs. 
This involves ${O(\text{volume}^2)}$ measurements. 
It would be interesting to compare with protocols involving sparse long-range measurements.

The measured gauge theory in  Sec.~\ref{sec:gaugemeasurement} has a rich phase diagram which could be studied with simulations.

\textit{Fixed points with even more symmetry.} At the $\mathcal{G}_n^+$--symmetric fixed point, the fields $\phi_\alpha\sim S_\alpha$ and $\Phi_{\alpha\beta}\sim S_\alpha S_\beta$ are united into a single ``superfield'' $\Psi_{\alpha\beta}$.
There may be a hierarchy of multicritical points, labelled by an integer $m$, in which fields with more replica indices  combine into a superfield with  $m$ indices. It could be interesting to study RG for these theories. However, these  multicritical points might only be relevant to measurement for $m=2$ and for ``$m\to \infty$'' which can perhaps be thought of as the theory mentioned above Eq.~\ref{eq:strongeridentity}.

\textit{Nonperturbative results.} Extending perturbative results for exponents to low dimensions is hampered by the nonanalyticity expected in ${d=4}$. Therefore nonperturbative techniques would be useful. To this end we  made a preliminary analysis of the field theory using the nonperturbative renormalization group (NPRG), retaining a potential of fourth order in the fields (eleven couplings). Some results are shown in appendix \ref{a:NPRG}. Unfortunately, while this scheme works as expected close to 6 dimensions, the fixed point disappears already in  dimension $d=5.6$. According to Ref.~\cite{AnMesterhazyStephanov2016}, a similar problem arises for the Lee-Yang fixed point (another cubic theory), where it is known to be an artifact of the scheme. To cure it  one needs to implement the gradient expansion. We leave this as a challenge for the NPRG community. 

\textit{High-order results.} High-order results will be given in~\cite{DavletbaevaKompanietsWiese2026}. 
For ${n<1}$ one might anticipate that the epsilon expansion can be continued down to low dimensions directly, without any nonanalyticity in $d=4$.

\smallskip

\textit{Note added:} After completing this work we learned of a related independent study of the measurement transition in the critical Ising model by Patil, P\"utz, Trebst, Zhu and Ludwig \cite{PatilPutzTrebstZhuLudwig2026}, which reaches similar conclusions regarding the role of enlarged replica symmetry at the critical point. Otherwise, the approaches used in in the present paper and in Ref.~\cite{PatilPutzTrebstZhuLudwig2026} are largely complementary.

\begin{acknowledgments}
KW thanks M.V.\ Kompaniets for a lecture on the $\ca R'$ operation, and D.A.\ Davletbaeva and M.V.\ Kompaniets for independently verifying the RG results. 
We thank A. Choudhury for collaboration on other approaches to long-range models, and J. Jacobsen, S. Rychkov, B. Delamotte and G. Tarjus for discussions.
AN and AD are supported by the European Union  (ERC, STAQQ, 101171399). Views and opinions expressed are however those of the authors only and do not necessarily reflect those of the European Union or the European Research Council Executive Agency. Neither the European Union nor the granting authority can be held responsible for them.
\end{acknowledgments}

\appendix

\section{Review of the replica formalism}
\label{a:Review_definitions}

For the reader's convenience, here we review a few important definitions and properties of the measurement problem as defined in \cite{NahumJacobsen2025}.

For a lattice system with degrees of freedom $S$ and Hamiltonian $H(S)$
(we absorb $1/k_BT$ into the Hamiltonian)
the Gibbs probability distribution is $P(S)=e^{-H(S)}/Z$. Now, given a spin configuration, the conditional probability of getting a measurement outcome $M$ is written as $P(M|S)$. Here, $M=\{M_{ij}\}$ is the set of outcomes for a set of bonds $\mathcal{B}=\{(i,j)\}$. For example, if we consider  a probability of Gaussian type, then $M_{ij}\in \mathbb{R}$ and
\begin{align}
    P(M|S)=\mathcal{N}_M\exp[-\Gamma\sum_{(i,j)\in\mathcal B}(M_{ij}-S_{i}S_j)^2].
\end{align}
Here, $\mathcal{N}_M$ is a normalisation factor and $\Gamma=1/2\Delta^2$, where $\Delta$ is the standard deviation of the measurement errors.  ${\Gamma\to \infty}$ is the limit of perfectly precise measurements, while ${\Gamma\to 0}$ corresponds to measurements that reveal no information about the system. 
The formulas below hold for more general $P(M|S)$.

We further define the joint probability distribution as $e^{-H(S,M)}/Z=P(S,M)=P(S)P(M|S)$. In the Gaussian example,
\be\label{eq:appjoint}
H(S,M) = H(S) + \Gamma \sum_{(i,j)\in \mathcal{B}}
(M_{ij} - S_i S_j)^2 
\ee
up to an additive constant.

Once a measurement outcome $M$ is obtained, 
the observer's knowledge of the spin configuration is updated to the conditioned distribution, or Bayesian posterior, ${P(S|M)=e^{-H(S, M)}/Z(M)}$, where 
\begin{align}
    Z(M)=\int_S e^{-H(S, M)}
\end{align} 
is the partition function given the measurement outcome $M$.

Expectation values conditioned on the measurement data $M$ are denoted $\<\bullet\>_M$, and are computed from the posterior distribution via $\int_S P(S|M)(\bullet)$.

On the other hand, $\mathbb{E}$ denotes the average over all possible $M$; i.e.\ an average over different instances of this process. 
Physically, this would mean averaging \textit{both} over the ``true'' configuration, which was measured, \textit{and} over the measurement errors obtained in measuring this configuration. 
Formally, it is simply an average using ${P(M) = \int_S P(S,M)}$. The formulae above imply that $P(M) = Z(M)/Z$.

As an example, the measurement-averaged $l$-th moment of some observable $O$ can be written as
\begin{align}\label{eq:EOl}
&    \mathbb{E} [\<O\>^l_M]=\int_M P(M)\left [\int_S O(S)P(S|M)\right]^l\nonumber\\
&
    =\frac{1}{Z}\int_M Z(M)\left[\frac{1}{Z(M)^l}\int_{S^1,...,S^l}\prod_{\alpha=1}^l O(S^\alpha)e^{-H(S^\alpha,M)}\right] \nonumber\\
    &
    =\frac{1}{Z}\int_M Z(M)^{1-l}\int_{S^1,..,S^l} O(S^1)...O(S^l)e^{-\sum_{\alpha=1}^l H(S^\alpha,M)}
\end{align}
where $\alpha=1,..,l$  label replicas (in the main text we used subscripts, here we use superscripts). Now we can replace the exponent  ${1-l}$ of the $Z(M)$ factor with ${n-l}$, so long as we analytically continue to   ${n= 1}$ at the end. 
This gives $n$ replicas in total, and we have
\begin{align}
    \mathbb{E}[\<O\>^l_M]=\lim\limits_{n\to 1}\frac{1}{Z}\int_{M,S^1,..,S^n} O^{\alpha_1}...O^{\alpha_l}e^{-\sum_{\alpha=1}^n H(S^\alpha,M)}.
    \label{eq:Moment_rep_int}
\end{align}
We use $O^{\alpha}\equiv O(S^\alpha)$ as a shorthand notation, and $\{\alpha_i\}_{i=1}^l$ can be chosen to be any set of $l$ replicas among the $n$ of them. 
Finally, note that $\lim_{n\to 1} Z_n = Z$ if we define
\be
Z_n = \int_{M, S^1, \ldots, S^n} e^{-\sum_{\alpha=1}^n H(S^\alpha,M)}. 
\ee
Therefore we can replace (\ref{eq:Moment_rep_int})
with 
\begin{align}
    \mathbb{E}[\<O\>^l_M]=\lim\limits_{n\to 1}\frac{1}{Z_n}\int_{M,S^1,..,S^n} O^{\alpha_1}...O^{\alpha_l}e^{-\sum_{a=1}^n H(S^\alpha,M)}
    \label{eq:Moment_rep_int2}
\end{align}
The quantity inside the limit on the right-hand side is now an expectation value in the $n$-replica theory.

We can now integrate over $M$ to obtain an effective Hamiltonian $H(S^1,\ldots, S^n)$ for $S^1,\ldots, S^n$ which involves interactions between these replicas. 
For Gaussian measurements, this requires only a Gaussian integral over $M$, and leads to  pairwise replica interactions $\propto - (S^\alpha_i S^\alpha_j)(S^\beta_i S^\beta_j)$. This replica Hamiltonian was given in Sec.~\ref{sec:latticeprotocols}. 
We may write, for example, 
\begin{align}
    \mathbb{E}[\<O\>^l_M]=\lim\limits_{n\to 1}
    \< O^1 O^2 \ldots O^l\>,
\end{align}
where the expectation on the right-hand side is evaluated using the replica Hamiltonian.

This replica Hamiltonian describes more than we might at first have expected. 

First,  we are free to consider more general averages $\mathbb{E}$ than (\ref{eq:EOl}). In particular, 
the quantity inside $\mathbb{E}$ can depend explicitly on $M$.
We might worry that such correlations went beyond the ones that can be captured by $H(S^1,\ldots, S^n)$;
however, this is not the case. 
For example, we may check that, in the case of Gaussian measurements, a single insertion of $M_{ij}$ on some bond into  $\mathbb{E}$ is equivalent to an insertion involving the spin operators,
\be
M_{ij} \rightarrow \f{1}{n} \sum_{\alpha=1}^n S^\alpha_i S^\alpha_j.
 \ee

Physically, there is also a distinction between  the \textit{true} measured sample (which we denote by $S^\text{true}$) and a sample from the posterior measure. 
For example, we might be interested in comparing the value of an observable $O^\text{true}$ in the true configuration with our estimate of it $\<O\>_M$ obtained using the Bayesian posterior.
In the present setting, $O$ could be a spin difference $S_i S_j =\pm 1$, in which case a natural measure of the success of inference is the overlap
\be
\mathbb{E} \left[ O^\text{true} \<O\>_M \right] = 
\mathbb{E} \left[ (S^{\text{true}}_{i} S^{\text{true}}_{j}) \< S_i S_j\>_M \right].
\ee
(We have extended the definition of $\mathbb{E}$ to allow averaging over both $S^\text{true}$ and $M$, using the joint distribution $P(S^\text{true}, M)$ defined above Eq.~\ref{eq:appjoint}.)
The average above involves two ``kinds'' of sample, the true sample $S^\text{true}$, and a sample $S$ from the posterior measure.

An important point is that, from the point of view of the replica theory, these two kinds of sample are interchangeable \cite{ZdeborovaKrzakala2016}. 
If desired, we are free to identify one of the replicas (say $S^1$) with the true configuration $S^\text{true}$, and another (say $S^2$) with the posterior sample.
This can be seen from a straightforward extension of the first line of (\ref{eq:EOl}). Say that we average $S^\text{true}$, $M$ and the posterior sample $S$. The natural ``physical'' picture is to imagine first sampling $S^\text{true}$; then sampling $M$ using a measurement of $S_\text{true}$; then sampling $S$ from the posterior distribution $P(\bullet|M)$. This gives the joint distribution
\be
P(S^\text{true}) P(M|S^\text{true}) P(S|M).
\ee
But using $P(S)P(M|S)=P(M)P(S|M)$, this is also equal to 
\be
P(M)
P(S^\text{true}|M) P(S|M),
\ee
making apparent the symmetry between $S^\text{true}$ and $S$.
As a result, 
\be
\mathbb{E} \left[ O^\text{true} \<O\>_M \right] = 
\mathbb{E} \left[ \<O\>_M^2 \right].
\ee
This symmetry does however rely on the fact that we are considering ``optimal'' Bayesian inference, i.e.\ we are assuming that in performing the inference task (constructing the posterior distribution) we use the correct form of $H$ and of $P(S|M)$. If the inferrer has incorrect assumptions about either of these, then inference is ``nonoptimal'' and the symmetry between $S^\text{true}$ and a posterior sample is broken. This also reduces the permutational symmetry of the replica theory.

Note that when $l=1$, in the $n\to 1$ limit the contribution of measurement vanishes and we simply get
\be
\mathbb{E}[\<O\>_M]=\<O\>,
\ee
where the right-hand side denotes the  conventional expectation value in the Gibbs state without any measurement.
This apparently trivial constraint is a key feature of the replica CFTs describing the measurement problems: single-replica observables reduce to  observables in the conventional thermodynamic ensemble.

\section{SR calculation}
\subsection{Summary of the RG procedure}
\label{Summary of the RG procedure}
We use  dimensional regularization and minimal subtraction in a massless scheme \cite{Zinn,Vasilev2004,KleinertSchulte-FrohlindeBook}.
All diagrams are calculated with momentum $p=1$ entering into the external legs. 
They are normalized s.t.
\begin{equation}\label{A1}
     \triagdiag \equiv \frac1{\epsilon}, \quad \epsilon = 6-d, 
\end{equation}
where the momentum enters and exits at two distinct vertices. 
To streamline the calculations, we evaluate diagrams under 
the $\ca R'$-operation \cite{Vasilev2004}. Contrary to subtraction via the original Bogoliubov $\ca R$-operation \cite{BogoliubovParasiuk1957}, this procedure does not render a diagram divergence free; instead it renders its momentum dependence finite. 
It also allows to freely choose where momenta enter into a diagram and where they exit, as long as this choice of momenta does not induce IR divergencies. It is today the standard procedure to perform high-order calculations, both for quartic theories \cite{KompanietsPanzer2017, Schnetz2018} and for cubic ones \cite{BorinskyGraceyKompanietsSchnetz2021,KompanietsPikelner2021}. 
Note that in these authors usually use 
\be
d = 4- 2 \varepsilon, \quad d= 6 - 2 \varepsilon . 
\ee
In contrast we use $d = 6-\epsilon$, implying $\epsilon = 2 \varepsilon$. The natural normalization \eqK{A1} is then with $\varepsilon$, which induces factors of $2^{\ell-1}$, where $\ell $ is the loop order.

\subsection{Diagrams for the SR calculation}

\textit{Vertex renormalization.}
\be
\ca R' ~\triagdiag  = \frac1{\epsilon}
\ee
\be
\label{RprimediagD1}
\ca R' ~\diagDone  = \diagDoneTwoLegs - \triagdiagCTtwoLegs = -\frac1{2\epsilon^2} + \frac1{8\epsilon}
\ee
Remark that under the original $\ca R$ operation the leading divergence $1/(2\epsilon^2)$ is canceled. The combination written here, with its   momentum dependence  reads
\be
\ca R' \diagDfive = \left[ \frac1{2\epsilon^2} + \frac1{8\epsilon}\right] p^{-2 \epsilon} - \left[\frac1{\epsilon^2}\right] p^{\
-\epsilon}  
\ee
The second term has a different $p$-dependence, since the 1-loop counter term, the box in \Eq{RprimediagD1}, is $1/\epsilon$, and the remaining diagram then scales as $p^{-\epsilon}$, as written. Taylor expanding in $p$, we find that the term of order $\ln p $ is  finite, as stated in Sec.~\ref{Summary of the RG procedure}.

The remaining diagrams at 2-loop order are
\begin{align}
&\ca R' \diagDfive \nn\\
&= \diagDfive - \propdiag \times \triagdiagCTWF \nn\\
&=  \frac{1}{6 \epsilon ^2}-\frac{7}{72 \epsilon }.
\\
& \ca R' \diagDsix = \frac{1}{2 \epsilon }. 
\end{align}

\textit{Propagator renormalization.}
\bea
 {\propdiag}  &=& \frac{p^{2-\epsilon }}{(\epsilon -3) \epsilon }\\
 &=& -\frac{1}{3\epsilon } p^2 \left[ 1+\frac{\epsilon}{3}   \big(1-3 \log (p)\big)+\ca O (\epsilon ^2 ) \right]  \nn
 \\
\ca R' \partial_p^2  {\propdiag} &=& -\frac1{3\epsilon}
\eea
\begin{align}
&  \ca R'  \partial_p^2\diagDfour _p \nn\\
&=  \partial_{p^2} \diagDfour   -  \partial_{p^2}\propdiag \times \propdiagCTm  \nn\\
&= \frac{-1}{18 \epsilon ^2}+\frac{11}{216 \epsilon } 
\end{align}
\begin{align}
   &\ca R'  \partial_{p^2}  \diagDthree  
   =  \partial_{p^2}  \diagDthree 
   =  \frac{1}{3 \epsilon ^2}-\frac{1}{9 \epsilon }.
\end{align}

\textit{Mass renormalization.}
\begin{align}
\label{78}
&\ca R' \partial_{m^2} \propdiag =- 2 \ca R'  \triagdiag  = -\frac2\epsilon .\\
&\ca R'  \partial_{m^2}  \diagDthree = -4 \ca R'  \diagDone - \ca R' \diagDsix = \frac{2}{\epsilon ^2}-\frac{1}{\epsilon } .
\label{79}\\
&\ca R' \partial_{m^2}  \diagDfour = -3 \ca R'\diagDfive -2\ca R' \diagDone \nn\\
& = \frac{1}{2 \epsilon ^2}+\frac{1}{24 \epsilon }.
\label{80}
\end{align}

\subsection{Graphical representation of  2-loop corrections}

The corrections to the 3-point function are 
\be\label{delta-S3}
\delta S^{(3)} = \frac 1{3!} \smalltriagdiag + \frac 1{2}  \smalldiagDone
+  \frac 1{4} \smalldiagDfive  +  \frac 1{12} \smalldiagDsix + \ca O(\lambda^7)
\ee
Corrections to the quadratic part of the action read
\be\label{delta-S1}
-\delta S^{(1)}_{p^2} = \frac12 \smallpropdiag_{p^2} +  \frac 1{4} \smalldiagDthree_{p^2}
+  \frac 1{4}  \smalldiagDfour_{p^2} + \ca O(\lambda^6)
\ee
A similar expression holds for the correction to the mass. 
\be
-\delta S^{(1)}_{m^2} = \frac12 \smallpropdiag_{m^2} +  \frac 1{4} \smalldiagDthree_{m^2}
+  \frac 1{4}  \smalldiagDfour_{m^2} + \ca O(\lambda^6)\Big|_{m^2}
\ee
The corresponding diagrams are given in \Eqs{78}-(\ref{80}).

\subsection{Group-theory factors}

\textit{One-loop order.}
We consider the corrections to $\lambda_1$ and $\lambda_2$. 
The first line is   for $n$ uncoupled copies of Lee-Yang, with which we checked our calculations. 
\bea
\ca G( \smalltriagdiag|\textstyle  \Phi_a^3) &=& \lambda_0^3 \\
\ca G(\smalltriagdiag|\textstyle 
\Phi_\alpha \Phi_\beta \Phi_{\alpha\beta} ) &=&3 \left[  \lambda _1^3+\lambda _2 \lambda _1^2 (n-2) \right] \\
\ca G(\smalltriagdiag| \textstyle 
\Phi_{\alpha\beta} \Phi_{\beta\gamma} \Phi_{\alpha\gamma} ) &=& \lambda _1^3+\lambda _2^3 (n-2).
\eea
For the wave-function renormalization, we have 
\bea
\ca G(\smallpropdiag|(\nabla\Phi_\alpha)^2) &=& \half \lambda _0^2\\
\ca G(\smallpropdiag|(\nabla\Phi_\alpha)^2) &=& \lambda _1^2 (n-1) \\
\ca G(\smallpropdiag|(\nabla\Phi_{\alpha\beta})^2) &=&  \frac{1}{2} \left[\lambda _1^2+\lambda _2^2 (n-2)\right]
\eea

\textit{Two-loop order.}
\begin{align}
&\ca G(\smalldiagDone  |\textstyle   \Phi_a^3 ) = \lambda_0^5 \\
&\ca G(\smalldiagDone | \Phi_\alpha \Phi_\beta \Phi_{\alpha\beta} ) =\lambda _1^2 \big[\lambda _1^3 (n+1)+5 \lambda _2 \lambda _1^2 (n-2) \nn\\& \qquad 
+2 \lambda
   _2^2 \lambda _1 (n-2)^2+\lambda _2^3 (n-2)^2\big] \\
&\ca G(\smalldiagDone  |\Phi_{\alpha\beta} \Phi_{\beta\gamma} \Phi_{\alpha\gamma}  )= \lambda _1^5+\lambda _2 \lambda _1^4 (n-2)+\lambda _2^2 \lambda _1^3 (n-2)  \nn\\ &
 \qquad+\lambda
   _2^5 (n-2)^2.
\end{align}
\bea
\ca G(\smalldiagDthree|(\nabla\Phi_\alpha)^2) &=&   \lambda _0^2\\
\ca G(\smalldiagDthree|(\nabla\Phi_\alpha)^2) &=& 2 \lambda _1^3 (n-1) \left[\lambda _1+\lambda _2 (n-2)\right]  \\
\ca G(\smalldiagDthree|(\nabla\Phi_{\alpha\beta})^2) &=&  \lambda _1^4+2 \lambda _2 \lambda _1^3 (n-2)+\lambda _2^4 (n-2)^2\nn\\
\eea
\bea
\ca G(\smalldiagDfour|(\nabla\Phi_\alpha)^2) &=&   \lambda _0^2\\
\ca G(\smalldiagDfour|(\nabla\Phi_\alpha)^2) &=& 2 \lambda _1^2 (n-1) \left[\lambda _1^2 n+\lambda _2^2 (n-2)\right]\nn\\ \\
\ca G(\smalldiagDfour|(\nabla\Phi_{\alpha\beta})^2) &=&   2 \big[\lambda _1^4 (n-1)+\lambda _2^2 \lambda _1^2 (n-2) \nn\\
&& ~~+\lambda _2^4
   (n-2)^2\big]
\eea
\begin{align}
&\ca G(\smalldiagDfive  |\textstyle \sum_\alpha \Phi_a^3 ) = \lambda_0^5  \\
&\ca G(\smalldiagDfive | \Phi_\alpha \Phi_\beta \Phi_{\alpha\beta} ) =  2 \lambda _1^2 \big[\lambda _1^3 (2 n-1) \nn\\
&\qquad +\lambda _2 \lambda _1^2 (n-2)
   (n+1)+\lambda _2^2 \lambda _1 (n-2) \nn\\
&\qquad +2 \lambda _2^3 (n-2)^2\big]   \\
&\ca G(\smalldiagDfive  |\Phi_{\alpha\beta} \Phi_{\beta\gamma} \Phi_{\alpha\gamma}  )= 
2 \big[\lambda _1^5 (n-1)+\lambda _2^3 \lambda _1^2 (n-2) \nn\\
&\qquad +\lambda _2^5
   (n-2)^2\big]
\end{align}
\bea
\ca G(\smalldiagDsix  |\textstyle \sum_\alpha \Phi_a^3 ) &=& \lambda_0^5  \\
\ca G(\smalldiagDsix | \Phi_\alpha \Phi_\beta \Phi_{\alpha\beta} ) &=& 6 \lambda _1^3 \left[\lambda _1^2 (n-1)+2 \lambda _2^2 (n-2)\right] \nn\\ \\
\ca G(\smalldiagDsix  |\Phi_{\alpha\beta} \Phi_{\beta\gamma} \Phi_{\alpha\gamma}  )&=& 6 \lambda _1^4 \lambda _2+2 \lambda _2^5 (3 n-8)
\eea

\subsection{RG functions and critical exponents}
\label{app:RGfunctionsandexponents}

\textit{The $\beta$-functions.}
To obtain the  $\beta$-functions, one needs to evaluate 
the corrections for the 3-point vertex in \Eq{delta-S3} and to the 2-point vertex in \Eq{delta-S1}. The corrections to the couplings $\lambda_i$ and critical exponents $\eta_i$ are obtained via 
\bea
\delta \lambda_1 &:=& 2 ( \delta S^{(3)} |  \Phi_\alpha \Phi_\beta \Phi_{\alpha\beta}  ) \\
\delta \lambda_2 &:=& 6 ( \delta S^{(3)} |  \Phi_{\alpha\beta} \Phi_{\beta\gamma} \Phi_{\alpha\gamma}   )
\\
\delta \eta_1 &:=& 2 ( \delta S^{(1)} | (\nabla\Phi_\alpha)^2) \\
\delta \eta_2 &:=& 2 ( \delta S^{(1)} | (\nabla\Phi_{\alpha\beta})^2)
\eea
The effective renormalized couplings are
\bea
f_1(\lambda_1,\lambda_2) &:=& \frac{\lambda_1 -\delta \lambda_1}{(1-\delta \eta_1)(1-\delta \eta_2)^{1/2}} \\
f_2(\lambda_1,\lambda_2) &:=& \frac{\lambda_2 -\delta \lambda_2}{(1-\delta \eta_2)^{3/2}}
\eea
The unintuitive  signs are due to the fact that under the $\mathcal R'$ operation, we get the bare coupling in terms of the renormalized one; and similarly for the anomalous exponents $\eta$. 

Let us now evaluate the $\beta$-functions. To this aim we vary the efffective coupling with respect to a momentum RG subtraction scale $\mu$, 
\bea
0 &=& \mu \partial_\mu\big[  f_1(\lambda_1,\lambda_2) \mu^{-\epsilon/2} \big]\nn \\
0 &=& \mu \partial_\mu \big[f_2(\lambda_1,\lambda_2) \mu^{-\epsilon/2}\big]
\eea
This implies 
\bea
\frac\epsilon 2 f_1(\lambda_1,\lambda_2) &=& \Big[\beta_1 (\lambda_1,\lambda_2)\partial_{\lambda_1} + \beta_2 (\lambda_1,\lambda_2)\partial_{\lambda_2}\Big]f_1(\lambda_1,\lambda_2) \nn \\
\frac\epsilon 2 f_1(\lambda_1,\lambda_2) &=& \Big[\beta_1 (\lambda_1,\lambda_2)\partial_{\lambda_1} + \beta_2 (\lambda_1,\lambda_2)\partial_{\lambda_2}\Big]f_2(\lambda_1,\lambda_2)
\nn\\
\eea
Solving for $\beta_i(\lambda_1,\lambda_2)$ yields the $\beta$-functions in the main text, see \Eqs{beta1-all-n} and \eqK{beta2-all-n}.  Specified to $n=1$ this is given in \Eqs{beta1-n=1} and \eqK{beta2-n=1}.

\textit{$\omega$ exponents.}
The correction-to-scaling exponent $\omega$ is discussed in section \ref{sec:correctionscaling}. Here we give this exponent for $n<1$, evaluated at the isotropic 1-loop fixed point. 
\begin{eqnarray}\label{a:SRomega1}
    \omega_1^{n<1} &=&  -\epsilon
   +\left(\frac{5}{18}+\frac{6}{n{-}1}-\frac{4}{(n{-}1)^2}\right) \epsilon ^2 + ...\\
    \omega_2^{n<1} &=& -\frac{(n-5)
   \epsilon }{3 (n-1)} \nn\\
   && +\frac{(n (n (7 n+292)-955)+368)
   \epsilon ^2}{54 (n{-}1)^3}  +...
   \label{a:SRomega2}
\end{eqnarray}

\textit{$\eta$ exponent.}
Results for $\eta$ are given in section \ref{s:Anomalous field dimensions: eta}. Details and higher-loop results will be given in \cite{DavletbaevaKompanietsWiese2026}.

\textit{$\nu$ exponents.}
Results for $\nu$ at 2-loop order are given in section \ref{sec:massterms}. We defer higher-order corrections to 
\cite{DavletbaevaKompanietsWiese2026}. 

\section{LR calculation}
\label{a:LR}
\subsection{Diagrams for  LR calculation}

\begin{widetext}
\begin{align}\label{diagD1LRnorm}
&\ca D_1 = {\diagDoneTwoLegs} =    \frac{\Gamma (\alpha )\Gamma (\epsilon ) \Gamma
    (\frac{\alpha }{2}-\epsilon  ) \Gamma  (\frac{3    \alpha }{2}-\epsilon  )}{\epsilon ^2 \Gamma    \left(\frac{\epsilon }{2}\right) \Gamma  (\frac{\alpha    -\epsilon }{2} ) \Gamma  (\frac{3 (\alpha -\epsilon  )}{2} ) \Gamma  (\alpha +\frac{\epsilon }{2} )}
= \frac{1}{2 \epsilon ^2}+\frac{\psi\left(\frac{3 \alpha  }{2}\right) -\psi(\alpha )  -\psi\left(\frac{\alpha }{2}\right) -\gamma_{\rm E} }{4 \epsilon } 
+ ...
\end{align}

\be
\label{RprimediagD1LR}
\ca R' \ca D_1 = \ca R' ~\diagDone  = \diagDoneTwoLegs + \triagdiagCTtwoLegs = -\frac1{2\epsilon^2}  +\frac{\psi (\frac{3 \alpha  }{2} ) -\psi(\alpha )  -\psi (\frac{\alpha }{2} )-\gamma_{\rm E} }{4 \epsilon } + ...
\ee
\be
\ca D_5 = {\diagDfiveAmp }  = -\frac{3\ 2^{\alpha -2} \Gamma (\alpha )   \Gamma (\frac{\epsilon    +1}{2}) \Gamma
   (\frac{1}{2} (\alpha -\epsilon +1)) \Gamma (\frac{3 \alpha }{2}-\epsilon
   )^2}{\epsilon ^2 \sin (\frac{\pi}{2}   (\alpha -\epsilon ))  \Gamma (\frac{\epsilon }{2}) \Gamma (\frac{3 \alpha
   }{2}-\frac{3 \epsilon }{2}+1) \Gamma (\frac{\alpha -\epsilon }{2}) \Gamma
   (2 \alpha -\epsilon ) \Gamma (\frac{\alpha +\epsilon }{2})}
   = \frac{ \Gamma  (-\frac{\alpha }{2} ) \Gamma  (\frac{3 \alpha
   }{2} ) \Gamma  (\frac{\alpha +1}{2} )^2}{4^{2-\alpha } \pi  \Gamma (2 \alpha )\epsilon} + ...
\ee
\end{widetext}
\begin{align}
\ca D_6   =  {\diagDsix} 
= \frac{\Gamma  (\frac{\alpha }{2} )^3 \Gamma  (\frac{3
   \alpha }{2} )}{4 \Gamma (\alpha )^3}\frac1\epsilon +...
\end{align}
Neither $\ca D_5$ nor $\ca D_6$     have counter terms, thus 
$\ca R' \ca D_5= \ca D_5 +\ca O({\epsilon})$ and 
$\ca R' \ca D_6= \ca D_6+  \ca O({\epsilon})$. 
We define $\ca C_1$, $\ca C_5$ and $\ca C_6$ via 
\be
\ca R' \ca D_1 =-\frac1{2\epsilon^2}+  \frac{\ca C_1}\epsilon, 
\quad \ca R' \ca D_5 = \frac{\ca C_5}\epsilon, 
\quad \ca R' \ca D_6 = \frac{\ca C_6}\epsilon. 
\ee
As a function of $\alpha$
\bea
\ca C_1 &=& \frac{\psi (\frac{3 \alpha  }{2} ) -\psi(\alpha )  -\psi (\frac{\alpha }{2} )-\gamma_{\rm E} }{4   }, \\
\ca C_5 &=& \frac{ \Gamma  (-\frac{\alpha }{2} ) \Gamma  (\frac{3 \alpha
   }{2} ) \Gamma  (\frac{\alpha +1}{2} )^2}{4^{2-\alpha } \pi  \Gamma (2 \alpha )}, \\
\ca C_6 &=& \frac{\Gamma  (\frac{\alpha }{2} )^3 \Gamma  (\frac{3
   \alpha }{2} )}{4 \Gamma (\alpha )^3} .
\eea
As a function of $d$, obtained by setting $\alpha\to d/3$
\bea
\ca C_1 &=& \frac{\psi (\frac{d  }{2} ) -\psi(\frac d 3 )  -\psi (\frac{d }{6} )-\gamma_{\rm E} }{4   }, \\
\ca C_5 &=& \frac{\Gamma  (-\frac{d}{6} ) \Gamma  (\frac{d}{2} ) \Gamma
    (\frac{d+3}{6} )^2}{4^{2-\frac{d}{3}}  \pi  \Gamma \left(\frac{2 d}{3}\right)}, \\
\ca C_6 &=& \frac{\Gamma  (\frac{d}{6} )^3 \Gamma  (\frac{d}{2} )}{4    \Gamma  (\frac{d}{3} )^3}.
\eea

\subsection{$\beta$-function and stability}
\label{s:LR-beta}

\begin{widetext}
 In the full $(\lambda_1, \lambda_2)$ plane, the beta functions to 2-loop order are:
\bea
\beta_1(\lambda_1,\lambda_2)&=& \frac{
   \epsilon_1 }{2} \lambda _1 +\lambda _1^3+\lambda _2 \lambda _1^2 (n -2)  \nn\\
   &&+ 2 \lambda _1^2 \Big[\lambda _1^3 \left(\mathcal{C}_1 (n+1)+\mathcal{C}_5 (2 n-1)+\mathcal{C}_6 (n-1)\right)+\lambda _2
   \lambda _1^2 (n-2) \left(5 \mathcal{C}_1+\mathcal{C}_5 (n+1)\right) \nn\\
   && \qquad\quad  +\lambda _2^2 \lambda _1 (n-2) \left(\mathcal{C}_5+2
   \mathcal{C}_6+2 \mathcal{C}_1 (n-2)\right)+\left(\mathcal{C}_1+2 \mathcal{C}_5\right) \lambda _2^3 (n-2)^2\Big]\\
\beta_2(\lambda_1,\lambda_2)&=&    \frac{\epsilon_2}2\lambda_2 + \lambda _1^3+\lambda _2^3 (n-2) \nn\\
&& +2 \mathcal{C}_6 \lambda _2 \Big[3 \lambda _1^4+\lambda _2^4 (3 n-8)\Big]+6 \mathcal{C}_1 \Big[\lambda _1^5+\lambda _2
   \lambda _1^4 (n-2)+\lambda _2^2 \lambda _1^3 (n-2)+\lambda _2^5 (n-2)^2\Big] \nn\\
&& +6 \mathcal{C}_5 \Big[\lambda _1^5
   (n-1)+\lambda _2^3 \lambda _1^2 (n-2)+\lambda _2^5 (n-2)^2\Big]
\eea
At $n=0$:
\bea
\beta_1(\lambda_1,\lambda_2)&=& \frac{
   \epsilon_1 }{2} \lambda _1 +\lambda _1^3-2\lambda _2 \lambda _1^2    \nn\\
   && +2 \left(\mathcal{C}_1-\mathcal{C}_5-\mathcal{C}_6\right) \lambda _1^5-4 \left(5 \mathcal{C}_1+\mathcal{C}_5\right)
   \lambda _2 \lambda _1^4+4 \left(4 \mathcal{C}_1-\mathcal{C}_5-2 \mathcal{C}_6\right) \lambda _2^2 \lambda _1^3+8
   \left(\mathcal{C}_1+2 \mathcal{C}_5\right) \lambda _2^3 \lambda _1^2\\
\beta_2(\lambda_1,\lambda_2)&=&    \frac{\epsilon_2}2\lambda_2 + \lambda _1^3-2\lambda _2^3   \nn\\
&& +2 \mathcal{C}_6 \lambda _2 \left(3 \lambda _1^4-8 \lambda _2^4\right)-6 \mathcal{C}_5 \left(\lambda _1^5+2 \lambda _2^3
   \lambda _1^2-4 \lambda _2^5\right)+6 \mathcal{C}_1 \left(\lambda _1^5-2 \lambda _2 \lambda _1^4-2 \lambda _2^2 \lambda
   _1^3+4 \lambda _2^5\right)
\eea
At $n=1$:
\bea
\beta_1(\lambda_1,\lambda_2)&=& \frac{
   \epsilon_1 }{2} \lambda _1 +\lambda _1^3 - \lambda _2 \lambda _1^2   \nn\\
   && +2 \lambda _1^2 \left[\left(2 \mathcal{C}_1+\mathcal{C}_5\right) \lambda _1^3-\left(5 \mathcal{C}_1+2 \mathcal{C}_5\right)
   \lambda _2 \lambda _1^2-\left(-2 \mathcal{C}_1+\mathcal{C}_5+2 \mathcal{C}_6\right) \lambda _2^2 \lambda
   _1+\left(\mathcal{C}_1+2 \mathcal{C}_5\right) \lambda _2^3\right]\\
\beta_2(\lambda_1,\lambda_2)&=&    \frac{\epsilon_2}2\lambda_2 + \lambda _1^3 - \lambda _2^3   \nn\\
&& +6 \mathcal{C}_1 \lambda _1^5+6 \left(\mathcal{C}_6-\mathcal{C}_1\right) \lambda _2 \lambda _1^4-6 \mathcal{C}_1 \lambda
   _2^2 \lambda _1^3-6 \mathcal{C}_5 \lambda _2^3 \lambda _1^2+2 \left[3 \left(\mathcal{C}_1+\mathcal{C}_5\right)-5
   \mathcal{C}_6\right] \lambda _2^5
\eea
At $n=2$:
\bea
\beta_1(\lambda_1,\lambda_2)&=& \frac{
   \epsilon_1 }{2} \lambda _1 +\lambda _1^3+\lambda _2 \lambda _1^2 (n -2)  +2 \left(3 \left(\mathcal{C}_1+\mathcal{C}_5\right)+\mathcal{C}_6\right) \lambda _1^5\\
\beta_2(\lambda_1,\lambda_2)&=&    \frac{\epsilon_2}2\lambda_2 + \lambda _1^3+\lambda _2^3 (n-2) + 6 \left(\mathcal{C}_1+\mathcal{C}_5\right) \lambda _1^5+6 \mathcal{C}_6 \lambda _2 \lambda _1^4-4 \mathcal{C}_6 \lambda
   _2^5
\eea
In $d=1$
\bea
\beta_1(\lambda_1,\lambda_2)&=& \frac{
   \epsilon_1 }{2} \lambda _1 +\lambda _1^3+\lambda _2 \lambda _1^2 (n -2)  \nn\\
   &&+ \lambda _1^2 \Big[\lambda _2 \lambda _1^2 \Big(n (18.3353\, -1.0267 n)-32.5638\Big)+1.40833 \lambda _2^3 (2  -  n)^2\nn\\
   && \qquad +\lambda
   _1^3 (9.3582 n-3.46146)+\lambda _2^2 \lambda _1 (n (6.92344 n-12.8207)-2.05236)\Big]\\
\beta_2(\lambda_1,\lambda_2)&=&    \frac{\epsilon_2}2\lambda_2 + \lambda _1^3+\lambda _2^3 (n-2) \nn\\
&&+\lambda _1^5 (13.4652\, -3.08009 n)+\lambda _2^3 \lambda _1^2 (6.16017\, -3.08009 n)+\lambda _2 \lambda _1^4 (10.3852
   n+3.07932) \nn\\
&&   +\lambda _2^2 \lambda _1^3 (10.3852 n-20.7703)+\lambda _2^5 \Big[n (7.30507 n-5.37065)-34.3787\Big]
\eea
\end{widetext}
We now consider $\epsilon_1 = \epsilon_2 = \epsilon$, and $n=1$.
The $\beta$-functions for $\lambda_1=\lambda_2$ are given in 
\Eqs{beta-LR-iso1} or \eqK{beta-LR-iso2}. 
We find that for all relevant dimensions $0<d\le6$ or $0<\alpha\le 2$, the diagonal fixed point $\lambda_1=\lambda_2$ is stable. 
The corection-to-scaling exponents are 
\begin{eqnarray}
\omega_1 &=&  2 \epsilon \nn\\
\omega_2 &=&   \frac{2}3  \frac{\sqrt{2 \epsilon } \Gamma
   (\frac{d}{3})^{3/2}}{ \Gamma
   (\frac{d}{6})^{3/2} \sqrt{\Gamma
   (\frac{d}{2})}}. 
\end{eqnarray}
To get the result as a function of $\alpha$ instead of $d$, one replaces $d\to 3\alpha$. $\omega_2/\sqrt\epsilon$ varies continuously between $\alpha=0$ and $\alpha=2$.

\section{NPRG}
\label{a:NPRG}

\begin{figure*}
    \centering
    \includegraphics{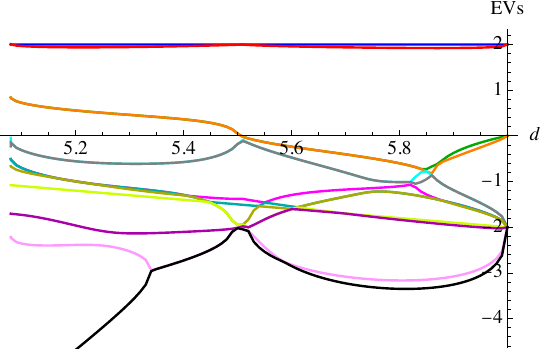}\Fig{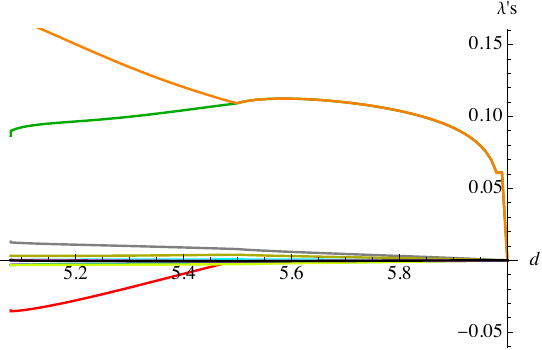}
    \caption{The NPRG eigenvalues (left) and couplings (right) in the LPA' scheme with all operators with up to four fields (9 couplings).}
    \label{NPRG-LPAprime}
\end{figure*}

We studied the problem also in the NPRG scheme using the LPA' approximation. Our ansatz (not shown) contains all terms up to order four in the fields, i.e.~11 couplings. 
We then numerically integrate the flow equations. 
The results are shown in Fig.~\ref{NPRG-LPAprime}. 
In the left plot, one sees the RG eigenvalues; in the right one the couplings. As one can see, 
NPRG works down to dimension $d=5.6$; then the fixed points move into the complex plane. 
We expect this is an artifact of NPRG, as it is very similar to what happens when the Yang-Lee problem, i.e.\ scalar $\phi^3$ theory (with an imaginary coupling), is treated similarly. 

Ref.~\cite{AnMesterhazyStephanov2016} has shown that artifacts for Yang-Lee can be cured by using the gradient expansion to second order.
This can be understood as follows: power-counting at the Gaussian fixed point shows that the field has dimension ${[\phi]_\mu = \frac{d-2}2 = 2-\frac{\epsilon}2}$. This implies that in $d=6$ an operator with two additional derivatives is as relevant as an operator with one additional power of $\phi$. 
The NPRG in the LPA' scheme expands in powers of fields, 
thus it neglects  the fact that at subleading order $\phi^4$ will mix with $\phi^2 \nabla^2 \phi$.
It would therefore be interesting to redo the analysis using the gradient expanion. We leave this to future studies. 

\section{Details of simulation}
\label{a:Sim_details}

First we consider computations involving moments of ${\Phi = L^{-2} \sum_i S_{1i}S_{2i}}$. Formally, 
\begin{align}
  &  \mathbb{E}\<\Phi^{2l}\>_M=\int_M P(M)     \sum_{S_1, S_2} 
\frac{ \exp(-\sum_{a=1}^2 \mathcal{H}_M [S^a])}{Z_M^2} \Phi^{2l} 
\end{align}
with $Z_M=\sum_S \exp(-\mathcal{H}_M[S])$, 
where 
\be
\label{eq:heffapp}
\mathcal{H}_M[S] = \mathcal{H}[S] - {\sum_{ij}}' \ln P(M_{ij}|S_iS_j)
\ee 
is the effective Hamiltonian in a given measurement realization $M=\{M_{ij}\}$, 
and $P(M)=Z_M/Z$.
In Eq.~\ref{eq:heffapp}, the prime indicates that we sum only once over each measured bond, and $P(M_{ij}|S_{i} S_{j})$ is the conditional probability that defines the given measurement protocol. We may also write
\be
P(M) = \f{1}{Z} \sum_S e^{-\mathcal{H}[S]} {\prod_{ij}}' P(M_{ij}|S_iS_j),
\ee
which reflects how $M$ is sampled by starting with an equilibrated spin configuration.

In our simulations, we start with a uniformly random spin configuration and evolve it under Metropolis dynamics with the pure Ising Hamiltonian $\mathcal{H}[S]$ for a time $T\sim 10\tau_c$, where $\tau_c\approx 2.5L^{2.1}$ for a system of length $L$ (one time step is a sweep of the lattice, i.e.\ $L^2$ attempted single-spin updates). 
From this equilibrated spin configuration $S_T$ (say), we choose a measurement outcome for each bond with the conditional probability $P(M_{ij}|S_{Ti} S_{Tj})$.

Next, we take two copies of this configuration ${S^1=S^2=S_T}$ and evolve them independently with the effective Hamiltonian $\mathcal{H}_M(S)$. After an initial waiting for a time $T_{\text{eq}}$ (equilibration time), we sample 
$\Phi$ at time intervals $t_0$, obtaining samples $\Phi^{(u)}$ for ${u = 1, \ldots, n}$.
This gives an estimate 
$\< \Phi^{2l} \>_M^\text{est} =\frac{1}{n}\sum_{u=1}^n (\Phi^{(u)})^{2l}$
which would converge to $\< \Phi^{2l} \>_M$ in the limit of large $n$.

We repeat this process and average over different measurement outcomes to obtain the full Monte Carlo estimate
\be
\<\Phi^{2l}\>_\text{MC}=\frac{1}{m}\sum_{M} \< \Phi^{2l} \>_M^\text{est} .
\ee
Here $m$ is the number of measurement outcomes considered.

\textit{Timescales.}
In the binary measurement protocol, effective autocorrelation timescales for $\Phi^2$ are estimated as  ${\tilde\tau_{\exp, \Phi^2}\sim 0.6L^{3.1}}$, ${\tilde \tau_{\text{int}, \Phi^2}\sim 0.14L^{3.1}}$.
We have taken ${1/\tilde \tau_{\text{exp}}=\lim \limits_{t \to \infty}-\frac{1}{t}\ln \overline{\rho(t)}}$ and ${\tilde \tau_{\text{int}}}$ is the corresponding integrated autocorrelation time obtained from $\overline{\rho(t)}$. Here, ${\overline{\rho(t)}=\mathbb{E} [\rho(t)_M]}$ where $\rho(t)_M$ is the autocorrelation function given a measurement realization $M$.
(There is considerable variation in the equilibration times between different samples.)
These time-scales are of a similar order for the Gaussian protocol as well.
Note that the timescale for ``even" quantities (invariant under global spin flip of a single replica) is much smaller than that of the ``odd" quantities. 

We take the waiting time to be $T_{\text{eq}}=20\tau_{\exp, \Phi^2}$, and ${t_0=4\tau_{\text{int},\Phi^2}}$. 

\textit{Statistical error.}
Consider an observable $\mathcal O$ which is a function of a pair of replicas, e.g.\ $\Phi^{2l}$ as above. 
Given a measurement $M$, the estimated value is $\<\mathcal O\>_{M}^\text{est}=\<\mathcal O\>_M+\eta_{M}$, where $\<\mathcal O\>_M$ is the true value and $\eta_M \sim \mathcal N(0, \frac{\sigma_M}{\sqrt{n}})$ is the statistical error, which is approximately Gaussian for large $n$. 
Here, $n$ is the number of samples of $\mathcal O$ taken per measurement, and $\sigma_M^2$ is a variance that can depend on $M$.
The full numerical  average is 
\begin{align}
\<\mathcal O\>_\text{MC} & =\frac{1}{m}\sum_{M}\<\mathcal O\>_{M}^\text{est}
 =\frac{1}{m}\sum_M(\<\mathcal O\>_M+\eta_M)
\\
& =\<\mathcal O\>_\text{true}+\mathcal N(0,\frac{\sigma}{\sqrt m})+\mathcal N(0,\frac{\tilde \sigma}{\sqrt{mn}}).
\end{align}
Here $\sigma^2$ is the variance of $\<\mathcal O\>_M$ and $\tilde \sigma^2={\overline{\sigma_M^2}}$. Note that the error vanishes at large $m$ even if $n$ is held fixed.

The statistical errors for $A$ are calculated from the error-propagation method, and the error bars in Fig.~\ref{fig: Binders} show the upper bound of statistical error. 

\textit{Other observables.} Expectation values using larger numbers of replicas (Sec.~\ref{sec:simmethod}) up to ${n=6}$ are computed similarly to the above.
The typical correlator (\ref{eq:typdefn}) cannot be computed straightforwardly using real replicas: it is necessary to take the average, over $M$, of a non-linear function of a conditioned observable

\be
\f{1}{m} \sum_M 
\ln |\<S_i S_j\>_M^\text{est}|.
\ee
In this case, it is important that $n$ is large, or we will obtain a \textit{systematic} error of order $1/n$. In order to control this, we compare the values of this quantity for $n=200$ and $400$, and they have a relative difference of less than $1\%$. In our simulation, we take $n=400$, which ensures a relative error within this order. In all our simulations, we average over more than $2\times 10^4$ measurements.

\begin{figure}[h]
    \fig{0.9}{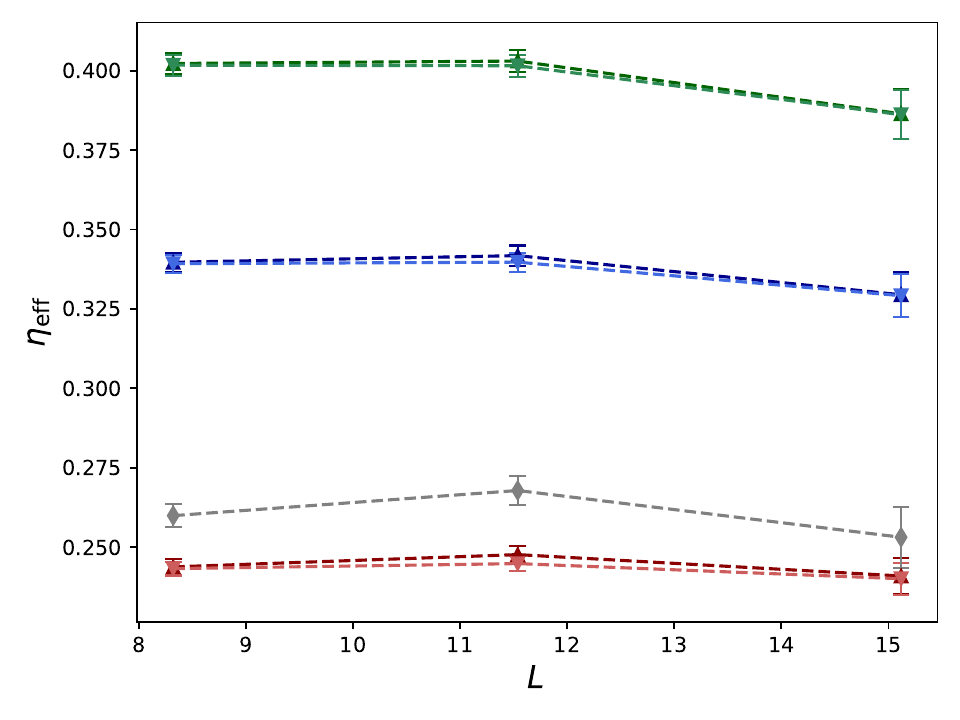}
    \caption{This shows the variation of exponents obtained by fitting three consecutive $ L$ iteratively. The $x$-axis shows $(L_1L_2L_3)^{1/3}$ for those three lengths. The color codes are the same as in Fig.~\ref{fig:Moments}.}
    \label{fig: Exponents_variation}
\end{figure}

\textit{Finite-size error in exponents}.
Fig.~\ref{fig: Exponents_variation} shows the finite size estimates of the multiscaling exponents mentioned in Sec.~\ref{sec:gaussiansim}.

%

\end{document}